\documentclass{lmcs} %


\usepackage{hyperref}
\usepackage{xcolor}
\usepackage{latexsym, amsmath, amsfonts, textcomp, mathrsfs}
\usepackage{graphics,graphicx,color,wrapfig}
\usepackage{stmaryrd, amssymb}%
\usepackage{mathtools}
\usepackage{subcaption}
\usepackage{listings}
\definecolor{vscText}{HTML}{333333}
\definecolor{vscKeyword}{HTML}{267F99}
\definecolor{vscString}{HTML}{006400}
\definecolor{vscFunction}{HTML}{A055A0}
\definecolor{vscType}{HTML}{267F99}
\definecolor{vscNumber}{HTML}{FF8800}
\definecolor{vscComment}{HTML}{008000}
\definecolor{vscMyFunction}{HTML}{795E26}
\lstdefinestyle{vscode}{
  basicstyle=\ttfamily\scriptsize\color{vscText},
  keywordstyle=\color{vscKeyword},
  stringstyle=\color{vscString},
  commentstyle=\color{vscComment}\itshape,
  identifierstyle=\color{vscText},
  emph={FILE,printf,fgetc}, emphstyle=\color{vscFunction},
  emph={[2]count_a}, emphstyle={[2]\color{vscMyFunction}\bfseries},
  numbers=left,
  numberstyle=\tiny\color{gray},
  stepnumber=1,
  numbersep=5pt,
  frame=none,
  tabsize=2,
  showstringspaces=false,
  breaklines=true,
  keepspaces=true,
  aboveskip=4pt,
  belowskip=4pt,
  literate=
    *{0}{{{\color{vscNumber}0}}}1
     {1}{{{\color{vscNumber}1}}}1
     {2}{{{\color{vscNumber}2}}}1
     {3}{{{\color{vscNumber}3}}}1
     {4}{{{\color{vscNumber}4}}}1
     {5}{{{\color{vscNumber}5}}}1
     {6}{{{\color{vscNumber}6}}}1
     {7}{{{\color{vscNumber}7}}}1
     {8}{{{\color{vscNumber}8}}}1
     {9}{{{\color{vscNumber}9}}}1
}
\usepackage{pifont}
\usepackage{proof}
\usepackage{enumitem}
\usepackage{yfonts}
\usepackage[bbgreekl]{mathbbol}
\usepackage{tikz}
\usetikzlibrary{calc,arrows.meta,backgrounds}
\usepackage{nicefrac}
\usepackage{comment}

\def\preord{\preceq} %
\newcommand{\pjoin}{\oplus} %
\newcommand{\Cont}{k\mbox{-}ALip\tuple{\tudelta{\eta}{},\tudelta{\rho}{}}} %
\newcommand{\Contpar}[3]{#1\mbox{-}ALip\tuple{#2,#3}}
\def\+Bound{\pjoin\mbox{-Bound}} %
\def\*Bound{\ast\mbox{-Bound}} %

\newcommand{\quotes}[1]{``#1''} %
\newcommand{\bound}{\uplus}
\newcommand{\dcount}[1]{\delta^{#1}_{\!\scriptstyle siz}} %
\newcommand{\Int}{\ensuremath{\mathsf{Int}}} %
\newcommand{\Sign}{\ensuremath{\mathsf{Sign}}} %
\newcommand{\Par}{\ensuremath{\mathsf{Par}}} %
\newcommand{\Count}{size} %
\newcommand{\dpath}[1]{\delta^{\scriptsize #1}_{\!\scriptstyle pat}} %
\newcommand{\dupath}[1]{\delta^{\scriptsize #1}_{\!\scriptstyle upat}} %
\newcommand{\dincl}[1]{\delta^{\scriptsize #1}_{\!\scriptstyle \subseteq}} %
\newcommand{\deucl}{\tudelta{}{2}} %
\newcommand{\progp}[1]{\mbox{\sf #1}}
\definecolor{darkgreen}{rgb}{0.0, 0.5, 0.0}
\definecolor{darkred}{rgb}{0.55, 0.0, 0.0}
\newcommand{\cmark}{\ding{51}}
\newcommand{\xmark}{\ding{55}}
\newcommand{\cross}{\color{red} \xmark}
\newcommand{\valid}{\color{darkgreen} \cmark}

\newcommand{\li}{\ar@{-}} %
\newcommand{\lp}{\ar@{.}} %
\newcommand{\fp}{\ar@{.>}} %

\makeatletter
\newcommand{\todo}{\@ifnextchar[{\@TODO}{\@@TODO}}
\long\def\@TODO[#1]#2{\bgroup\color{red}\ifmmode\else\itshape\fi TODO[#1]: #2\egroup}
\long\def\@@TODO#1{\bgroup\color{red}\ifmmode\else\itshape\fi TODO: #1\egroup}
\makeatother

\newcommand{\nat}{\mathbb{N}}
\newcommand{\real}{\mathbb{R}}
\newcommand{\raz}{\mathbb{Q}}

\def\Imp{{\mbox{\sf Prog}}}

\newcommand{\uco}{\mbox{\sl uco}}

\newcommand{\id}{ id }

\newcommand{\ra}{\rightarrow}

\newcommand{\Ra}{\Rightarrow}
\newcommand{\La}{\Leftarrow}
\newcommand{\Lra}{\Leftrightarrow}

\def\definit{\mbox{\raisebox{0ex}[1ex][1ex]{$\stackrel{\mbox{\tt\tiny def}}{\; =\;}$}}}
\def\lradef{\mbox{\raisebox{0ex}[1ex][1ex]{$\stackrel{\mbox{\tt\tiny def}}{\; \Longleftrightarrow\;}$}}}

\newcommand{\hide} [1]{}

\def\odd{\mbox{\sl Odd\/}}
\def\even{\mbox{\sl Even\/}}

\def\comp{\mathrel{\hbox{\footnotesize${}\!{\circ}\!{}$\normalsize}}}

\def\defemb#1#2{\expandafter\def\csname #1\endcsname
	{\relax\ifmmode #2\else\hbox{$#2$}\fi}}

\def\2c-math#1#2{{\par\medskip\noindent ${#1}$
		\par\smallskip
		\noindent\hspace*{\fill} ${#2}$}
	\\[10pt]}

\def\cal{\mathcal}
\defemb{aF}{{\cal F}^\cA}
\defemb{sF}{{\cal F}^{\star}}
\defemb{cA}{{\cal A}}
\defemb{cB}{{\cal B}}
\defemb{cC}{{\cal C}}
\defemb{cD}{{\cal D}}
\defemb{cE}{{\cal E}}
\defemb{cF}{{\cal F}}
\defemb{cG}{{\cal G}}
\defemb{cH}{{\cal H}}
\defemb{cI}{{\cal I}}
\defemb{cJ}{{\cal J}}
\defemb{cK}{{\cal K}}
\defemb{cL}{{\cal L}}
\defemb{cM}{{\cal M}}
\defemb{cN}{{\cal N}}
\defemb{cO}{{\cal O}}
\defemb{cP}{{\cal P}}
\defemb{cQ}{{\cal Q}}
\defemb{cR}{{\cal R}}
\defemb{cS}{{\cal S}}
\defemb{cT}{{\cal T}}
\defemb{cU}{{\cal U}}
\defemb{cV}{{\cal V}}
\defemb{cW}{{\cal W}}
\defemb{cY}{{\cal Y}}
\defemb{cX}{{\cal X}}
\defemb{cZ}{{\cal Z}}
\defemb{cHys}{{\cH_{\cY}^{\cS}}}
\defemb{cHy}{{\cH_{\cY}}}
\defemb{ld}{\dashv}
\defemb{rd}{\vdash}
\defemb{cGH}{{\cal GH}}
\defemb{cGWP}{{\cal GWP}}
\defemb{tL}{{\tt L}}
\defemb{tH}{{\tt H}}
\defemb{tT}{{\tt T}}
\defemb{tS}{{\tt S}}
\defemb{tP}{{\tt P}}
\defemb{tQ}{{\tt Q}}
\defemb{tR}{{\tt R}}
\defemb{tB}{{\tt B\:}}
\defemb{tA}{{\tt A\:}}
\newcommand{\tC}{C}
\newcommand{\tAA}{A}

\newcommand{\sset}[2]{\left\{~#1  \left |
	\begin{array}{l}#2\end{array}
	\right.     \right\}}

\newcommand{\false}{\mbox{\texttt{ff}}}
\newcommand{\true}{\mbox{\texttt{tt}}}

\newcommand{\tuple}[1]{\langle #1 \rangle}

\defemb{tD}{{\tt D\:}}

\defemb{tX}{{\tt X}}

\newcommand{\f}{\ar@{->}} %

\def\cal{\mathcal}
\defemb{aF}{{\cal F}^\cA}
\defemb{sF}{{\cal F}^{\star}}
\defemb{cA}{{\cal A}}
\defemb{cB}{{\cal B}}
\defemb{cC}{{\cal C}}
\defemb{cD}{{\cal D}}
\defemb{cE}{{\cal E}}
\defemb{cF}{{\cal F}}
\defemb{cG}{{\cal G}}
\defemb{cH}{{\cal H}}
\defemb{cI}{{\cal I}}
\defemb{cJ}{{\cal J}}
\defemb{cK}{{\cal K}}
\defemb{cL}{{\cal L}}
\defemb{cM}{{\cal M}}
\defemb{cN}{{\cal N}}
\defemb{cO}{{\cal O}}
\defemb{cP}{{\cal P}}
\defemb{cQ}{{\cal Q}}
\defemb{cR}{{\cal R}}
\defemb{cS}{{\cal S}}
\defemb{cT}{{\cal T}}
\defemb{cU}{{\cal U}}
\defemb{cV}{{\cal V}}
\defemb{cW}{{\cal W}}
\defemb{cY}{{\cal Y}}
\defemb{cX}{{\cal X}}
\defemb{cZ}{{\cal Z}}
\defemb{cHys}{{\cH_{\cY}^{\cS}}}
\defemb{cHy}{{\cH_{\cY}}}
\defemb{ld}{\dashv}
\defemb{rd}{\vdash}
\defemb{cGH}{{\cal GH}}
\defemb{cGWP}{{\cal GWP}}
\defemb{tL}{{\tt L}}
\defemb{tH}{{\tt H}}
\defemb{tT}{{\tt T}}
\defemb{tS}{{\tt S}}
\defemb{tP}{{\tt P}}
\defemb{tQ}{{\tt Q}}
\defemb{tR}{{\tt R}}
\defemb{tB}{{\tt B\:}}
\defemb{tA}{{\tt A\:}}

\DeclareSymbolFont{italics}{OT1}{cmr}{m}{it}

\DeclareMathSymbol{a}{\mathalpha}{italics}{"61}
\DeclareMathSymbol{b}{\mathalpha}{italics}{"62}
\DeclareMathSymbol{c}{\mathalpha}{italics}{"63}
\DeclareMathSymbol{d}{\mathalpha}{italics}{"64}
\DeclareMathSymbol{e}{\mathalpha}{italics}{"65}
\DeclareMathSymbol{f}{\mathalpha}{italics}{"66}
\DeclareMathSymbol{g}{\mathalpha}{italics}{"67}
\DeclareMathSymbol{h}{\mathalpha}{italics}{"68}
\DeclareMathSymbol{i}{\mathalpha}{italics}{"69}
\DeclareMathSymbol{j}{\mathalpha}{italics}{"6A}
\DeclareMathSymbol{k}{\mathalpha}{italics}{"6B}
\DeclareMathSymbol{l}{\mathalpha}{italics}{"6C}
\DeclareMathSymbol{m}{\mathalpha}{italics}{"6D}
\DeclareMathSymbol{n}{\mathalpha}{italics}{"6E}
\DeclareMathSymbol{o}{\mathalpha}{italics}{"6F}
\DeclareMathSymbol{p}{\mathalpha}{italics}{"70}
\DeclareMathSymbol{q}{\mathalpha}{italics}{"71}
\DeclareMathSymbol{r}{\mathalpha}{italics}{"72}
\DeclareMathSymbol{s}{\mathalpha}{italics}{"73}
\DeclareMathSymbol{t}{\mathalpha}{italics}{"74}
\DeclareMathSymbol{u}{\mathalpha}{italics}{"75}
\DeclareMathSymbol{v}{\mathalpha}{italics}{"76}
\DeclareMathSymbol{w}{\mathalpha}{italics}{"77}
\DeclareMathSymbol{x}{\mathalpha}{italics}{"78}
\DeclareMathSymbol{y}{\mathalpha}{italics}{"79}
\DeclareMathSymbol{z}{\mathalpha}{italics}{"7A}

\DeclareMathSymbol{A}{\mathalpha}{italics}{"41}
\DeclareMathSymbol{B}{\mathalpha}{italics}{"42}
\DeclareMathSymbol{C}{\mathalpha}{italics}{"43}
\DeclareMathSymbol{D}{\mathalpha}{italics}{"44}
\DeclareMathSymbol{E}{\mathalpha}{italics}{"45}
\DeclareMathSymbol{F}{\mathalpha}{italics}{"46}
\DeclareMathSymbol{G}{\mathalpha}{italics}{"47}
\DeclareMathSymbol{H}{\mathalpha}{italics}{"48}
\DeclareMathSymbol{I}{\mathalpha}{italics}{"49}
\DeclareMathSymbol{J}{\mathalpha}{italics}{"4A}
\DeclareMathSymbol{K}{\mathalpha}{italics}{"4B}
\DeclareMathSymbol{L}{\mathalpha}{italics}{"4C}
\DeclareMathSymbol{M}{\mathalpha}{italics}{"4D}
\DeclareMathSymbol{N}{\mathalpha}{italics}{"4E}
\DeclareMathSymbol{O}{\mathalpha}{italics}{"4F}
\DeclareMathSymbol{P}{\mathalpha}{italics}{"50}
\DeclareMathSymbol{Q}{\mathalpha}{italics}{"51}
\DeclareMathSymbol{R}{\mathalpha}{italics}{"52}
\DeclareMathSymbol{S}{\mathalpha}{italics}{"53}
\DeclareMathSymbol{T}{\mathalpha}{italics}{"54}
\DeclareMathSymbol{U}{\mathalpha}{italics}{"55}
\DeclareMathSymbol{V}{\mathalpha}{italics}{"56}
\DeclareMathSymbol{W}{\mathalpha}{italics}{"57}
\DeclareMathSymbol{X}{\mathalpha}{italics}{"58}
\DeclareMathSymbol{Y}{\mathalpha}{italics}{"59}
\DeclareMathSymbol{Z}{\mathalpha}{italics}{"5A}

\newcommand{\integer}{\mathbb{Z}}

\defemb{tK}{{\tt K}}

\def\Stm{\mbox{\sf Stm}}

\newcommand{\tdelta}[1]{\delta_{\scriptstyle {\tt #1}}}
\newcommand{\tudelta}[2]{\delta^{\scriptstyle #1}_{\scriptstyle {\tt #2}}}

\newcommand{\tpreord}[1]{\preord_{\scriptstyle {\tt #1}}}

\def\prealz{\real_{\geq 0}}

\newcommand{\skipc}{\mbox{\bf skip}}

\newcommand{\ignore}[1]{}

\newcommand{\fffc}{\ar@/_/} %

\definecolor{dkred}{rgb}{.6,0,0}

\renewcommand{\ell}{q}

\def\code{\mbox{\tt c}}

\def\Aexp{{\sf a}}

\def\Bexp{{\sf b}}

\DeclareMathOperator{\imp}{\Imp}

\newcommand{\prog}{\mbox{\sf P}}

\newcommand{\pMem}{\mathbb{M}}
\newcommand{\pmem}{\mathbb{m}}
\newcommand{\pmemSet}{\wp(\pMem)}
\newcommand{\collSem}[1]{\llbracket{#1}\rrbracket}%

\newcommand{\func}[2]{{{#1}\rightarrow{#2}}}
\newcommand{\ofType}[2]{{#1} \colon {#2}}

\newcommand{\sut}{.\;}

\newcommand{\judge}[4]{#1\,\vdash[#2]\,#3\,[#4]}%

\newcommand{\qedex}{\hfill{\scriptstyle \blacksquare}} %
\newcommand{\qeddef}{\hfill{\scriptstyle \blacktriangleleft}} %
\newcommand{\fqeddef}{\tag*{$\blacktriangleleft$}}

\theoremstyle{plain} %

\begin{document}

\title{Abstract Lipschitz Continuity}

\author[M.~Campion]{Marco Campion\lmcsorcid{0000-0002-1099-3494}}[a]
\author[I.~Mastroeni]{Isabella Mastroeni\lmcsorcid{0000-0003-1213-536X}}[b]
\author[M.~Pasqua]{Michele Pasqua\lmcsorcid{0000-0002-9475-4836}}[b]
\author[C.~Urban]{Caterina Urban\lmcsorcid{0000-0002-8127-9642}}[c]

\address{Sorbonne Université, LIP6, Paris, France}	%
\email{marco.campion@lip6.fr}  %

\address{University of Verona, Department of Computer Science, Italy}	%
\email{\{isabella.mastroeni, michele.pasqua\}@univr.it}  %

\address{Inria \& ENS | PSL, Paris, France}	%
\email{caterina.urban@inria.fr}  %

\begin{abstract}
  \noindent We introduce \emph{Abstract Lipschitz Continuity} (ALC), an extensional
  property of functions, in particular of program semantics, requiring that
  bounded differences between properties of two inputs yield proportionally
  bounded differences between properties of the corresponding outputs. ALC
  explicitly discerns between two complementary forms of approximation: \emph{property-based} (or \emph{qualitative})
  approximations, which retain a property of the data and are formalized by upper
  closure operators as in abstract interpretation, and \emph{quantitative}
  approximations, which measure the closeness of data and are formalized by
  pre-metrics, a weakening of metrics that covers the distances arising in program
  analysis. Keeping the two forms of approximation explicit has two main benefits.
  First, it allows us to precisely relate ALC to established extensional properties:
  ALC is a sufficient condition for partial completeness in abstract interpretation,
  for partial abstract non-interference in language-based security and, with a null
  Lipschitz constant, for abstract robustness in machine learning. Second, it supports
  reasoning about ALC of program semantics by induction on the syntax of programs: we present a sound
  deductive system, parametric in the property-based approximations and in the pre-metric of interest,
  that reduces a proof of ALC for a program to proofs for its basic commands and to
  some explicit side conditions, which isolate what makes ALC
  compositional across sequential composition, non-deterministic choice and
  iteration.
\end{abstract}

\maketitle

\section*{Introduction}\label{sec:introduction}
\emph{Lipschitz continuity} (LC) is a fundamental property in mathematical analysis, which provides a strong quantitative guarantee: the output of a Lipschitz continuous function changes at most linearly with respect to changes in its input, the slope being the Lipschitz constant. Its utility is wide-ranging: it is a cornerstone of convergence guarantees in optimization~\cite{bookNesterox,JiT20} and a key tool for certifying the robustness of machine learning models, particularly in adversarial settings~\cite{FazlyabRHMP19,HuangZSKA21,ZhangJH022}. In program analysis, LC provides a measure of \emph{program robustness} for code operating on uncertain inputs~\cite{gulwani10,gulwani12,Gulwani11}, and Lipschitz-style or continuity-based notions of robustness have been studied, e.g., for string transducers~\cite{SamantaDC13,HenzingerOS14}, for numerical programs~\cite{GoubaultP13}, and, as function sensitivity, in type systems~\cite{sensitivity}. In all these settings, however, perturbations are measured on concrete data, through a distance between the values themselves.

In this work, we argue that making the \emph{properties} of data explicit provides a natural and complementary perspective. Indeed, in complex systems such as machine learning models or programs, small changes in the \emph{raw representation} of data may correspond to negligible or irrelevant changes in the \emph{properties of interest}, and, conversely, small changes in the raw data may cause relevant changes in their properties. Moreover, properties, rather than values, are what several established notions are about: static analyses compute properties of program states in an abstract domain, whose imprecision has been measured through distances~\cite{campion2022popl,CampionSAS23,logozzo2009pseudo,PierroW00a}; information-flow policies specify which properties of the inputs may flow to an observer of properties of the outputs~\cite{GM:ANI2018}; and robustness requirements for classifiers may tolerate changes of the output class within a given property~\cite{GMP24}. These notions, however, are either qualitative, as they require properties to be preserved exactly, or, in their partial variants, they only bound the effect of indistinguishable inputs. By reasoning quantitatively on how differences between properties of inputs propagate to differences between properties of outputs, we can capture robustness and stability in a way that more closely aligns with the intended behavior of such systems, with applications in areas ranging from program verification~\cite{Cousot2004Watermarking,GM:ANI2018,rival2020book,Schoepe2015Opacity} to machine learning robustness~\cite{AlhamoudHAG23,GMP24,LiWSS24}.

\begin{figure}[t]
    \begin{minipage}{.58\linewidth}
\begin{lstlisting}[style=vscode, language=C,numbers=none]
<?php

$id = $argv[0];
$q = "SELECT * FROM users WHERE id = $id;";
$result = pg_query($conn, $q);

?>
\end{lstlisting}
    \end{minipage}%
    \begin{minipage}{.42\linewidth}
\begin{lstlisting}[style=vscode, language=C,numbers=none]
int count-a(FILE *F) {
  int c, nc = 0;
  while ((c=fgetc(F)) != EOF) {
    if (c == 'a') nc++;
  }
  return nc;
}
\end{lstlisting}
    \end{minipage}
    \caption{SQL injection (left) and counting pattern program (right).}\label{fig:example}
\end{figure}

For instance, in the context of security, applications such as code injection and file integrity call for an explicit distinction between the closeness of data and their properties.
SQL injection vulnerabilities arise from an unchecked interaction between untrusted input and the execution of a SQL query, which allows an attacker to execute arbitrary queries on the database. A trivial example is given on the left of Figure~\ref{fig:example}: an attacker can dump the whole \texttt{users} table by injecting the value \texttt{"3 OR 1 = 1"} in the unsanitized input \texttt{argv[0]}, which turns the condition of the query \texttt{q} into a tautology. In this scenario, we could require that inputs with a similar number of Boolean operators select a similar number of tuples: a legitimate identifier contains no Boolean operator and selects at most one tuple, whereas the injected value above contains a single operator and selects the whole table. In other words, we aim at bounding the distance between \emph{properties} of the outputs by the distance between \emph{properties} of the inputs, rather than comparing concrete values.

As another example, malware detection tools look for malicious patterns, e.g., signatures or byte sequences, in the files under inspection. A simple detector is given on the right of Figure~\ref{fig:example}: it counts the occurrences of {\em ``a''\/}, standing for a malicious pattern, in a file. Here we could require the detector to be robust. Suppose that the count is observed only up to a fixed granularity, namely through the interval of width $10$ containing it ($0$--$9$, $10$--$19$, \dots), which plays the role of the verdict of the detector. Then, modifying a few bytes of a file should move this interval by a proportionally bounded amount, whereas merely reordering the bytes of the file should not move it at all: an attacker cannot make the verdict switch between distant intervals by tampering with a file only slightly. Again, the relevant distances are between properties of data, namely the content of a file up to the order of its bytes and the interval containing a count (Example~\ref{ex:byte-histogram}). The choice of the property matters as well: the length of a file would not be a suitable input property, since files of the same length, such as {\em ``aaaa''\/} and {\em ``bbbb''\/}, may contain very different numbers of patterns.

In principle, these requirements can be expressed by LC, provided that the property of interest is incorporated into the distance, e.g., by comparing files through their edit distance up to the order of their bytes, rather than through the plain edit distance, which would also accept detectors whose count changes when the bytes of a file are merely reordered. However, such a tailored distance hides the property within the distance itself: the property cannot be combined with a different distance, nor the distance with a different property, and the resulting requirement is hardly comparable with other established notions in the literature, which are typically expressed in terms of abstractions of data.

We introduce \emph{Abstract Lipschitz Continuity} (ALC for short), a property that generalizes LC from data to their properties (Section~\ref{sec:alc}). ALC requires that \emph{differences between the properties} of inputs lead to \emph{proportionally bounded differences between the properties} of the corresponding outputs: the distance between the properties of two outputs is at most $k$ times the distance between the properties of the corresponding inputs, for an abstract Lipschitz constant $k$. In particular, inputs with the same property must be mapped to outputs at distance $0$, as required, e.g., for files that only differ in the order of their bytes in the example above.
Our formalization is grounded in the theory of \emph{abstract interpretation}~\cite{CC77,CC79}. We model \emph{property-based} (or \emph{qualitative}) approximations by \emph{upper closure operators}, which abstract concrete values to the set of values sharing a common property, e.g., all the files with the same bytes, up to their order, or all the strings with the same number of Boolean operators.
We model \emph{quantitative} approximations by \emph{pre-metrics}, a weakening of metrics that only requires each element to be at distance $0$ from itself, and thus covers the distances arising in program analysis, which in general are neither symmetric nor satisfy the triangle inequality. Unlike LC, ALC is parametrized by both forms of approximation and keeps them explicitly apart, which is what allows us to relate ALC to other extensional program properties, and to reason about ALC of programs by induction on their syntax.

Thanks to this explicit separation, we relate ALC of program semantics to other extensional properties that capture input/output relations in comparable yet distinct ways, and show that ALC is a \emph{sufficient condition} for three of them (Section~\ref{sec:alc4prog}). In abstract interpretation, ALC implies (\emph{partial}) \emph{completeness}~\cite{campion2022popl,CampionSAS23,GRS00}: when the distance satisfies mild assumptions, any program whose semantics is ALC w.r.t.\ an abstraction admits a \emph{complete} best correct approximation on it, i.e., an analysis on that abstraction that introduces no imprecision; when ALC is known for an input abstraction different from the one used by the analysis, the imprecision of the analysis is at most $k$ times the largest distance, measured through the former abstraction, between an input and its approximation in the latter, and it is null when the latter is more precise. In language-based security, ALC implies (\emph{partial}) \emph{abstract non-interference}~\cite{CampionMU25,GM:ANI2018}: inputs with the same property yield outputs with the same property, or with properties at a bounded distance. In machine learning, ALC with a null constant implies (\emph{abstract}) \emph{robustness}~\cite{GMP24,MarzMF25} for any perturbation radius, and it is, in general, strictly stronger.

Finally, we propose a sound \emph{deductive system} for proving ALC of programs by induction on their syntax (Section~\ref{sec:proof}), parametric in the input and output abstractions and in the pre-metric of interest. Unlike Lipschitz continuous functions, ALC functions are not closed under composition in general, since the output abstraction of a program need not match the input abstraction of the next one. The rules of the system make explicit the side conditions that recover compositionality: a compatibility, through the distance, between consecutive abstractions for sequential composition, a bound on how joins affect the output distance for non-deterministic choice, and either a continuity condition on the distance or a bound on the number of iterations for Kleene iteration. A proof of ALC for a program is thus reduced to proofs for its basic commands and to these side conditions.

\paragraph{Related Work.} Section~\ref{sec:alc4prog} studies in detail the relationship between ALC and (partial) completeness~\cite{campion2022popl,CampionSAS23,GRS00}, (partial) abstract non-interference~\cite{CampionMU25,goguenmes82,GM:ANI2018}, which generalizes approximate non-interference~\cite{DHW02}, and (abstract) robustness~\cite{GMP24,MarzMF25}.

Our deductive system (Section~\ref{sec:proof}) generalizes the proof system of Chaudhuri et al.~\cite{gulwani10,Gulwani11,gulwani12} for proving the Lipschitz continuity of programs: when the property-based approximations are the identity, it recovers (an extension, due to the $(\textbf{star})$ rule, of) the system in~\cite{gulwani12}. Lipschitz-style robustness has also been studied for string transducers~\cite{SamantaDC13,HenzingerOS14}, w.r.t.\ similarity functions generalizing, e.g., the Hamming and Levenshtein distances, and for numerical programs~\cite{GoubaultP13}, where abstract interpretation is used to bound the discontinuity errors due to unstable tests in finite-precision implementations. Type systems track Lipschitz constants as function sensitivity~\cite{sensitivity}, and de Amorim et al.~\cite{AmorimGHK19} encode relational properties, such as approximate differential privacy, as sensitivity w.r.t.\ suitable distances, possibly violating the triangle inequality. In all these works, perturbations are measured directly on concrete values, whereas ALC measures them between properties of data, formalized by upper closure operators.

The notion of software doping, technically introduced in~\cite{DArgenioBBFH17}, pursues an effect similar to ALC, with the aim of detecting programs that intentionally deviate from their expected behavior. The authors propose a Lipschitz-style condition for defining software doping which is tailored to the Hausdorff lifting. This is then further generalized via a functional modulus of continuity. Their work, like ours, discerns a property of interest, expressed through the Hausdorff lifting, while, in our approach, we discern a property that can be formalized as a closure operator (or Galois connection).
Other potential related works are cited in Section~\ref{sec:concl}, since a better understanding of the relation with ALC deserves further research.

\paragraph{Structure of the Paper.} The rest of the paper is organized as follows.
\begin{itemize}
    \item Section~\ref{sec:background} introduces the programming language of interest and its semantics, and recalls property-based approximations via upper closure operators, quantitative approximations via pre-metrics, and their combination.
    \item Section~\ref{sec:alc} defines Lipschitz continuity over pre-metric spaces and ALC, and illustrates ALC on examples.
    \item Section~\ref{sec:alc4prog} shows that ALC is a sufficient condition for partial completeness, partial abstract non-interference and abstract robustness.
    \item Section~\ref{sec:proof} presents the deductive system for proving ALC of programs and proves it sound.
    \item Section~\ref{sec:concl} concludes and discusses future work.
\end{itemize}

\medskip
This article is an extended version of~\cite{CMPU2026}. With respect to the conference version, the whole presentation has been thoroughly revised and extended, with many new examples illustrating the main results, the complete proofs of all the results, and the correction of some statements and examples. A new figure, Figure~\ref{fig:relations}, summarizes the relations between ALC and the other extensional properties established in Section~\ref{sec:alc4prog}. Moreover, Section~\ref{sec:alc4prog} contains two groups of new results. First, we relate ALC to partial completeness and partial abstract non-interference when the latter are expressed w.r.t.\ an input abstraction different from the one for which ALC is known, as is often the case when the abstraction of interest is dictated by an existing analysis or by a security policy.
Second, we show that $0$-ALC is equivalent to abstract robustness when the latter is required for every perturbation radius, whereas it is, in general, strictly stronger than abstract robustness for a fixed radius.

\section{Background}\label{sec:background}

This section fixes the notation and recalls the notions on which the rest of the paper relies. We first introduce the programming language of interest and its semantics (Section~\ref{sec:prog}), which plays the role of the function whose continuity properties we study. We then recall the two forms of approximation that ALC explicitly keeps apart: \emph{property-based} (or \emph{qualitative}) approximations (Section~\ref{sec:uco}), formalized by upper closure operators as customary in abstract interpretation, and \emph{quantitative} approximations (Section~\ref{sec:pre-metric}), formalized by pre-metrics. Finally, we show how the two can be combined into a single distance between properties, and which axioms of the underlying distance survive the combination (Section~\ref{sec:combining}).

\subsection{Program Syntax and Semantics}\label{sec:prog}
We consider programs written in the language $\imp$ of regular commands~\cite{BruniGGR23,ohearn20}:
\begin{align*}
            \imp \ni \prog ::=& \; \code \mid \prog \,;\, \prog \mid \prog \pjoin \prog \mid \prog^{\ast}
\end{align*}
which is general enough to cover deterministic imperative languages as well as other programming paradigms that include, e.g.,
non-deterministic computations (and, with a suitable semantics, probabilistic ones).
The term $\prog \,;\, \prog$ denotes sequential composition, $\prog \pjoin \prog$ non-deterministic choice and $\prog^{\ast}$ is the Kleene closure where  $\prog$ can be executed any finite number of times (including zero).
For our purposes, we instantiate the basic commands $\code \in \Stm$ with the following operations:
\begin{align*}
    \Stm \ni \code ::=& \; \skipc \mid x:=\Aexp \mid \Bexp?
\end{align*}
Informally, these correspond to: the no-operation, variable assignments, and Boolean tests, respectively.
We assume a standard grammar for arithmetic expressions in $\mathsf{AExp}$ and Boolean expressions in $\mathsf{BExp}$, defined as follows:
\begin{align*}
    \mathsf{AExp} \ni \Aexp  ::=& \; x \in \mathbb{X} \mid n \in \mathbb{V} \mid \Aexp + \Aexp \mid \Aexp -\Aexp \mid \Aexp*\Aexp\\
	\mathsf{BExp} \ni \Bexp ::=& \; \true \mid \false \mid \Aexp=\Aexp \mid \Aexp < \Aexp \mid \Aexp \leq \Aexp \mid \Aexp > \Aexp \mid \Aexp \geq \Aexp \mid \Bexp \vee \Bexp \mid \neg\Bexp
\end{align*}
Variables range over a denumerable set $\mathbb{X}$ while values range over a denumerable set $\mathbb{V}$ (e.g., integer or natural numbers).

As an example, a standard deterministic imperative language (like $\texttt{Imp}$ in~\cite{winskel}) can be defined using guarded branching and loop commands as syntactic sugar as follows~\cite{Kozen97}:
\begin{align*}
    \progp{if}~\Bexp~\progp{then}~\prog_1~\progp{else}~\prog_2 &\;\definit\; (\Bexp ?; \prog_1) \pjoin (\neg \Bexp ?; \prog_2) \\
    \progp{while}~\Bexp~\progp{do}~\prog_1 &\;\definit\; (\Bexp ? ; \prog_1)^\ast ; \neg \Bexp ?
\end{align*}

\begin{figure}[t]
		\centering
		\begin{align*}
			\collSem{\prog_1 \,;\, \prog_2}c &\smash{\definit} \collSem{\prog_2} \comp \collSem{\prog_1}c \\
			\collSem{\prog_1 \pjoin \prog_2}c &\smash{\definit} \collSem{\prog_1}c \mathrel{\vee} \collSem{\prog_2}c \\
			\collSem{\prog^{\ast}}c &\smash{\definit} \textstyle\bigvee \{\collSem{\prog}^n c ~|~ n \in \mathbb{N}\}
		\end{align*}
	\caption{Semantics of $\imp$.}\label{fig:syntax-semantics}
\end{figure}

Suppose we have a semantics $\ofType{\collSem{\code}}{\func{C}{C}}$ for basic commands $\code \in \Stm$ on a complete lattice $\tuple{C, \tpreord{}, \vee, \land ,\top ,\bot}$,
where $\tpreord{}$ is the partial order (i.e., a reflexive, antisymmetric and transitive binary relation), $\vee$ is the least upper bound, $\land$ the greatest lower bound, $\top$ is the supremum of $C$ and $\bot$ is the infimum of $C$.
Then, the \emph{semantics for programs} $\ofType{\collSem{\cdot}}{\func{\imp}{\func{C}{C}}}$, also called the \emph{concrete semantics},
is inductively defined on program syntax as shown in Figure~\ref{fig:syntax-semantics}, where $\circ$ denotes function composition ($f \circ g \equiv \lambda x.\, f(g(x))$), and $f^n$ for natural $n$ denotes the composition of $f$ $n$-times (formally, $f^0(x) \definit x$, $f^{n+1}(x) \definit f(f^{n}(x)) $).

A notable instance of such concrete semantics is the \emph{collecting semantics}, central in abstract interpretation~\cite{CC77,CC79}, defined over a domain of program memories $\pMem$. The aim of the collecting semantics consists in expressing the set of all possible program states that could occur at some program point.
Formally, consider the complete lattice $\tuple{\pmemSet, \subseteq, \cup, \cap , \pMem , \varnothing}$ of program memories,
where $\pmem \in \pMem$ maps variables to values, namely $\ofType{\pmem}{\func{\mathbb{X}}{\mathbb{V}}}$, and $\varnothing$ is the empty set.
We can define a collecting big-step semantics $\ofType{\collSem{\prog}}{\func{\pmemSet}{\pmemSet}}$ for a program $\prog \in \imp$
as the standard predicate transformer semantics on sets of program memories $\wp(\pMem)$, collecting all possible program memories that reach \emph{the end} of the program.
Assume a big-step evaluation semantics $\mathrel{\Downarrow_\Aexp}$ for arithmetic expressions and $\mathrel{\Downarrow_\Bexp}$
for Boolean expressions. Given an input $S \in \pmemSet$, namely a set of program memories, the semantics of basic commands is defined as follows:
\begin{align*}
    \collSem{\skipc}S &\definit S \\
    \collSem{x:=\Aexp}S &\definit \{ \pmem[x \mapsfrom v] ~|~ \pmem \in S \wedge \pmem \mathrel{\Downarrow_\Aexp}\! v \} \\
    \collSem{\Bexp?}S &\definit \{ \pmem \in S ~|~ \pmem \mathrel{\Downarrow_\Bexp}\! \true \}
\end{align*}
It is easy to note that the collecting semantics for basic commands is monotone (i.e., for $S_1, S_2 \in \pmemSet$ such that $S_1 \subseteq S_2$, it holds that $\collSem{\code}S_1 \subseteq \collSem{\code}S_2$) by construction on the powerset lattice of program memories, and so $\collSem{\prog}$ is also monotone for any program in $\imp$ since both composition and the join operations preserve monotonicity.
\begin{exa} \label{ex:R}
    Let us consider the following program
    \[
         \progp{R} = (x>0? \,;\, x:=x-1) \pjoin (x\leq 0? \,;\, x:=x+1)
    \]
    which decrements all positive values by $1$ and increments all non-positive values by $1$.
    For simplicity, here and in all the examples of the paper, we assume that $x$ is the only program variable, so that program memories can be identified with integer values and $\pmemSet \cong \wp(\integer)$.
    Let us consider the Kleene closure of $\progp{R}$, $\progp{R}^\ast$, and its collecting semantics $\collSem{\progp{R}^\ast} : \wp(\integer) \ra \wp(\integer)$, which collects all the values reachable from the input set $S \subseteq \integer$ by iterating $\progp{R}$ any finite number of times, i.e., $\collSem{\progp{R}^\ast} S = \bigcup_{n \in \nat} \collSem{\progp{R}}^n S$. Every value $s > 0$ generates the descending sequence $s, s-1, \dots, 1, 0$, after which the computation alternates between $0$ and $1$, while every value $s \leq 0$ generates the ascending sequence $s, s+1, \dots, 0, 1$. Hence, for any non-empty $S \subseteq \integer$, the collecting semantics of $\progp{R}^\ast$ is given by:
    \[
        \collSem{\progp{R}^\ast} S = \{ z \in \integer \mid \min(\inf S, 0) \leq z \leq \max(\sup S, 1) \}
    \]
    where $\inf S \in \integer \cup \{-\infty\}$ and $\sup S \in \integer \cup \{+\infty\}$, while $\collSem{\progp{R}^\ast} \varnothing = \varnothing$. For instance, $\collSem{\progp{R}^\ast} \{3\} = \{0,1,2,3\}$ and $\collSem{\progp{R}^\ast} \{-2,2\} = \{-2,-1,0,1,2\}$. Note that the output is always a set of consecutive integers (an interval), determined only by the infimum and the supremum of the input set.
    $\qedex$
\end{exa}
We will extensively use this semantics in the examples of Section~\ref{sec:alc4prog} and~\ref{sec:proof}.

\subsection{Abstractions and Distances}
In program analysis approximations are fundamental for simplifying reasoning while preserving essential properties. Following~\cite{CampionMU25}, we consider \emph{qualitative} or \emph{property-based} approximations, formalized by \emph{upper closure operators}, and \emph{quantitative} approximations, formalized by \emph{pre-metrics}. Their combination yields a general approximation parametrized by both an abstraction and a distance.

\subsubsection{Property-based Approximations via Upper Closure Operators.} \label{sec:uco}
Qualitative or property-based approximations preserve certain \emph{properties} of the approximated data.
Property-based approximations are at the heart of \emph{abstract interpretation}~\cite{CC77,CC79}, which offers a general methodology for approximating computations by evaluating functions (e.g., program semantics) over an abstract domain $\tAA$ instead of the concrete domain $\tC$.
This approach is especially valuable when exact analysis is computationally infeasible or undecidable, trading precision for decidability.
In this setting, $\tAA$ is referred to as an \emph{abstraction} of $\tC$ whenever there is a
\emph{Galois Connection} (GC), or a \emph{Galois Insertion} (GI), between the two domains. More formally, given a partially ordered set (poset, for short) $\tuple{\tC,\tpreord{\tC}}$, called the \emph{concrete domain}, and a poset $\tuple{\tAA,\tpreord{\tAA}}$, called the \emph{abstract domain,} a GC is a pair $\tuple{\alpha,\gamma}$ where $\alpha \colon \tC \to \tAA$ is the (monotone) abstraction function, relating concrete elements to their abstract counterparts, sometimes referred to as the lower adjoint, and $\gamma \colon \tAA \to \tC$ is the (monotone) concretization function, relating abstract elements to their concrete counterparts, also referred to as the upper adjoint, both satisfying the following condition:
\[
    \forall a \in \tAA\sut \forall c \in \tC \sut \alpha(c) \tpreord{\tAA} a \Lra c \tpreord{\tC} \gamma(a).
\]
A GC is a GI when $\alpha \circ \gamma = \id$, namely when composing the abstraction function with the concretization gives rise to the identity function $\id \definit \lambda x.\, x$.

GIs, which are the focus of this paper, can be equivalently formulated in terms of \emph{upper closure operators}.
\begin{defi}[Upper Closure Operator~\cite{CC79}] \label{def:uco}
    Let $\tuple{\tC,\tpreord{\tC}}$ be a poset. A function $\rho \colon \tC \to \tC$ is an \emph{upper closure operator} (uco or closure, for short) when it satisfies the following properties $\forall c, c' \in \tC$:
    \begin{itemize}[itemsep=5pt,leftmargin=40mm,topsep=8pt]
        \item[({\textit{monotonicity}})] $c \tpreord{\tC} c' \Rightarrow \rho(c) \tpreord{\tC} \rho(c')$;
        \item[({\textit{extensivity}})] $c \tpreord{\tC} \rho(c)$;
        \item[({\textit{idempotence}})] $\rho(\rho(c)) = \rho(c)$. $\qeddef$
\end{itemize}
\end{defi}
Ucos are uniquely determined by the set of their fixpoints: $\rho(C) = \{c \in \tC \mid \rho(c) = c\}$.
For instance, the composition $\gamma \comp \alpha$ is an uco of $\tC$.
The set of all ucos on a poset $\tC$ is denoted by $\uco(\tC)$. An uco $\rho \in \uco(\tC)$ over a complete lattice $\tuple{\tC, \tpreord{}}$ is said to be \emph{additive} when it preserves arbitrary joins, i.e., $\rho(\bigvee S) = \bigvee_{c \in S} \rho(c)$ for all $S \subseteq \tC$.
\begin{figure}[t]
    \centering
    \begin{tikzpicture}[scale=1, every node/.style={scale=1},
    elem/.style={inner sep=1.5pt},
    edge/.style={line width=.6pt, gray!80!black}]
    \node[elem] (bot)  at (0,0)      {$\varnothing$};
    \node[elem] (neg)  at (-2.2,1.3) {$\integer_{<0}$};
    \node[elem] (zero) at (0,1.3)    {$\{0\}$};
    \node[elem] (pos)  at (2.2,1.3)  {$\integer_{>0}$};
    \node[elem] (leq)  at (-2.2,2.6) {$\integer_{\leq 0}$};
    \node[elem] (neq)  at (0,2.6)    {$\integer_{\neq 0}$};
    \node[elem] (geq)  at (2.2,2.6)  {$\integer_{\geq 0}$};
    \node[elem] (top)  at (0,3.9)    {$\integer$};
    \draw[edge] (bot) -- (neg);  \draw[edge] (bot) -- (zero);  \draw[edge] (bot) -- (pos);
    \draw[edge] (neg) -- (leq);  \draw[edge] (neg) -- (neq);
    \draw[edge] (zero) -- (leq); \draw[edge] (zero) -- (geq);
    \draw[edge] (pos) -- (neq);  \draw[edge] (pos) -- (geq);
    \draw[edge] (leq) -- (top);  \draw[edge] (neq) -- (top);  \draw[edge] (geq) -- (top);
\end{tikzpicture}
    \caption{Hasse diagram of the domain of signs $\Sign(\wp(\integer))$, ordered by inclusion.}
    \label{fig:sign}
\end{figure}
\begin{exa}[Sign Closure] \label{ex:sign}
    As an example, the closure $\Sign \in \uco(\wp(\integer))$ abstracts a set of integers by only recording the signs of its elements, i.e., whether the set contains negative values, the value $0$ and positive values. Formally, let $\integer_{<0}$, $\{0\}$ and $\integer_{>0}$ be the three \emph{blocks} of the partition of $\integer$ according to the sign; then $\Sign(S)$ is the union of the blocks that intersect $S$:
    \[
        \Sign(S) \definit \textstyle\bigcup \{ B \in \{\integer_{<0}, \{0\}, \integer_{>0}\} \mid B \cap S \neq \varnothing \}.
    \]
    For instance, $\Sign(\{-3,-1\}) = \integer_{<0}$, $\Sign(\{0,4\}) = \integer_{\geq 0}$ and $\Sign(\{-1,1\}) = \integer_{\neq 0}$. The set of fixpoints of $\Sign$ consists of all the unions of blocks:
    \[
        \Sign(\wp(\integer)) = \{\varnothing, \integer_{<0}, \{0\}, \integer_{>0}, \integer_{\leq 0}, \integer_{\neq 0}, \integer_{\geq 0}, \integer\},
    \]
    that is, the classical domain of signs~\cite{CC79}, whose Hasse diagram is depicted in Figure~\ref{fig:sign}.
    Note that $\Sign$ is additive since it is defined element-wise.
    $\qedex$
\end{exa}
Henceforth, we formally model property-based approximation through ucos:
\begin{defi}[Property-based Approximation~\cite{CampionMU25}] \label{def:semapp}
    Given a poset $\tuple{\tC,\tpreord{\tC}}$ and an uco $\rho \in \uco(\tC)$, an element $x \in \tC$ is \emph{approximated (or abstracted) by the property} $\rho(x)$. The set $\{y \in \tC \mid \rho(y) = \rho(x) \}$ represents all elements in $\tC$ sharing the same property as $x$ according to $\rho$.
    $\qeddef$
\end{defi}
\begin{exa}[Interval Approximation] \label{ex:qual}
    Let $\Int \in \uco(\wp(\integer))$ be the interval approximation~\cite{CC76}, mapping
    a set of integers $S \in \wp(\integer)$ to the smallest interval
    \[
        [l,u] \definit \{i \in \integer \mid l \leq i \leq u \}, \qquad l \in \integer \cup \{-\infty\}, \qquad u \in \integer \cup \{+\infty\}, \qquad l \leq u
    \]
    such that $S \subseteq [l,u]$. Explicitly, $\Int(S) \definit [\inf S, \sup S]$ for every non-empty $S \in \wp(\integer)$, where the infimum and the supremum are taken in $\integer \cup \{-\infty,+\infty\}$, and $\Int(\varnothing) \definit \varnothing$ (the empty interval).
    For instance, the set of integers $\{0,1,4\}$ can be approximated by the interval $[0,4]$ through the interval approximation $\Int$. Moreover, the set
    \[
        \{\{0,4\}, \{0,1,4\}, \{0,2,4\}, \{0,3,4\}, \{0,1,2,4\}, \{0,2,3,4\}, \{0,1,3,4\}, \{0,1,2,3,4\}\}
    \]
     contains all sets of integers $S$ such that $\Int(S) = [0,4]$.
     Note that $\Int$ is not additive since the smallest interval containing a union may contain elements that belong to neither of the smallest intervals containing the two sets: for instance, $\Int(\{0\} \cup \{4\}) = \{0,1,2,3,4\}$, whereas $\Int(\{0\}) \cup \Int(\{4\}) = \{0,4\}$.
     $\qedex$
\end{exa}
\noindent
In the following, we will often abbreviate with $\rho f$ the composition of any two functions $\rho \circ f$.

\subsubsection{Quantitative Approximations via Pre-Metrics.} \label{sec:pre-metric}
Quantitative approximations preserve \emph{closeness} of the approximated data, where here the meaning of \quotes{closeness} is
measured through a distance function between elements of a set. We model distance functions using \emph{pre-metrics}.
Let $\real^\infty \definit \real \cup \{\infty\}$ be the set of real numbers extended with the infinity symbol $\infty$,
such that for all $r \in \real$, $r < \infty$, and
let $\real_{\geq 0}$ be the restriction of $\real$ to values greater or equal to $0$.
Hence, $\prealz^\infty \definit \{ r \in \real \mid r \geq 0 \} \cup \{\infty\}$ and, similarly, $\nat^\infty \definit \nat \cup \{\infty\}$.
We extend the arithmetic operations of $\real_{\geq 0}$ to $\prealz^\infty$ by letting $r + \infty = \infty + r = \infty$ and $r \cdot \infty = \infty \cdot r = \infty$ for all $r \in \prealz^\infty$, including $r = 0$.\footnote{Distances with infinite values are customary in metric geometry, where a metric space with infinite distances canonically splits into subspaces at finite distance, separated from one another by infinite distances~\cite[Section~1.1]{BuragoBI01}, as well as in the theory of generalized metric spaces of Lawvere~\cite{Lawvere73}, which take values in $[0,\infty]$. In this setting, Lipschitz continuity is defined by the same inequality as in the finite case~\cite[Definition~1.4.1]{BuragoBI01}, so that, for every constant $k > 0$, two elements at infinite distance impose no constraint on a Lipschitz continuous function. The convention $0 \cdot \infty = \infty$, rather than the measure-theoretic $0 \cdot \infty = 0$, extends this reading to the degenerate constant $k = 0$, consistently with the definition of the least Lipschitz constant as a supremum of ratios $\nicefrac{\delta(f(x),f(y))}{\delta(x,y)}$, where $\nicefrac{r}{\infty} = 0$ (cf.\ Section~\ref{sec:alc}).}
\begin{defi}[Pre-Metric~\cite{CampionSAS23}] \label{def:pre-metric}
    Given a non-empty set $L$, a \emph{pre-metric} is a binary function $\delta \colon L\times L \to \prealz^\infty$ satisfying the only axiom $\forall x,y \in L$:
    \begin{itemize}[itemsep=5pt,leftmargin=40mm,topsep=8pt]
        \item[(\textit{if-identity})] $x=y \;\Ra\; \delta(x,y)=0$.
    \end{itemize}
    The pair $\tuple{L,\delta}$ is called a \emph{pre-metric space}. $\qeddef$
\end{defi}
\begin{figure}[t]
    \centering
    \resizebox{\textwidth}{!}{%
    \begin{tabular}{rcccccccc}
     & \hspace*{5pt}\textbf{pre-}\hspace*{5pt} & \hspace*{2pt}\textbf{quasipseudo-}\hspace*{2pt} & \hspace*{2pt}\textbf{pseudosemi-}\hspace*{2pt} & \hspace*{2pt}\textbf{quasisemi-}\hspace*{2pt} & \hspace*{2pt}\textbf{semi-}\hspace*{2pt} & \hspace*{2pt}\textbf{quasi-}\hspace*{2pt} & \hspace*{2pt}\textbf{pseudo-}\hspace*{2pt} & \hspace*{5pt}\textbf{metric}\hspace*{5pt} \\[.75em]
    (\textit{if-identity}) & \valid & \valid & \valid & \valid & \valid & \valid & \valid & \valid  \\
    (\textit{iff-identity}) & \cross & \cross & \cross & \valid & \valid & \valid & \cross & \valid  \\
    (\textit{symmetry}) & \cross & \cross & \valid & \cross & \valid & \cross & \valid & \valid  \\
    (\textit{triangle-inequality}) & \cross & \valid & \cross & \cross & \cross & \valid & \valid & \valid  \\[.75em]
    Example & & $\dincl{\Int}$ & & &  & $\dincl{}, \dpath{}$ & $\dcount{}$ & $\deucl, \delta_{\mathrm{H}}$  \\[.5em]
    Reference & & Ex.~\ref{ex:inclusion-dist-combined} & & &  & Ex.~\ref{ex:inclusion-dist},~\ref{ex:path-len} & Ex.~\ref{ex:counting} & Ex.~\ref{ex:scale-star}
    \end{tabular}%
    }
    \caption{Metrics and their weakenings.}
    \label{fig:pre-metrics}
\end{figure}

We will occasionally use the subscript $\tdelta{L}$ in cases where the set $L$ may not be immediately clear from the context.
The same convention will be adopted for orderings $\tpreord{}$.
Due to the only requirement, namely the (\textit{if-identity}) axiom, pre-metrics are among the least restrictive formal notions of distance. However, depending on what we want to measure and on which domain, a distance function may satisfy more axioms. For instance, \emph{metrics} could be considered as a very restrictive formal notion of distance, since they further satisfy the following axioms $\forall x,y,z \in L$:
\begin{itemize}[itemsep=5pt,leftmargin=40mm,topsep=8pt]
    \item[({\textit{iff-identity}})] $x=y \;\Lra\; \delta(x,y)=0$;
    \item[({\textit{symmetry}})] $\delta(x,y)=\delta(y,x)$;
    \item[({\textit{triangle-inequality}})] $\delta(x,y) \leq \delta(x,z) + \delta(z,y)$.
\end{itemize}

A classic example of metric is the Euclidean distance $\deucl(x,y) \definit |x-y|$ measuring the distance between two real values $x,y \in \real$ as the absolute value of their difference.

A distance that is not required to satisfy (\textit{symmetry}) is called a \emph{quasi-metric}, while a distance that is not required to satisfy the $\La$ implication of (\textit{iff-identity}) is called a \emph{pseudo-metric}. \emph{Semi-metrics} are not required to satisfy the (\textit{triangle-inequality}). By composing the words pseudo-, quasi- and semi- we obtain different distance flavors by simply keeping the axioms that are satisfied by all the combined words. For instance, a quasisemi-metric is a pre-metric that additionally satisfies the (\textit{iff-identity}), while a pseudosemi-metric only satisfies (\textit{symmetry}) besides (\textit{if-identity}). Note that all these classes are inclusive: e.g., every metric is also a quasi-metric and a pseudo-metric.\footnote{The terminology is not uniform in the literature. We follow the classical one of~\cite{Wilson31} for semi-metrics, i.e., distances not required to satisfy the (\textit{triangle-inequality}); other authors (e.g.,~\cite{DezaDeza}) use the word semi-metric for what we call pseudo-metric, and quasi-semi-metric (or hemi-metric) for what we call quasipseudo-metric.} Figure~\ref{fig:pre-metrics} summarizes the above distance notions and their properties. The last two rows display the distance symbol and the example in which the distance is defined and used for the first time.
As we will see in the next sections, understanding the type of distance function we are manipulating is essential for proving some implications between properties of programs (cf. Section~\ref{sec:alc4prog}).

From this point forward, whenever we say that a function $\delta$ is a distance, we assume that it satisfies, at least, the axiom of a pre-metric.

Definition~\ref{def:pre-metric} is general enough to be instantiated with several practical notions of distance proposed in the abstract interpretation literature (e.g.,~\cite{campion2022popl,LiewCL24,logozzo2009pseudo,PierroW00a,sotin:inria-00457324}). The following examples illustrate two of these notions, used extensively throughout the rest of the paper. For additional examples of pre-metrics and their applications in domains used within the context of program analysis, we refer to~\cite{CampionSAS23}.

\begin{exa}[Size Distance] \label{ex:counting}
    Consider the powerset $\wp(L)$ of a set $L$. We write $\Count(S) \in \nat^\infty$ for the number of elements in $S \in \wp(L)$, which is $\infty$ when $S$ is infinite.
    We define the size distance
    \[
        \dcount{}\colon \wp(L) \times \wp(L) \ra \prealz^\infty
    \]
    between two sets $S_1, S_2 \in \wp(L)$ as follows:
    \[
        \dcount{}(S_1,S_2) \;\definit\;
        \left\{
        \begin{array}{ll}
          0 & \mbox{if $S_1 = S_2$,}\\
          |\Count(S_2) - \Count(S_1)| & \mbox{otherwise}
        \end{array}
        \right.
    \]
    where we assume $|\infty - \infty| = \infty$. In other words, $\dcount{}$ calculates the absolute value of the difference in their size.
    Note that $\dcount{}$ is a pseudo-metric since it does not satisfy the (\textit{iff-identity}) axiom:
    two sets may have the same size yet be different.
    In program analysis, $\dcount{}$ could be used to count, for instance, the number of spurious elements added by an abstract computation with respect to the abstraction of a concrete computation.
    For instance, if $[0,0]$ is the (interval abstraction of the) strongest numerical invariant of a program variable $x$ at certain program point, while $[0,10]$ is the abstract invariant generated by an abstract interpretation over $\Int$, then $\dcount{}(\{0\},\{0,1,\dots , 10\}) = 10$ indicates that the abstract interpretation added $10$ more spurious values with respect to the (interval abstraction of the) concrete execution.
    $\qedex$
\end{exa}

\begin{exa}[Inclusion Distance] \label{ex:inclusion-dist}
    Given the poset $\tuple{\wp(L),\subseteq}$, suppose we are interested in defining a distance that calculates the number of more elements one set $S_2 \in \wp(L)$ has with respect to another $S_1 \in \wp(L)$, but only in the case $S_1 \subseteq S_2$, otherwise it simply returns $\infty$.
    Formally, we define the inclusion distance
    \[
        \dincl{} \colon \wp(L) \times \wp(L) \ra \nat^\infty
    \]
    between two sets $S_1, S_2 \in \wp(L)$ as follows:
    \[
        \dincl{}(S_1,S_2) \;\definit\;
        \left\{
        \begin{array}{ll}
          \Count(S_2 \smallsetminus S_1) & \mbox{if $S_1 \subseteq S_2$,}\\
          \infty & \mbox{otherwise}
        \end{array}
        \right.
    \]
    For instance, over $\wp(\integer)$, $\dincl{}(\{0,1,4\}, \{0,1,4,10\}) = 1$ because $\{0,1,4,10\}$ has one more integer than $\{0,1,4\}$ namely the number $10$,
    while $\dincl{}(\{0,1,4\}, \{1,4,10\}) = \infty$ because $\{0,1,4\} \not\subseteq \{1,4,10\}$.
    This is another distance, like $\dcount{}$, that could be used in program analysis to count the number of spurious elements added by an abstract computation with respect to the abstraction of the concrete execution.
    Note that $\dincl{}$ may differ from $\dcount{}$ even between comparable sets: $\dincl{}(\integer_{>0}, \integer_{\geq 0}) = 1 \neq \infty = \dcount{}(\integer_{>0}, \integer_{\geq 0})$.
    The pair $\tuple{\wp(L),\dincl{}}$ forms a quasi-metric space: (\textit{iff-identity}) holds since $\dincl{}(S_1,S_2) = 0$ if and only if $S_1 \subseteq S_2$ and $S_2 \smallsetminus S_1 = \varnothing$, i.e., $S_1 = S_2$, while (\textit{triangle-inequality}) holds since $\Count(S_3 \smallsetminus S_1) = \Count(S_3 \smallsetminus S_2) + \Count(S_2 \smallsetminus S_1)$ whenever $S_1 \subseteq S_2 \subseteq S_3$, and the right-hand side is $\infty$ otherwise. It is not a metric space because $\dincl{}$ does not satisfy (\textit{symmetry}). %
    $\qedex$
\end{exa}

\begin{exa}[Path-Length Distance] \label{ex:path-len}
    Counting elements, as $\dincl{}$ does, is not informative on abstract domains whose elements are infinite sets. For instance, in the domain of signs $\Sign(\wp(\integer))$ of Example~\ref{ex:sign} we have $\dincl{}(\integer_{>0},\integer_{\neq 0}) = \infty$, although $\integer_{\neq 0}$ is an immediate successor of $\integer_{>0}$ in Figure~\ref{fig:sign}. The \emph{path-length distance}~\cite{CampionSAS23} measures instead how many steps apart two elements are in the Hasse diagram of a poset. Given a poset $\tuple{L,\tpreord{L}}$, we define
    \[
        \dpath{} \colon L \times L \ra \nat^\infty
    \]
    by letting $\dpath{}(x,y)$ be the length of a shortest path from $x$ to $y$ in the Hasse diagram of $\tuple{L,\tpreord{L}}$ that only traverses edges upwards, i.e., the least $n \in \nat$ such that $x = z_0 \prec z_1 \prec \dots \prec z_n = y$, where each $z_{i+1}$ covers $z_i$ (namely $z_i \prec z_{i+1}$ and no element lies strictly between them), and $\dpath{}(x,y) \definit \infty$ when no such path exists, in particular when $x \not\tpreord{L} y$. This is the directed version, with unit weights, of the path-length distance of~\cite{CampionSAS23}, where paths can be traversed in both directions and edges may carry arbitrary weights. On the poset $\tuple{\wp(L),\subseteq}$, where each step adds a single element, $\dpath{}$ coincides with the inclusion distance $\dincl{}$ of Example~\ref{ex:inclusion-dist}. On the domain of signs $\Sign(\wp(\integer))$, ordered by inclusion, $\dpath{}(x,y)$ counts instead the number of blocks that have to be added to $x$ in order to obtain $y$: for instance, $\dpath{}(\integer_{>0},\integer_{\neq 0}) = 1$, $\dpath{}(\{0\},\integer) = 2$ and $\dpath{}(\varnothing,\integer) = 3$, whereas $\dpath{}(\integer_{<0},\integer_{\geq 0}) = \infty$ and $\dpath{}(\integer,\{0\}) = \infty$.
    For every poset $\tuple{L,\tpreord{L}}$, the pair $\tuple{L,\dpath{}}$ is a quasi-metric space: (\textit{iff-identity}) holds since a path of length $0$ exists only from $x$ to $x$; (\textit{triangle-inequality}) holds since the concatenation of an upward path from $x$ to $z$ with one from $z$ to $y$ is an upward path from $x$ to $y$; (\textit{symmetry}) fails as soon as some element of $L$ covers another one, since then $\dpath{}(x,y) = 1 \neq \infty = \dpath{}(y,x)$. $\qedex$
\end{exa}

We formally define quantitative approximations via pre-metrics as follows:

\begin{defi}[Quantitative Approximation~\cite{CampionMU25}] \label{def:quantapp}
    Given a pre-metric space  $\tuple{\tC, \tdelta{}}$ and a fixed constant $\varepsilon \in \real^\infty_{\geq 0}$, an element $x \in \tC$ is \emph{quantitatively approximated} by any element $y \in \tC$ such that $\tdelta{}(x,y) \leq \varepsilon$.
    The set $\{ y \in \tC \mid \tdelta{}(x,y) \leq \varepsilon \}$ represents all elements in $\tC$ that quantitatively approximate $x$ up to $\varepsilon$ according to $\tdelta{}$. $\qeddef$
\end{defi}
Note that, when $\tdelta{}$ is not symmetric, the order of the arguments matters: the set above is the \quotes{forward} ball of radius $\varepsilon$ centered in $x$.

\begin{exa} \label{ex:quant}
    Continuing Example~\ref{ex:qual}, we may approximate sets of integer numbers by the size distance $\dcount{}$ defined in Example~\ref{ex:counting}.
    For instance, the set $\{0,1,4\}$ can be quantitatively approximated by any set of integers whose maximum distance from it is at most $\varepsilon = 1$.
    Examples of such approximations include sets $\{0,1\}$ and $\{5,6,8,10\}$.  $\qedex$
\end{exa}

Here, the admitted noise concerns elements that are \quotes{close}, according to the chosen $\delta$, to the original one but that may not share a specific property identified by an uco $\rho$.

\subsubsection{Combining Property-based and Quantitative Approximations.} \label{sec:combining}
By \emph{combining} the two forms of approximation, we obtain a general approximation that incorporates a quantitative error within property abstractions, while still keeping the two types of approximations distinct.
Given $\rho \in \uco(C)$ and a pre-metric space $\tuple{C,\delta}$, we define  the distance $\tudelta{\rho}{} \colon  C \times C \ra \real^\infty_{\geq 0}$ as
\[
    \tudelta{\rho}{}(x,y) \definit \delta(\rho(x),\rho(y))
\]
which calculates the distance between the property-based approximations of $x$ and $y$.
Note that $\tudelta{\rho}{}$ only depends on the restriction of $\delta$ to the abstract domain $\rho(C)$. Hence, the definition and all that follows apply verbatim when $\delta$ is a pre-metric on $\rho(C)$ only, such as the path-length distance of Example~\ref{ex:path-len} on the domain of signs.

\begin{defi}[General Approximation~\cite{CampionMU25}] \label{def:gen}
    Let $\tuple{C,\tpreord{}}$ be a poset and $\tuple{C, \tdelta{}}$ be a pre-metric space, and let $\rho \in \uco(C)$. An element $x \in C$ is approximated by the property $\rho$ \emph{and} quantitatively approximated up to $\varepsilon \in \real^\infty_{\geq 0}$ by $\tdelta{}$, by any element $y \in \tC$ such that $\tudelta{\rho}{}(x,y) \leq \varepsilon$.
    The set $\{ y \in C \mid \tudelta{\rho}{}(x,y) \leq \varepsilon \}$ incorporates both forms of approximation.
    $\qeddef$
\end{defi}

\begin{exa} \label{ex:inclusion-dist-combined}
    By considering the interval approximation $\Int \in \uco(\wp(\integer))$ from Example~\ref{ex:qual} and the inclusion distance $\dincl{}$ from Example~\ref{ex:inclusion-dist}, we can combine the two forms of approximation, namely $\dincl{}$ and $\Int$, into $\dincl{\Int}$: this new distance calculates the number of more elements between two comparable interval abstractions rather than considering the original input sets. Continuing Example~\ref{ex:inclusion-dist}, the set $\{0,1,4\}$ can then be qualitatively and quantitatively approximated by $\dincl{\Int}$ and $\varepsilon = 1$ in any set in $\{S \in \wp(\integer) \mid \dincl{\Int}(\{0,1,4\}, S) \leq 1 \}$, namely in any set $S$ whose interval abstraction $\Int(S)$ is $[0,4]$, $[-1,4]$ or $[0,5]$ (e.g., $\{0,4\}$, $\{-1,2,4\}$ or $\{0,3,5\}$).
    $\qedex$
\end{exa}

Clearly, when considering the identity function $\id \in \uco(C)$ as abstraction (i.e., $\forall x \in C \sut \id(x) \definit x$), it holds that $\tudelta{\id}{}(x,y) = \delta(x,y)$ for any $x,y \in C$ and pre-metric $\delta$. However, for $\rho \neq \id$, combining $\delta$ with $\rho$ preserves some axioms of $\delta$ but may lose others.

\begin{prop}\label{prop:comb-dist}
Let $\tuple{C,\tpreord{}}$ be a poset, $\tuple{C,\delta}$ a pre-metric space, and $\rho \in \uco(C)$.
The following hold:
\begin{enumerate}[label={(\roman*)},itemsep=3pt,leftmargin=10mm,topsep=5pt]
    \item $\tudelta{\rho}{}$ satisfies (\textit{if-identity}), i.e., $\tuple{C,\tudelta{\rho}{}}$ is a pre-metric space;
    \item $\tudelta{\rho}{}$ satisfies (\textit{symmetry}), resp.\ (\textit{triangle-inequality}), if and only if the restriction of $\delta$ to $\rho(C)$ does; in particular, whenever $\delta$ does;
    \item if $\rho \neq \id$ then $\tudelta{\rho}{}$ does \emph{not} satisfy (\textit{iff-identity}), even when $\delta$ does;
\end{enumerate}
\end{prop}
\begin{proof}
(i) If $x = y$ then $\tudelta{\rho}{}(x,y) = \delta(\rho(x),\rho(y)) = \delta(\rho(x),\rho(x)) = 0$ by (\textit{if-identity}) of $\delta$.

(ii) Assume that $\delta$ satisfies (\textit{symmetry}) on $\rho(C)$. Since $\rho(x),\rho(y) \in \rho(C)$ for all $x,y \in C$, we have $\tudelta{\rho}{}(x,y) = \delta(\rho(x),\rho(y)) = \delta(\rho(y),\rho(x)) = \tudelta{\rho}{}(y,x)$. Conversely, assume that $\tudelta{\rho}{}$ satisfies (\textit{symmetry}) and let $x,y \in \rho(C)$. By idempotence of $\rho$, we have $\rho(x) = x$ and $\rho(y) = y$, hence $\delta(x,y) = \delta(\rho(x),\rho(y)) = \tudelta{\rho}{}(x,y) = \tudelta{\rho}{}(y,x) = \delta(\rho(y),\rho(x)) = \delta(y,x)$. The proof for (\textit{triangle-inequality}) is analogous: if $\delta$ satisfies it on $\rho(C)$ then, for all $x,y,z \in C$, $\tudelta{\rho}{}(x,y) = \delta(\rho(x),\rho(y)) \leq \delta(\rho(x),\rho(z)) + \delta(\rho(z),\rho(y)) = \tudelta{\rho}{}(x,z) + \tudelta{\rho}{}(z,y)$; conversely, if $\tudelta{\rho}{}$ satisfies it then, for all $x,y,z \in \rho(C)$, $\delta(x,y) = \tudelta{\rho}{}(x,y) \leq \tudelta{\rho}{}(x,z) + \tudelta{\rho}{}(z,y) = \delta(x,z) + \delta(z,y)$. Finally, if $\delta$ satisfies an axiom on the whole $C$, then it trivially satisfies it on $\rho(C) \subseteq C$.

(iii) Since $\rho \neq \id$, there exists $x \in C$ with $\rho(x) \neq x$; let $y = \rho(x)$, so that $x \neq y$. By idempotence of $\rho$, $\rho(y) = \rho(\rho(x)) = \rho(x)$, hence $\tudelta{\rho}{}(x,y) = \delta(\rho(x),\rho(y)) = \delta(\rho(x),\rho(x)) = 0$ by (\textit{if-identity}) of $\delta$, despite $x \neq y$.
\end{proof}

Consequently, $\tudelta{\rho}{}$ is a pseudo-metric whenever $\delta$ is a metric or a pseudo-metric, a quasipseudo-metric whenever $\delta$ is a quasi-metric or a quasipseudo-metric, a pseudosemi-metric whenever $\delta$ is a semi-metric or a pseudosemi-metric, and just a pre-metric whenever $\delta$ is a quasisemi-metric. In other words, when we combine a non-identity abstraction $\rho$ with the pre-metric $\delta$, then $\tudelta{\rho}{}$ preserves (\textit{if-identity}), (\textit{symmetry}) and (\textit{triangle-inequality}), but always loses (\textit{iff-identity}) on $C$.

\begin{exa} \label{ex:inclusion-dist-int-axioms}
    Recall from Example~\ref{ex:inclusion-dist} that $\dincl{}$ is a quasi-metric forming a quasi-metric space $\tuple{\wp(\integer),\dincl{}}$, i.e., it satisfies (\textit{iff-identity}) and (\textit{triangle-inequality}) but not (\textit{symmetry}); by Proposition~\ref{prop:comb-dist}, $\dincl{\Int}$ keeps (\textit{triangle-inequality}) but loses (\textit{iff-identity}). For instance, $\dincl{\Int}(\{0,2\},\{0,1,2\}) = \dincl{}([0,2],[0,2]) = 0$ even though $\{0,2\} \neq \{0,1,2\}$. Hence $\tuple{\wp(\integer),\dincl{\Int}}$ forms a quasipseudo-metric space. On the other hand, the restriction of $\dincl{\Int}$ to the abstract domain $\Int(\wp(\integer))$ of intervals \emph{does} satisfy (\textit{iff-identity}) and is thus again a quasi-metric: two intervals at $\dincl{}$-distance $0$ are, by definition of $\dincl{}$, necessarily the same set.
    $\qedex$
\end{exa}

\section{Abstract Lipschitz Continuity}\label{sec:alc}

In mathematical analysis, Lipschitz continuity is a strong form of uniform continuity
that establishes a quantitative relationship between perturbations of the input of a
function and the resulting perturbations of its output: the latter are bounded by a
linear function of the former, whose slope is the Lipschitz constant. Classically, it
is defined for functions between metric spaces, and it underlies, e.g., the notion of
program robustness of Chaudhuri et al.~\cite{gulwani10,Gulwani11,gulwani12}, function
sensitivity in differential privacy~\cite{sensitivity}, and certified robustness of
neural networks~\cite{FazlyabRHMP19,HuangZSKA21,JiT20,ZhangJH022}. In our setting,
however, the distances of interest (cf.\ Section~\ref{sec:background}) are in general
neither symmetric nor satisfying the triangle inequality or the identity of
indiscernibles, and the input and output spaces need not carry the same distance.
We therefore define Lipschitz continuity for functions between (possibly distinct)
pre-metric spaces.

\begin{defi}[Lipschitz Continuity]\label{def:lc}
    Let $\tuple{C,\tdelta{C}}$ and $\tuple{D,\tdelta{D}}$ be pre-metric spaces denoting the input and output spaces, respectively.
    Let $k\in \real_{\geq 0}$. A function $f\colon C\ra D$ satisfies \emph{$k$-Lipschitz Continuity} ($k$-LC for short) w.r.t. $\tuple{\tdelta{C}, \tdelta{D}}$ when the following inequality holds:
    \[
        \forall x, y \in C \sut\; \tdelta{D}(f(x),f(y)) \,\leq\, k \tdelta{C}(x,y).
    \]
    A function $f$ satisfies \emph{Lipschitz Continuity} (LC) w.r.t. $\tuple{\tdelta{C}, \tdelta{D}}$ when there exists $k\in \real_{\geq 0}$ such that $f$ satisfies $k$-Lipschitz continuity w.r.t. $\tuple{\tdelta{C}, \tdelta{D}}$.
    $\qeddef$
\end{defi}
Any such $k$ is called a Lipschitz constant of $f$: it bounds the rate of change of the
output of $f$, i.e., $\nicefrac{\delta_{\scriptscriptstyle {\tt D}}(f(x),f(y))}{\delta_{\scriptscriptstyle {\tt C}}(x,y)} \leq k$
for all $x,y \in C$ with $\tdelta{C}(x,y) \neq 0$ (recall that $\nicefrac{r}{\infty} = 0$),
while inputs at distance $0$ must be mapped to outputs at distance $0$.

Note that a pre-metric $\delta$ on $C$ still induces a topology, whose open sets are
the $U \subseteq C$ such that every $x \in U$ has a forward ball
$\{y \in C \mid \delta(x,y) < r\}$, $r > 0$, contained in $U$, and every $k$-LC
function is continuous w.r.t.\ the topologies induced by $\tdelta{C}$ and $\tdelta{D}$.
However, balls need not be open, limits need not be unique, and the results of metric
analysis that depend on the metric axioms (e.g., Banach's fixed point theorem) are not
available in general. What the rest of the paper relies on are two properties of
Lipschitz constants that hold in any pre-metric setting: the set of admissible
constants of a function is upward closed, and Lipschitz continuous functions are
closed under composition, with the product of the constants as a constant of the
composition. These are precisely the properties exploited by the deductive system
of Section~\ref{sec:proof}, while the results of Section~\ref{sec:alc4prog} will
require additional axioms on the distances at hand.

Definition~\ref{def:lc} of $k$-Lipschitz continuity can be equivalently formulated as follows:
\begin{prop} \label{prop:lc}
    Let $\tuple{C,\tdelta{C}}$ and $\tuple{D,\tdelta{D}}$ be pre-metric spaces and
    $k \in \real_{\geq 0}$. A function $f\colon C\ra D$ satisfies $k$-LC w.r.t.\
    $\tuple{\tdelta{C},\tdelta{D}}$ if and only if
    \[
        \forall x,y\in C \sut \forall\varepsilon \in \real_{\geq 0} \sut
        \tdelta{C}(x,y)\leq\varepsilon  \Ra\; \tdelta{D}( f(x), f(y))\leq k\varepsilon.
    \]
\end{prop}
\begin{proof}
    ($\Ra$) Let $x,y \in C$ and $\varepsilon \in \real_{\geq 0}$ with $\tdelta{C}(x,y) \leq
    \varepsilon$. Then $\tdelta{D}(f(x),f(y)) \leq k\tdelta{C}(x,y) \leq k\varepsilon$.

    ($\La$) Let $x,y \in C$. If $\tdelta{C}(x,y) = \infty$ then $\tdelta{D}(f(x),f(y)) \leq
    \infty = k \tdelta{C}(x,y)$, by the conventions of Section~\ref{sec:background}.
    Otherwise, $\varepsilon = \tdelta{C}(x,y) \in \real_{\geq 0}$ satisfies
    $\tdelta{C}(x,y) \leq \varepsilon$, hence $\tdelta{D}(f(x),f(y)) \leq k\varepsilon =
    k\tdelta{C}(x,y)$.
\end{proof}

When approximations are introduced to the input of a function, they propagate through its computations, affecting the output. Understanding how approximations evolve during computations provides insight into the behavior of the function and, consequently, into program executions when the function represents program semantics~\cite{gulwani10,gulwani12,ZhangNEURIPS2022,luxburg2004distance,CampionPOPL2026}.

\begin{figure}[t]
	\begin{minipage}[t]{.5\linewidth}
		\centering
	      \begin{tikzpicture}[scale=1.1, every node/.style={scale=1}]
	\coordinate (in-point) at (0,0);
	\coordinate (in-above) at ($(in-point)+(0,.75)$);
	\coordinate (in-below) at ($(in-point)-(0,.75)$);
	\coordinate (out-point) at (3,0);
	\coordinate (out-above) at ($(out-point)+(0,1.25)$);
	\coordinate (out-below) at ($(out-point)-(0,1.25)$);
	\phantom{\draw ($(in-point)-(.75,1.75)$) rectangle ($(out-point)+(.75,1.75)$);}
	\filldraw[draw=white,left color=teal!20, right color=magenta!15,shading angle=90] (in-above) to[looseness=1,out=0,in=-160] (out-above) -- (out-below) to[looseness=1,out=160,in=0] (in-below) -- cycle;
	\draw[dashed,shorten <=6pt,shorten >=4pt] (in-above) to[looseness=1,out=0,in=-160] (out-above);
	\draw[dashed,shorten <=6pt,shorten >=4pt] (in-below) to[looseness=1,out=0,in=160] (out-below);
	\draw[teal,line width=.75pt] ($(in-above)-(.1,0)$) -- ($(in-above)+(.1,0)$);
	\draw[teal,line width=.75pt] (in-above) -- ($(in-point)+(0,.1)$);
	\draw[teal,line width=.75pt] ($(in-point)-(0,.1)$) -- (in-below);
	\draw[teal,line width=.75pt] ($(in-below)-(.1,0)$) -- ($(in-below)+(.1,0)$);
	\draw[magenta,line width=.75pt] ($(out-above)-(.1,0)$) -- ($(out-above)+(.1,0)$);
	\draw[magenta,line width=.75pt] (out-above) -- ($(out-point)+(0,.1)$);
	\draw[magenta,line width=.75pt] ($(out-point)-(0,.1)$) -- (out-below);
	\draw[magenta,line width=.75pt] ($(out-below)-(.1,0)$) -- ($(out-below)+(.1,0)$);
	\filldraw[draw=white,fill=gray!20,rounded corners=4pt,rotate=120] ($(in-point)-(.35,.35)$) -- ($(in-point)-(.45,0)$) -- ($(in-point)-(.35,-.35)$) -- ($(in-point)+(0,.25)$) -- ($(in-point)+(.35,.35)$) -- ($(in-point)+(.45,0)$) -- ($(in-point)+(.35,-.25)$) -- ($(in-point)+(0,-.45)$) -- cycle;
	\filldraw[draw=white,fill=gray!20,rounded corners=4pt,rotate=-30] ($(out-point)-(.35,.35)$) -- ($(out-point)-(.45,0)$) -- ($(out-point)-(.35,-.35)$) -- ($(out-point)+(0,.25)$) -- ($(out-point)+(.35,.35)$) -- ($(out-point)+(.45,0)$) -- ($(out-point)+(.35,-.25)$) -- ($(out-point)+(0,-.45)$) -- cycle;
	\filldraw[black] (in-point) circle (1.5pt);
	\filldraw[black] (out-point) circle (1.5pt);
	\draw[-latex,shorten <=4pt,shorten >=4pt,line width=1pt] (in-point) -- node[midway,above,yshift=-2pt] {$f$} (out-point);
	\filldraw[gray!50] ($(in-point)+(.1,.2)$) circle (1pt);
	\filldraw[gray!50] ($(in-point)+(-.2,.2)$) circle (1pt);
	\filldraw[gray!50] ($(in-point)+(0,-.25)$) circle (1pt);
	\filldraw[gray!50] ($(in-point)+(.25,-.15)$) circle (1pt);
	\filldraw[gray!50] ($(out-point)+(.25,.05)$) circle (1pt);
	\filldraw[gray!50] ($(out-point)+(-.2,.2)$) circle (1pt);
	\filldraw[gray!50] ($(out-point)+(-.15,-.2)$) circle (1pt);
	\filldraw[gray!50] ($(out-point)+(.2,-.2)$) circle (1pt);
	\node[teal] at ($(in-above)+(0,.2)$) {$\scriptstyle \varepsilon^\prime$};
	\node[magenta] at ($(out-above)+(0,.2)$) {$\scriptstyle k\varepsilon^\prime > 0$};
	\node[anchor=east,gray] at ($(in-point)-(.35,0)$) {$\scriptstyle \eta$};
	\node[anchor=west,gray] at ($(out-point)+(.35,0)$) {$\scriptstyle \rho$};
	\node[teal] at ($(in-point)-(0,1.5)$) {\scriptsize \textsc{Input}};
	\node[magenta] at ($(out-point)-(0,1.5)$) {\scriptsize \textsc{Output}};
\end{tikzpicture}
	\end{minipage}%
	\begin{minipage}[t]{.5\linewidth}
		\centering
	      \begin{tikzpicture}[scale=1.1, every node/.style={scale=1}]
	\coordinate (in-point) at (0,0);
	\coordinate (in-above) at ($(in-point)+(0,.75)$);
	\coordinate (in-below) at ($(in-point)-(0,.75)$);
	\coordinate (out-point) at (3,0);
	\coordinate (out-above) at ($(out-point)+(0,.1)$);
	\coordinate (out-below) at ($(out-point)-(0,.1)$);
	\phantom{\draw ($(in-point)-(.75,1.75)$) rectangle ($(out-point)+(.75,1.75)$);}
	\filldraw[draw=white,left color=teal!20, right color=white,shading angle=90] (in-above) to[looseness=1,out=0,in=-190] (out-above) -- (out-below) to[looseness=1,out=190,in=0] (in-below) -- cycle;
	\draw[dashed,shorten <=6pt,shorten >=4pt] (in-above) to[looseness=1,out=0,in=-190] (out-above);
	\draw[dashed,shorten <=6pt,shorten >=4pt] (in-below) to[looseness=1,out=0,in=190] (out-below);
	\draw[teal,line width=.75pt] ($(in-above)-(.1,0)$) -- ($(in-above)+(.1,0)$);
	\draw[teal,line width=.75pt] (in-above) -- ($(in-point)+(0,.1)$);
	\draw[teal,line width=.75pt] ($(in-point)-(0,.1)$) -- (in-below);
	\draw[teal,line width=.75pt] ($(in-below)-(.1,0)$) -- ($(in-below)+(.1,0)$);
	\filldraw[draw=white,fill=gray!20,rounded corners=4pt,rotate=120] ($(in-point)-(.35,.35)$) -- ($(in-point)-(.45,0)$) -- ($(in-point)-(.35,-.35)$) -- ($(in-point)+(0,.25)$) -- ($(in-point)+(.35,.35)$) -- ($(in-point)+(.45,0)$) -- ($(in-point)+(.35,-.25)$) -- ($(in-point)+(0,-.45)$) -- cycle;
	\filldraw[draw=white,fill=gray!20,rounded corners=4pt,rotate=-30] ($(out-point)-(.35,.35)$) -- ($(out-point)-(.45,0)$) -- ($(out-point)-(.35,-.35)$) -- ($(out-point)+(0,.25)$) -- ($(out-point)+(.35,.35)$) -- ($(out-point)+(.45,0)$) -- ($(out-point)+(.35,-.25)$) -- ($(out-point)+(0,-.45)$) -- cycle;
	\filldraw[black] (in-point) circle (1.5pt);
	\filldraw[black] (out-point) circle (1.5pt);
	\draw[-latex,shorten <=4pt,shorten >=4pt,line width=1pt] (in-point) -- node[midway,above,yshift=-2pt] {$f$} (out-point);
	\filldraw[gray!50] ($(in-point)+(.1,.2)$) circle (1pt);
	\filldraw[gray!50] ($(in-point)+(-.2,.2)$) circle (1pt);
	\filldraw[gray!50] ($(in-point)+(0,-.25)$) circle (1pt);
	\filldraw[gray!50] ($(in-point)+(.25,-.15)$) circle (1pt);
	\filldraw[gray!50] ($(out-point)+(.25,.05)$) circle (1pt);
	\filldraw[gray!50] ($(out-point)+(-.2,.2)$) circle (1pt);
	\filldraw[gray!50] ($(out-point)+(-.15,-.2)$) circle (1pt);
	\filldraw[gray!50] ($(out-point)+(.2,-.2)$) circle (1pt);
	\node[teal] at ($(in-above)+(0,.2)$) {$\scriptstyle \varepsilon^\prime$};
	\node[magenta] at ($(out-above)+(0,.5)$) {$\scriptstyle \varepsilon = 0$};
	\node[anchor=east,gray] at ($(in-point)-(.35,0)$) {$\scriptstyle \eta$};
	\node[anchor=west,gray] at ($(out-point)+(.35,0)$) {$\scriptstyle \rho$};
	\node[teal] at ($(in-point)-(0,1.5)$) {\scriptsize \textsc{Input}};
	\node[magenta] at ($(out-point)-(0,1.5)$) {\scriptsize \textsc{Output}};
\end{tikzpicture}
	\end{minipage}
	\caption{Controlled input/output general approximations (left) and suppression of input general approximation (right).}\label{fig:error-prop-forms}
\end{figure}
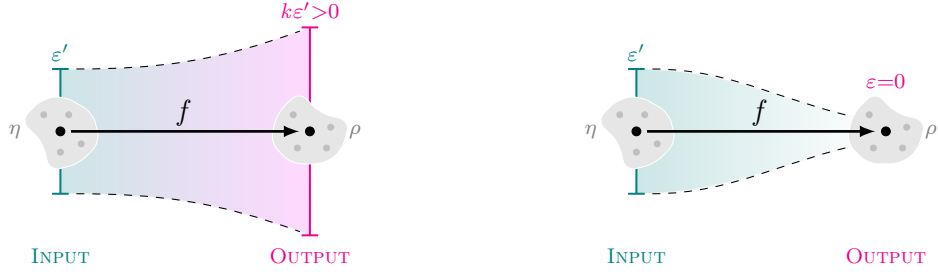

In this work, we generalize the definition of Lipschitz continuity (Definition~\ref{def:lc}), which relies on quantitative approximations (Definition~\ref{def:quantapp}), to work with general approximations (Definition~\ref{def:gen}), \emph{explicitly discerning property-based approximations (i.e., ucos) from quantitative approximations (i.e., distances)}. This yields the novel notion of \textit{abstract Lipschitz continuity}, which enforces a \emph{controlled (linear) error propagation} from a \emph{general approximation of the input} to a \emph{general approximation of the output} of a function computation (Figure~\ref{fig:error-prop-forms}).
\begin{defi}[Abstract Lipschitz Continuity] \label{def:alc}
Let $\tuple{C,\tpreord{C}}$ and $\tuple{D,\tpreord{D}}$ be the input and output domains (posets), respectively. Let
$\tuple{C,\tdelta{C}}$ and $\tuple{D,\tdelta{D}}$ be pre-metric spaces, and let $\eta\in\uco(C)$, $\rho\in\uco(D)$ be the ucos on the input and output domains, respectively. Consider a constant $k \in \real_{\geq 0}$.
 A function $f\colon C\ra D$ satisfies \emph{Abstract $k$-Lipschitz Continuity} ($k$-ALC, for short) w.r.t.\ $\tuple{\tdelta{C}^\eta,\tdelta{D}^\rho}$ when:
 \[
    \forall x,y\in C\sut\; \tdelta{D}^\rho(f(x),f(y))\,\leq\, k\tdelta{C}^\eta(x,y).
 \]
A function $f$ satisfies \emph{Abstract Lipschitz Continuity} (ALC) whenever there exists $k\in \real_{\geq 0}$ such that $f$ satisfies abstract $k$-Lipschitz continuity.
$\qeddef$
\end{defi}
Informally, $k$-ALC states that whenever $y$ is a general approximation of $x$ up to $\varepsilon$ w.r.t.\ $\eta$ and $\tdelta{C}$, then $f(y)$ is a general approximation of $f(x)$ up to $k\varepsilon$ w.r.t.\ $\rho$ and $\tdelta{D}$.

When $k$-ALC holds, $k$ is called an \emph{abstract Lipschitz constant} of $f$. The case
$k = 0$ deserves attention: by the assumption $0 \cdot \infty = \infty$ (Section~\ref{sec:pre-metric}), $0$-ALC requires
$\tdelta{D}^\rho(f(x),f(y)) = 0$ for all $x,y \in C$ at finite distance $\tdelta{C}^\eta(x,y)$,
i.e., $f$ completely suppresses any (finite) general approximation of its input
(Figure~\ref{fig:error-prop-forms}, right).

It is worth remarking that Definition~\ref{def:lc} can be instantiated to model Definition~\ref{def:alc} by incorporating the computation of the distance between property-based approximations within $\tdelta{C}$ and $\tdelta{D}$. Indeed, by Proposition~\ref{prop:comb-dist}(i), $\tudelta{\eta}{C}$ and $\tudelta{\rho}{D}$ are pre-metrics on $C$ and $D$, respectively, hence $f$ satisfies $k$-ALC w.r.t.\ $\tuple{\tdelta{C}^\eta,\tdelta{D}^\rho}$ exactly when it satisfies $k$-LC w.r.t.\ the pre-metric spaces $\tuple{C,\tdelta{C}^\eta}$ and $\tuple{D,\tdelta{D}^\rho}$. However, explicitly discerning qualitative approximations (i.e., ucos) from quantitative approximations (i.e., distances) as in Definition~\ref{def:alc}, has two main advantages. First it allows us to relate ALC to other extensional program properties (i.e., relative to the input/output behavior of programs) studied in the program analysis literature, such as abstract robustness~\cite{GMP24} and partial completeness~\cite{campion2022popl,CampionSAS23} (Section~\ref{sec:alc4prog}). Second, it allows reasoning on its validity over programs by induction on their syntax and on the chosen ucos (Section~\ref{sec:proof}).

As a consequence, every property of Lipschitz continuity that holds for arbitrary pre-metric spaces transfers to ALC, such as the equivalent formulation of Proposition~\ref{prop:lc}:

\begin{prop} \label{prop:alc}
Consider the premises of Definition~\ref{def:alc}.
 A function $f\colon C\ra D$ satisfies $k$-ALC w.r.t.\ $\tuple{\tdelta{C}^\eta,\tdelta{D}^\rho}$ if and only if:
 \[
    \forall x,y\in C\sut \forall\varepsilon \in \real_{\geq 0} \sut \tdelta{C}^\eta(x,y)\leq\varepsilon\ \Ra\ \tdelta{D}^\rho( f(x), f(y))\leq k\varepsilon.
\]
\end{prop}
\begin{proof}
    Immediate from Proposition~\ref{prop:lc} applied to the pre-metric spaces $\tuple{C,\tdelta{C}^\eta}$ and $\tuple{D,\tdelta{D}^\rho}$.
\end{proof}

\begin{exa}[Byte Histograms] \label{ex:byte-histogram}
Let $\Sigma=\{0,\ldots,255\}$ be the set of byte values.
A file $F\in\mbox{\sc Files}\definit\Sigma^\ast$ is a finite
sequence of bytes.
Its byte histogram
$\mathsf{hist}(F)\colon\Sigma\to\nat$ records the number of
occurrences of each byte. Two files have the same histogram
exactly when they differ only in the order of their bytes.

Let $C$ and $D$ be the sets of non-empty finite subsets of
$\mbox{\sc Files}$ and $\nat$, respectively, both ordered by
inclusion.
We consider the input abstraction
$\eta\in\uco(C)$ that forgets byte order:
\[
    \eta(\mathbf{F}) \definit
    \{ F'\in\mbox{\sc Files}
       \mid \exists F\in\mathbf{F}\sut
       \mathsf{hist}(F')=\mathsf{hist}(F) \}
\]
and, as input distance, we take the Hausdorff distance~\cite{DezaDeza} induced by
the edit distance $\mathsf{ed} \colon \mbox{\sc Files} \times \mbox{\sc Files} \to \nat$, which counts the minimum number
of single-byte insertions, deletions, and substitutions needed
to transform one file into another:
\[
    \tdelta{C}(\mathbf{F}_1,\mathbf{F}_2)
    \definit
    \max\Big\{
        \max_{F_1\in\mathbf{F}_1}
        \min_{F_2\in\mathbf{F}_2}\mathsf{ed}(F_1,F_2),
        \;
        \max_{F_2\in\mathbf{F}_2}
        \min_{F_1\in\mathbf{F}_1}\mathsf{ed}(F_1,F_2)
    \Big\}.
\]
The combined distance
$\tudelta{\eta}{C}(\mathbf{F}_1,\mathbf{F}_2)
=\tdelta{C}(\eta(\mathbf{F}_1),\eta(\mathbf{F}_2))$
therefore compares files up to byte reordering,
e.g., the files {\em ``ab''} and {\em ``ba''} have
edit distance $2$, but $\tudelta{\eta}{C}
    (\{\mbox{\em ``ab''}\},\{\mbox{\em ``ba''}\})=0$.

As output abstraction, we take the uco $\rho\in\uco(D)$
that approximates counts by intervals of a fixed width
$w\in\nat$, $w\geq 1$.
Let $B_i\definit\{iw,\ldots,iw+w-1\}$ for $i\in\nat$.
Then $\rho$ replaces each count with the interval containing it:
\[
    \rho(N)\definit
    \bigcup\{B_i\mid i\in\nat,\ N\cap B_i\neq\varnothing\}.
\]
For example, if $w=10$, then
$\rho(\{3\})=\rho(\{8\})=\{0,\ldots,9\}$,
while $\rho(\{12\})=\{10,\ldots,19\}$.
As output distance, we take the pseudo-metric
$\tdelta{D}(N_1,N_2)\definit|\max N_1-\max N_2|$.
For $w=10$, the distance between
$\rho(\{3\})$ and $\rho(\{8\})$ is $0$, whereas that between
$\rho(\{8\})$ and $\rho(\{12\})$ is $10$.

The semantics of \texttt{count-a} in
Figure~\ref{fig:example} (on the right) is the function
$\collSem{\mbox{\texttt{count-a}}}\colon C\to D$ defined by
    $\collSem{\mbox{\texttt{count-a}}}\mathbf{F}
    \definit
    \{\mathsf{hist}(F)(\mbox{\em ``a''})\mid F\in\mathbf{F}\}$,
collecting the number of occurrences of the malicious pattern {\em ``a''\/} in each file of $\mathbf{F}$.
It satisfies $w$-ALC w.r.t.\
$\tuple{\tudelta{\eta}{C},\tudelta{\rho}{D}}$.
Indeed, a single edit changes the number of occurrences of
{\em ``a''} by at most $1$, while byte reordering leaves it
unchanged. By the definition of Hausdorff distance, the maximum
counts in any $\mathbf{F}_1,\mathbf{F}_2\in C$ therefore differ
by at most $\tudelta{\eta}{C}(\mathbf{F}_1,\mathbf{F}_2)$.
Moreover, increasing a count by $1$ either leaves its interval
unchanged or moves it to the next interval, whose upper endpoint
is $w$ larger. Thus, applying $\rho$ increases the distance
between maximum counts by at most a factor $w$, giving
\[
    \tudelta{\rho}{D}
    (\collSem{\mbox{\texttt{count-a}}}\mathbf{F}_1,
     \collSem{\mbox{\texttt{count-a}}}\mathbf{F}_2)
    \leq w\,\tudelta{\eta}{C}(\mathbf{F}_1,\mathbf{F}_2).
\]
The factor $w$ is necessary: adding one {\em ``a''} to
$\mbox{\em ``a''}^{w-1}$ changes the count from $w-1$ to $w$.
After applying $\rho$, the corresponding maxima are $w-1$
and $2w-1$, whose difference is $w$.

Hence, ALC bounds how much the observed count can change under
file modifications, while ensuring that byte reordering leaves
it unchanged.
Retaining only file length would not suffice:
$\mbox{\em ``a''}^{w}$ and $\mbox{\em ``b''}^{w}$
have the same length, but their counts $w$ and $0$ satisfy
$\tudelta{\rho}{D}(\{w\},\{0\})=w$.
An abstraction identifying these files would therefore give
input distance $0$ and positive output distance, violating ALC
for every finite constant.
$\qedex$
\end{exa}

\begin{exa}\label{ex:R-ext}
    Consider the program $\progp{R} \in \imp$ of Example~\ref{ex:R}, the interval closure
    $\Int \in \uco(\wp(\integer))$ and the inclusion distance $\dincl{}$ of
    Example~\ref{ex:inclusion-dist}. We consider non-empty sets of integers only, i.e., the
    pre-metric space $\tuple{\wp(\integer) \smallsetminus \{\varnothing\}, \dincl{}}$, on
    which $\Int$ is still a uco and $\collSem{\progp{R}^\ast}$ is well defined, since
    $S \subseteq \collSem{\progp{R}^\ast} S$. Then $\collSem{\progp{R}^\ast}$ satisfies
    $1$-ALC w.r.t.\ $\tuple{\dincl{\Int},\dincl{\Int}}$. Indeed, let $\Int(S_1) = [a_1,b_1]$
    and $\Int(S_2) = [a_2,b_2]$. If $[a_1,b_1] \not\subseteq [a_2,b_2]$ then
    $\dincl{\Int}(S_1,S_2) = \infty$ and there is nothing to prove. Otherwise
    $a_2 \leq a_1 \leq b_1 \leq b_2$ and, by Example~\ref{ex:R},
    $\Int(\collSem{\progp{R}^\ast} S_i) = [\min(a_i,0), \max(b_i,1)]$ for $i \in \{1,2\}$,
    where $[\min(a_2,0), \max(b_2,1)] \smallsetminus [\min(a_1,0), \max(b_1,1)] \subseteq
    [a_2,b_2] \smallsetminus [a_1,b_1]$, so that
    $\dincl{\Int}(\collSem{\progp{R}^\ast}S_1, \collSem{\progp{R}^\ast}S_2) \leq
    \dincl{\Int}(S_1,S_2)$. For instance:
    \begin{align*}
        \dincl{\Int}(\collSem{\progp{R}^\ast}([2,6]),\collSem{\progp{R}^\ast}([0,7])) &= \dincl{}([0,6],[0,7]) = 1 \leq 3 = \dincl{\Int}([2,6],[0,7]) \\
        \dincl{\Int}(\collSem{\progp{R}^\ast}([-5,-2]),\collSem{\progp{R}^\ast}([-7,0])) &= \dincl{}([-5,1],[-7,1]) = 2 \leq 4 = \dincl{\Int}([-5,-2],[-7,0])
    \end{align*}
    The restriction to non-empty sets is necessary: $\dincl{\Int}(\varnothing,\{n\}) = 1$
    whereas $\dincl{\Int}(\collSem{\progp{R}^\ast}\varnothing, \collSem{\progp{R}^\ast}\{n\})
    = \dincl{}(\varnothing,[0,n]) = n+1$ for every $n > 0$.
    On the other hand, $\collSem{\progp{R}^\ast}$ does not satisfy $k$-ALC w.r.t.\
    $\tuple{\dcount{\Int},\dcount{\Int}}$ for any $k$: $\dcount{\Int}(\{5\},\{-5\}) = 0$,
    since $[5,5]$ and $[-5,-5]$ have the same size, whereas
    $\dcount{\Int}(\collSem{\progp{R}^\ast}\{5\},\collSem{\progp{R}^\ast}\{-5\}) =
    \dcount{}([0,5],[-5,1]) = 1$. This is a general phenomenon: since $k \cdot 0 = 0$,
    ALC forces inputs at distance $0$ to be mapped to outputs at distance $0$ (as
    observed after Definition~\ref{def:lc}), and $\dcount{}$ identifies all intervals
    of the same size. $\qedex$
\end{exa}

\section{ALC as a Sufficient Condition: Partial Completeness, Partial Abstract Non-Interference and Abstract Robustness} \label{sec:alc4prog}

By instantiating the function $f$ in Definition~\ref{def:alc} with the semantics of a program (e.g., the collecting semantics $\collSem{\prog} \colon C \to C$ of a program $\prog \in \imp$ defined in Section~\ref{sec:prog}), ALC becomes an \emph{extensional property} of programs, i.e., relative to their input/output behavior.
By explicitly discerning the role of qualitative and quantitative approximations in its definition, ALC can be related to other extensional property of programs in the literature that capture input/output relations in comparable yet distinct ways.
In this section, we show that ALC is a \emph{sufficient condition} for three such properties. The first two weaken classical notions by admitting a bounded imprecision $\varepsilon$ in the observed output properties:
\emph{partial completeness} (Section~\ref{sec:pcompl}), a weakening of completeness in abstract interpretation, and
\emph{partial abstract non-interference} (Section~\ref{sec:pani}), a weakening of abstract non-interference.
The third one is \emph{abstract robustness} (Section~\ref{sec:arob}), a weakening of the robustness of classifiers in machine learning, which instead requires the observed output properties to be invariant under perturbations of the inputs within a bounded radius $\varepsilon$.

When ALC and partial completeness or partial abstract non-interference are expressed w.r.t.\ the \emph{same} abstractions, ALC implies them for \emph{any} threshold $\varepsilon \in \real_{\geq 0}$, regardless of the abstract Lipschitz constant $k$; under mild additional assumptions on the distances, it even implies their exact counterparts, namely completeness of the best correct approximation and abstract non-interference. Intuitively, the quantitative bound provided by ALC on how output properties may vary leaves no room for imprecision when input properties are indistinguishable.
For abstract robustness, instead, the abstract Lipschitz constant must be null: $0$-ALC implies abstract robustness for any radius $\varepsilon \in \real_{\geq 0}$, whereas the converse does not hold for a fixed radius. Hence, $0$-ALC is, in general, strictly stronger than abstract robustness for a fixed radius.%

\looseness=-1
We will also relate ALC to partial completeness and partial abstract non-interference when they are expressed w.r.t.\ two \emph{different} input abstractions $\eta_1,\eta_2 \in \uco(C)$, a generalization that we do not consider for abstract robustness, which admits no imprecision in the observed output properties. In this case, the abstract Lipschitz constant $k$ does come into play: the threshold $\varepsilon$ guaranteed by ALC is proportional to $k$ and to how far $\eta_2$ is from $\eta_1$, as measured by the input distance. A recurring special case is when $\eta_2$ is \emph{more precise} than $\eta_1$, written $\eta_2 \tpreord{C} \eta_1$ as a slight abuse of notation for $\forall x \in C \colon \eta_2(x) \tpreord{C} \eta_1(x)$. The following lemma shows that, in this case, the composition of abstractions $\eta_1 \comp \eta_2$ is equivalent to $\eta_1$:
\begin{lem} \label{lem:eta1eta2}
Let $\tuple{C,\tpreord{C}}$ be a poset and $\eta_1\in\uco(C)$, $\eta_2\in\uco(C)$. If $\eta_2 \tpreord{C} \eta_1$, then
$\forall x \in C:\ \eta_1 \comp \eta_2(x) = \eta_1(x)$.
\end{lem}
\begin{proof}
$(\tpreord{C})$ Let us assume $\eta_2(x) \tpreord{C} \eta_1(x)$, for all input $x \in C$. Then, by monotonicity of $\eta_1$, we have $\eta_1 \comp \eta_2(x) \tpreord{C} \eta_1 \comp \eta_1(x)$. Thus, by idempotency of $\eta_1$, we have $\eta_1 \comp \eta_2 (x) \tpreord{C} \eta_1 (x)$. \\
$(\succeq_{\tiny {\tt C}})$ By extensivity of $\eta_2$, for all $x \in C$, we have $\eta_2(x) \succeq_{\tiny {\tt C}} x$. Thus, by monotonicity of $\eta_1$, we have $\eta_1 \comp \eta_2(x) \succeq_{\tiny {\tt C}} \eta_1(x)$.
\end{proof}

\begin{exa} \label{ex:eta1eta2}
Consider the poset $\langle
\wp(\integer),
\subseteq
\rangle$
and the abstractions $\Sign' \in \uco(\wp(\integer))$ and $\Int \in \uco(\wp(\integer))$, where $\Sign'$ is the simplified sign closure whose set of fixpoints is
\[
    \Sign'(\wp(\integer)) \definit \{\varnothing, \{0\}, \integer_{\leq 0}, \integer_{\geq 0}, \integer\}.
\]
Note that $\Sign'$ is coarser than the closure $\Sign$ of Example~\ref{ex:sign}, as it cannot express, e.g., $\integer_{\neq 0}$. The $\Int$ abstraction is more precise than $\Sign'$: for any $X \in \wp(\integer)$, we have $\Int(X) \subseteq \Sign'(X)$. Thus, by Lemma~\ref{lem:eta1eta2}, $\Sign' \circ \Int(X) = \Sign'(X)$. For instance, for $X = \{-1,1\}$, $\Sign'(\Int(X)) = \Sign'(\{-1,0,1\}) = \integer = \Sign'(X)$. Note that this does not hold for $\Sign$, which is incomparable with $\Int$: $\Int(\{-1,1\}) = \{-1,0,1\} \not\subseteq \integer_{\neq 0} = \Sign(\{-1,1\})$.

$\qedex$
\end{exa}

\begin{figure}[t]
	\begin{tikzpicture}[
    prop/.style={align=center,font=\small,inner ysep=5pt},
    impl/.style={-Implies,line width=.5pt,double distance=.75pt,shorten <= 2pt,shorten >= 3pt},
    thm/.style={midway,sloped,above,font=\scriptsize},
    abs/.style={sloped,below,scale=.7}
  ]

  \coordinate (kALC) at (0,0);
  \node[prop] (kALCtext) at (kALC) {Abstract $k$-Lipschitz Continuity};

  \coordinate (kpComp) at (-4,-2.35);
  \node[prop,anchor=east] (kpComptext) at (kpComp) {$k\varepsilon$-Partial Completeness};

  \coordinate (pComp) at ($(kALCtext.west)+(.9,-4.5)$);
  \node[prop] (pComptext) at (pComp) {$\varepsilon$-Partial Completeness};

  \coordinate (aRob) at ($(kALCtext)+(0,-2.95)$);
  \node[prop,text width=2cm] (aRobtext) at (aRob) {Abstract \\ Robustness};

  \coordinate (Comp) at ($(kALCtext.west)+(.9,-6.5)$);
  \node[prop] (Comptext) at (Comp) {Completeness};

  \coordinate (kpANI) at (4,-2.35);
  \node[prop,anchor=west] (kpANItext) at (kpANI) {$2k\varepsilon$-Partial ANI\vphantom{p}};

  \coordinate (pANI) at ($(kALCtext.east)-(1,4.5)$);
  \node[prop] (pANItext) at (pANI) {$\varepsilon$-Partial ANI\vphantom{p}};

  \coordinate (ANI) at ($(kALCtext.east)-(1,6.5)$);;
  \node[prop] (ANItext) at (ANI) {ANI\vphantom{p}};

  \begin{scope}[on background layer]
    \foreach \n in {kALCtext,kpComptext,pComptext,aRobtext,Comptext,kpANItext,pANItext,ANItext}
      \filldraw[fill=gray!15,draw=white,line width=2pt,rounded corners=3pt] (\n.south west) rectangle (\n.north east);
  \end{scope}

  \draw[impl] ($(kALCtext.south west)+(.9,0)$) -- node[midway,left,font=\scriptsize,yshift=-2pt] {Thm.\,\ref{thm:alc-pcompl}}
    node[pos=.1,right,scale=.7] {$\langle \tdelta{C}^{\rho}, \tdelta{C}^{\rho} \rangle$}
    node[pos=.9,right,scale=.7] {$\langle \rho \rangle$} (pComptext.north);
  \draw[impl] ($(kALCtext.south east)-(1,0)$) -- node[midway,right,font=\scriptsize,yshift=-2pt] {Thm.\,\ref{thm:alc-pani}}
    node[pos=.1,right,scale=.7] {$\langle \tdelta{C}^{\eta}, \tdelta{D}^{\rho} \rangle$}
    node[pos=.9,right,scale=.7] {$\langle \tdelta{C}^{\eta}, \tdelta{D}^{\rho} \rangle$} (pANItext.north);

  \draw[impl] (kALCtext.west) -- node[midway,above,font=\scriptsize,xshift=-6pt,yshift=-12pt] {\rotatebox{39}{Thm.\,\ref{th:p-comp-aulc}}}
    node[abs,pos=.25] {$\langle \tdelta{C}^{\eta_1}, \tdelta{D}^{\rho} \rangle$}
    node[abs,pos=.7] {$\langle \eta_2, \rho \rangle$} ($(kpComptext.north east)-(1,0)$);
  \draw[impl] (kALCtext.east) -- node[midway,above,font=\scriptsize,xshift=4pt,yshift=-12pt] {\rotatebox{-40}{Thm.\,\ref{thm:p-ani-aulc}}}
    node[abs,pos=.25] {$\langle \tdelta{1C}^{\eta_1}, \tdelta{D}^{\rho} \rangle$}
    node[abs,pos=.7] {$\langle \tdelta{2C}^{\eta_2}, \tdelta{D}^{\rho} \rangle$} ($(kpANItext.north west)+(1,0)$);
  \draw[impl] ($(kpComptext.south east)-(1,0)$) -- node[midway,left,font=\scriptsize,xshift=8pt,yshift=-2pt] {\rotatebox{-46}{Cor.\,\ref{cor:0-p-comp-aulc}}} (pComptext.north west);
  \draw[impl] ($(kpANItext.south west)+(1,0)$) -- node[midway,right,font=\scriptsize,xshift=-12pt,yshift=-4pt] {\rotatebox{36}{Cor.\,\ref{cor:0-p-ani-aulc}}} (pANItext.north east);

  \draw[impl] ($(kALCtext.west |- pComptext.south)+(.9,0)$) -- node[midway,left,font=\scriptsize] {Cor.\,\ref{cor:alc-compl}} (Comptext.north);
  \draw[impl] ($(kALCtext.east |- pANItext.south)-(1,0)$) -- node[midway,right,font=\scriptsize] {Cor.\,\ref{cor:alc-pani}} (ANItext.north);

  \node[anchor=west,scale=.7,xshift=1pt,yshift=-15pt] at (kALCtext.south) {$\langle \tdelta{C}^{id}, \tdelta{D}^{\rho} \rangle$};
  \node[anchor=west,scale=.7,xshift=1pt,yshift=15pt] at (aRobtext.north) {$\langle \rho \rangle$};
  \draw[impl] (kALCtext.south) -- node[midway,left] {\scriptsize Thm.\,\ref{thm:alc-ar}} node[midway,right,xshift=2pt,scale=.7] {$k = 0$} (aRobtext.north);
\end{tikzpicture}
	\caption{Relation between ALC and other extensional program properties.}
    \label{fig:relations}
\end{figure}
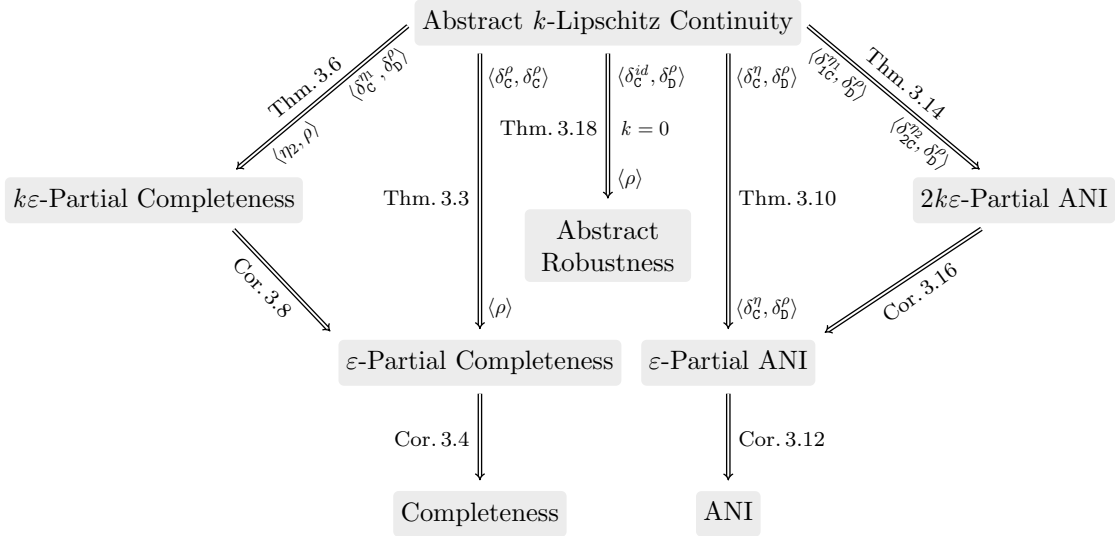

Fig.~\ref{fig:relations} summarizes all the relations between ALC, partial completeness, partial abstract non-interference and abstract robustness formalized in this section.

\subsection{Partial Completeness}
\label{sec:pcompl}
Partial completeness~\cite{campion2022popl,CampionSAS23} is a weakening of the standard notion of completeness in abstract interpretation.
In the classical abstract interpretation framework~\cite{CC77}, the computation of a concrete (possibly uncomputable) monotone function $f\colon C \to C$ over a poset $\tuple{C, \tpreord{C}}$ is replaced by an abstract (possibly computable) sound computation $f^\natural \colon A \to A$ over an abstract domain $\tuple{A = \rho(C), \tpreord{C}}$, where $\rho \in \uco(C)$.
For example, if $\prog$ is a program with a single variable then $f$ could be the collecting big-step semantics $\collSem{\prog} \colon \wp(\integer) \to \wp(\integer)$ defined in Section~\ref{sec:prog} over $\tuple{\wp(\integer), \subseteq}$, $\rho = \Int$ the interval abstract domain (Example~\ref{ex:qual}) and $\collSem{\prog}^\natural \colon \Int(\wp(\integer)) \to \Int(\wp(\integer))$ is the abstract interval semantics soundly computed on intervals.
An abstract function $f^\natural \colon A \to A$ is \emph{sound} when $\rho  f \tpreord{C} f^\natural \rho$, and is said to be
\emph{complete} when $\rho  f = f^\natural \rho$. In this case $f^\natural$ coincides on $A$ with the \emph{best correct approximation} (bca)~\cite{CC79} $\rho  f \rho$ of $f$, so that a complete abstract function exists if and only if $\rho f = \rho f \rho$, a condition that only depends on $\rho$ and $f$~\cite{GRS00}.
Completeness %
is the best possible scenario where no imprecision is introduced by the abstract computation of $f^\natural$.
In practice, however, completeness is rarely achieved.
For this reason, Campion et al. \cite{campion2022popl,CampionSAS23} introduced a weaker notion of completeness, called $\varepsilon$-\emph{partial completeness} which allows some degree of imprecision limited by $\varepsilon$. This imprecision is measured by pre-metrics $\tdelta{A} \colon A \times A \to \real^\infty_{\geq 0}$ that are \emph{$\tpreord{C}$-compatible} with the abstract domain, i.e., pre-metrics additionally satisfying~\cite{CampionSAS23}, for all $x,y,z \in A$:
\[
    x \tpreord{C} y \tpreord{C} z \;\Ra\; \tdelta{A}(x,y) \leq \tdelta{A}(x,z) \,\land\, \tdelta{A}(y,z) \leq \tdelta{A}(x,z).
\]

Partial completeness was originally defined as a local property (namely, defined on a strict subset of the input domain)~\cite{campion2022popl}. Here, for our purposes, we consider a {\em global} version of $\varepsilon$-partial completeness:
\begin{equation} \label{eq:partial-completeness}
    \forall x \in C\sut \tdelta{A}(\rho f(x), f^\natural \rho(x)) \leq \varepsilon
\end{equation}

To relate ALC w.r.t. $\tuple{\tudelta{\rho}{C},\tudelta{\rho}{C}}$ of a monotone function $f \colon C \to C$ to partial completeness,
we observe that, given a $\tpreord{C}$-compatible pre-metric  $\tdelta{C}$ defined over $C$, since $\rho(C) \subseteq C$ we can always restrict $\tdelta{C}$ to elements of $\rho(C)$, i.e., $\forall x,y \in \rho(C) \colon \tdelta{\rho(C)}(x,y) \definit \tdelta{C}(x,y)$ without losing the $\tpreord{C}$-compatible property.
With the restricted distance $\tdelta{\rho(C)}$ in hand, we can now state our first relation between ALC and partial completeness:

\begin{thm} \label{thm:alc-pcompl}
    Let $\tuple{C, \tpreord{C}}$ be a poset, let $\tuple{C, \tdelta{C}}$ be a $\tpreord{C}$-compatible pre-metric space, and let $\rho \in \uco(C)$.
    If a monotone function $f\colon \tC \ra \tC$ satisfies ALC w.r.t. $\tuple{\tudelta{\rho}{C},\tudelta{\rho}{C}}$, then the bca $\rho f \rho$ of $f$ satisfies $\varepsilon$-partial completeness w.r.t. $f$, for any $\varepsilon \in \real_{\geq 0}$. %
\end{thm}
\begin{proof}
We prove the following implication:
    \[
        [\forall x,y\in C\sut \tdelta{C}^\rho( f(x), f(y))\leq k\tdelta{C}^{\rho}(x,y)] \ \Rightarrow \ [\forall x \in C \sut \tdelta{\rho(C)}(\rho f(x), \rho f\rho (\rho(x))) \leq 0]
     \]
Let us assume $k$-ALC w.r.t.\ $\tuple{\tdelta{C}^\rho,\tdelta{C}^\rho}$, and let $x \in C$. Instantiating ALC with $y = \rho(x)$, we have $\tdelta{C}^\rho(f(x), f\rho(x)) \leq k\tdelta{C}^{\rho}(x,\rho(x))$. By idempotence of $\rho$ and (\textit{if-identity}), $\tdelta{C}^{\rho}(x,\rho(x)) = \tdelta{C}(\rho(x),\rho(x)) = 0$, hence $\tdelta{C}^\rho(f(x), f\rho(x)) \leq 0$. Then, by idempotence of $\rho$, by definition of $\tdelta{\rho(C)}$ (as $\rho f(x), \rho f\rho(x) \in \rho(C)$), and by definition of $\tdelta{C}^\rho$, we have:
    \[
        \tdelta{\rho(C)}(\rho f(x), \rho f \rho(\rho(x))) = \tdelta{\rho(C)}(\rho f(x), \rho f\rho(x)) = \tdelta{C}(\rho f(x), \rho f\rho(x)) = \tdelta{C}^\rho(f(x), f\rho(x)) \leq 0
    \]
Hence, the bca $\rho f \rho$ of $f$ satisfies $0$-partial completeness w.r.t.\ $f$, and thus $\varepsilon$-partial completeness holds for any $\varepsilon \in \real_{\geq 0}$.
\end{proof}
Monotonicity of $f$ is not used in the proof: for an arbitrary $f$, the same bound holds for $\rho f \rho$, which however is not guaranteed to be a sound approximation of $f$, and hence cannot be called the bca of $f$. In any case, monotonicity is not a restriction when $f$ is the collecting semantics $\collSem{\prog}$ of a program $\prog \in \imp$, which is monotone by construction (Section~\ref{sec:prog}).

Note that the combined distance $\tudelta{\rho}{C}$ never satisfies (\textit{iff-identity}) on $C$ (Proposition~\ref{prop:comb-dist}), whereas the restriction of $\tdelta{C}$ to $\rho(C)$ may regain it (cf.\ Example~\ref{ex:inclusion-dist-int-axioms}). In this case, ALC guarantees completeness:
\begin{cor} \label{cor:alc-compl}
    Let $\tuple{C,\tpreord{C}}$ be a poset, $\tuple{C, \tdelta{C}}$ a $\tpreord{C}$-compatible pre-metric space, and $\rho\in\uco(C)$, such that $\tdelta{\rho(C)}$ is a quasisemi-metric.
    If a monotone function $f\colon \tC \ra \tC$ satisfies ALC w.r.t. $\tuple{\tudelta{\rho}{C},\tudelta{\rho}{C}}$, then the bca $\rho f \rho$ of $f$ satisfies completeness w.r.t. $f$.
\end{cor}
\begin{proof}
By Theorem~\ref{thm:alc-pcompl}, $\rho f \rho$ satisfies $0$-partial completeness w.r.t.\ $f$, namely for all $x \in C$ we have
    $\tdelta{\rho(C)}(\rho f(x), \rho f\rho(x)) \leq 0$.
Since $\tdelta{\rho(C)}$ is a quasisemi-metric, it satisfies (\textit{iff-identity}), hence $\forall x \in C\sut \rho f(x) = \rho f\rho(x) = \rho f \rho(\rho(x))$, by idempotence of $\rho$. Thus, $\rho f \rho$ satisfies completeness w.r.t.\ $f$.
\end{proof}

Theorem~\ref{thm:alc-pcompl} and Corollary~\ref{cor:alc-compl} reveal a novel relationship between the ALC of a function $f$ and the partial completeness of its bca, particularly in the context of program analysis where $f$ represents a monotone semantics $\collSem{\prog}$ of a given program $\prog$. Notably, a proof of ALC of $\collSem{\prog}$ w.r.t. the distance $\tudelta{\rho}{C}$ implies that the imprecision introduced by approximating $\collSem{\prog}$ with its bca $\rho \collSem{\prog} \rho$ is bounded by any constant $\varepsilon$. Consequently, $\rho \collSem{\prog} \rho$ also satisfies $0$-partial completeness, namely \emph{no imprecision is generated by the bca according to $\tdelta{\rho(C)}$}. Thus, ALC of a program $\collSem{\prog}$ \emph{ensures the existence of an abstract interpreter capable of approximating the computation of $\collSem{\prog}$ over $\rho(C)$ with no imprecision} according to the imprecision measured by $\tdelta{\rho(C)}$. Furthermore, when the distance $\tdelta{\rho(C)}$ is a quasisemi-metric, then the bca $\rho \collSem{\prog} \rho$ is guaranteed to satisfy completeness, namely, \emph{no imprecision is generated by the bca}.
Another way to interpret Corollary~\ref{cor:alc-compl} (and, analogously, Theorem~\ref{thm:alc-pcompl}) is as follows: if a program $\collSem{\prog}$ does not admit a complete bca $\rho \collSem{\prog} \rho$, then $\collSem{\prog}$ \emph{cannot} satisfy ALC w.r.t.\ $\tuple{\tudelta{\rho}{C},\tudelta{\rho}{C}}$ for any $\tpreord{C}$-compatible $\tdelta{C}$ such that $\tdelta{\rho(C)}$ is a quasisemi-metric.

\begin{exa} \label{ex:alc-pcompl}
Consider again the program $\progp{R}$, its Kleene closure $\progp{R}^\ast$, and the interval closure $\Int \in \uco(\wp(\integer))$ from Example~\ref{ex:R}, and recall from Example~\ref{ex:R-ext} that, on non-empty sets of integers, the (monotone) collecting semantics $\collSem{\progp{R}^\ast}$ satisfies $1$-ALC w.r.t.\ $\tuple{\dincl{\Int},\dincl{\Int}}$.

By Theorem~\ref{thm:alc-pcompl}, the bca $\Int \collSem{\progp{R}^\ast} \Int$ then satisfies $0$-partial completeness w.r.t.\ $\collSem{\progp{R}^\ast}$, i.e., $\dincl{\Int}(\collSem{\progp{R}^\ast}S, \collSem{\progp{R}^\ast}\Int(S)) \leq 0$, for any non-empty $S \in \wp(\integer)$. Moreover, the restriction of $\dincl{}$ to $\Int(\wp(\integer))$ is a quasi-metric (Example~\ref{ex:inclusion-dist-int-axioms}), hence also a quasisemi-metric, so Corollary~\ref{cor:alc-compl} applies, and we can conclude that $\Int \collSem{\progp{R}^\ast} \Int$ is complete, namely it does not add any imprecision when approximating $\collSem{\progp{R}^\ast}$.

Consider instead the pseudo-metric space $\langle \wp(\integer), \dcount{} \rangle$ and the semantics $\collSem{\progp{R}}$ of $\progp{R}$. The bca $\Int \collSem{\progp{R}} \Int$ does not satisfy $0$-partial completeness w.r.t. $\dcount{\Int}$: given $X = \{-1,1\}$, we have that $\dcount{\Int} (\collSem{\progp{R}}X, \collSem{\progp{R}}\Int(X)) = \dcount{\Int} ([0,0], [0,1]) = 1 \neq 0$ and so $\collSem{\progp{R}}$ cannot satisfy ALC for $\langle \dcount{\Int}, \dcount{\Int} \rangle$. %
$\qedex$
\end{exa}

Theorem~\ref{thm:alc-pcompl} and Corollary~\ref{cor:alc-compl} use the same $\rho \in \uco(C)$ as both the input and the output abstraction of ALC (Definition~\ref{def:alc}), as well as to build the bca $\rho f \rho$.
In practice, however, the input abstraction for which ALC is known need not be the one through which an analysis approximates its input.
The latter is usually dictated by the analysis at hand, e.g., by the abstract domain of an existing analyser, whereas ALC may have been established  for an abstraction that makes the proof possible, e.g., one for which the basic commands are ALC; moreover, $f$ need not be ALC at all w.r.t.\ the input abstraction of the analysis (see Example~\ref{ex:th-p-comp-aulc}).
We therefore generalize the above relation by keeping $\rho$ as the output abstraction, but using two different input abstractions: $\eta_1$ for ALC, and $\eta_2$ for partial completeness.
Intuitively, replacing an input $x$ with its abstraction $\eta_2(x)$ introduces an input error, whose magnitude can be measured through $\eta_1$, and ALC propagates this error linearly to the output, as in the classical use of Lipschitz constants to bound the effect of input perturbations~\cite{Gulwani11,gulwani12}: the resulting bound on the imprecision of the bca $\rho f \eta_2$ is proportional to $k$.
To this end, we first extend~\eqref{eq:partial-completeness} to functions $f \colon C \to D$ between two posets $\tuple{C,\tpreord{C}}$ and $\tuple{D,\tpreord{D}}$, with an input abstraction $\eta \in \uco(C)$ and an output abstraction $\rho \in \uco(D)$. An abstract function $f^\natural \colon \eta(C) \to \rho(D)$ is sound when $\rho f \tpreord{D} f^\natural \eta$, and the bca of $f$ is $\rho f \eta$, which is sound whenever $f$ is monotone. Given a $\tpreord{D}$-compatible pre-metric $\tdelta{\rho(D)}$ on $\rho(D)$, $f^\natural$ satisfies \emph{$\varepsilon$-partial completeness} w.r.t.\ $f$ when
\begin{equation} \label{eq:partial-completeness-gen}
    \forall x \in C\sut \tdelta{\rho(D)}(\rho f(x), f^\natural \eta(x)) \leq \varepsilon
\end{equation}
Clearly, \eqref{eq:partial-completeness} is the special case where $D = C$ and $\eta = \rho$. As before, given a $\tpreord{D}$-compatible pre-metric $\tdelta{D}$ on $D$, we take as $\tdelta{\rho(D)}$ the restriction of $\tdelta{D}$ to $\rho(D)$.

\begin{thm} \label{th:p-comp-aulc}
Let $\tuple{C,\tpreord{C}}$ and $\tuple{D,\tpreord{D}}$ be posets, $\tuple{C,\tdelta{C}}$ a pre-metric space, $\tuple{D,\tdelta{D}}$ a $\tpreord{D}$-compatible pre-metric space, and $\eta_1,\eta_2\in\uco(C)$, $\rho\in\uco(D)$. Let $k\in \real_{\geq 0}$.
If a monotone function $f\colon C\ra D$ satisfies $k$-ALC w.r.t.\ $\tuple{\tdelta{C}^{\eta_1},\tdelta{D}^\rho}$, then the bca $\rho f \eta_2$ of $f$ satisfies $k\varepsilon$-partial completeness w.r.t.\ $f$, for $\varepsilon\definit\sup\sset{\tdelta{C}^{\eta_1}(x,\eta_2(x))}{x\in C}$.
\end{thm}
\begin{proof}
We prove the following implication:
\[
    [\forall x,y\in C\sut \tdelta{D}^\rho( f(x), f(y))\leq k\tdelta{C}^{\eta_1}(x,y)] \ \Rightarrow \ [\forall x \in C \sut \tdelta{\rho(D)}(\rho f(x), \rho f\eta_2(\eta_2(x))) \leq k\varepsilon]
\]
Let us assume $k$-ALC w.r.t.\ $\tuple{\tdelta{C}^{\eta_1},\tdelta{D}^\rho}$, and let $x \in C$. Instantiating ALC with $y = \eta_2(x)$, we have $\tdelta{D}^\rho(f(x), f\eta_2(x)) \leq k\tdelta{C}^{\eta_1}(x,\eta_2(x)) \leq k\varepsilon$, by definition of $\varepsilon$. Then, by idempotence of $\eta_2$, by definition of $\tdelta{\rho(D)}$ (as $\rho f(x), \rho f\eta_2(x) \in \rho(D)$), and by definition of $\tdelta{D}^\rho$, we have:
\[
    \tdelta{\rho(D)}(\rho f(x), \rho f\eta_2(\eta_2(x))) = \tdelta{\rho(D)}(\rho f(x), \rho f\eta_2(x)) = \tdelta{D}(\rho f(x), \rho f\eta_2(x)) = \tdelta{D}^\rho(f(x), f\eta_2(x)) \leq k\varepsilon
\]
Hence, the bca $\rho f \eta_2$ of $f$ satisfies $k\varepsilon$-partial completeness w.r.t.\ $f$.
\end{proof}

\looseness=-1
In this case, the bound on the output error for partial completeness depends on the largest distance between an input $x \in C$ and its abstraction $\eta_2(x)$ when observed through $\eta_1$.

\begin{exa} \label{ex:th-p-comp-aulc}
Consider again the collecting semantics
$\collSem{\progp{R}^\ast}$ on non-empty sets of integers
(Example~\ref{ex:R-ext}), the sign closure $\Sign$
(Example~\ref{ex:sign}), and the $\subseteq$-compatible path-length
distance $\dpath{}$ on the domain of signs $\tuple{\Sign(\wp(\integer)),\subseteq}$
(Example~\ref{ex:path-len}).
Note that, for any $A,B\in\Sign(\wp(\integer))$ with
$\varnothing\neq A\subseteq B$, we have $\dpath{}(A,B)\leq 2$.

Since
$\collSem{\progp{R}^\ast}S =
[\min(\inf S,0),\max(\sup S,1)]$,
we have
\begin{equation} \label{eq:R-star-sign}
    \Sign(\collSem{\progp{R}^\ast}S) =
    \begin{cases}
        \integer, & \text{if $S$ contains a negative value},\\
        \integer_{\geq 0}, & \text{otherwise}.
    \end{cases}
\end{equation}
Hence, $\collSem{\progp{R}^\ast}$ satisfies $1$-ALC w.r.t.\
$\tuple{\dpath{\Sign},\dpath{\Sign}}$.

Let now $\eta_2=\Par$ be the classical parity closure,
mapping a non-empty set of integers to $\even\definit\{2n\mid n\in\integer\}$ if all its elements are even, to $\odd\definit\{2n+1\mid n\in\integer\}$ if all its elements are odd, and to $\integer$ otherwise.
Since
$\Sign(S)\subseteq\Sign(\Par(S))$, we have
$\dpath{}(\Sign(S),\Sign(\Par(S)))\leq 2$.
This bound is attained for $S=\{2\}$.
Hence $\varepsilon=2$.
By Theorem~\ref{th:p-comp-aulc}, with
$\eta_1=\rho=\Sign$, $\eta_2=\Par$, and $k=1$,
the bca $\Sign\collSem{\progp{R}^\ast}\Par$ satisfies
$2$-partial completeness w.r.t.\
$\collSem{\progp{R}^\ast}$.

Observe that the bound is not tight: every parity abstraction
of a non-empty set contains negative values, so
$\Sign(\collSem{\progp{R}^\ast}\Par(S))=\integer$.
By~\eqref{eq:R-star-sign}, the actual imprecision is $0$
if $S$ contains a negative value, and $1$ otherwise. Note also that ALC cannot be established directly w.r.t.\ $\Par$: for any pre-metric $\tdelta{C}$, we have $\tdelta{C}^{\Par}(\{2\},\{-2\}) = \tdelta{C}(\even,\even) = 0$ by (\textit{if-identity}), whereas $\dpath{\Sign}(\collSem{\progp{R}^\ast}\{2\},\collSem{\progp{R}^\ast}\{-2\}) = \dpath{}(\integer_{\geq 0},\integer) = 1$ by~\eqref{eq:R-star-sign}, so $\collSem{\progp{R}^\ast}$ does not satisfy $k$-ALC w.r.t.\ $\tuple{\tdelta{C}^{\Par},\dpath{\Sign}}$ for any $k$. Theorem~\ref{th:p-comp-aulc} thus yields a guarantee for the parity analysis that ALC w.r.t.\ $\Par$ cannot provide.
$\qedex$
\end{exa}

When $\eta_2$ is more precise than $\eta_1$ (Lemma~\ref{lem:eta1eta2}), the bound of Theorem~\ref{th:p-comp-aulc} becomes $0$ and, as in Theorem~\ref{thm:alc-pcompl}, partial completeness holds for any $\varepsilon$:

\begin{cor} \label{cor:0-p-comp-aulc}
Let $\tuple{C,\tpreord{C}}$ and $\tuple{D,\tpreord{D}}$ be posets, $\tuple{C,\tdelta{C}}$ a pre-metric space, $\tuple{D,\tdelta{D}}$ a $\tpreord{D}$-compatible pre-metric space, and $\eta_1,\eta_2\in\uco(C)$, $\rho\in\uco(D)$, such that $\eta_2 \tpreord{C} \eta_1$.
If a monotone function $f\colon C\ra D$ satisfies ALC w.r.t.\ $\tuple{\tdelta{C}^{\eta_1},\tdelta{D}^\rho}$, then the bca $\rho f \eta_2$ of $f$ satisfies $\varepsilon$-partial completeness w.r.t.\ $f$, for any $\varepsilon \in \real_{\geq 0}$.
\end{cor}
\begin{proof}
By Lemma~\ref{lem:eta1eta2}, $\eta_1\eta_2(x) = \eta_1(x)$ for every $x \in C$, so $\tdelta{C}^{\eta_1}(x,\eta_2(x)) = \tdelta{C}(\eta_1(x),\eta_1\eta_2(x)) = \tdelta{C}(\eta_1(x),\eta_1(x)) = 0$ by (\textit{if-identity}). Hence $\sup\sset{\tdelta{C}^{\eta_1}(x,\eta_2(x))}{x\in C} = 0$, and, by Theorem~\ref{th:p-comp-aulc}, the bca $\rho f \eta_2$ of $f$ satisfies $0$-partial completeness w.r.t.\ $f$, and thus $\varepsilon$-partial completeness holds for any $\varepsilon \in \real_{\geq 0}$.
\end{proof}

\begin{exa} \label{ex:cor-0-p-comp-aulc}
Consider again the collecting semantics
$\collSem{\progp{R}^\ast}$ on non-empty sets of integers,
and the $\subseteq$-compatible path-length distance $\dpath{}$
(Example~\ref{ex:path-len}), but now on the simplified domain of signs
$\tuple{\Sign'(\wp(\integer)),\subseteq}$ of Example~\ref{ex:eta1eta2}.
Note that $\Sign'(\collSem{\progp{R}^\ast}S) = \Sign(\collSem{\progp{R}^\ast}S)$ as defined in~\eqref{eq:R-star-sign}.
Hence, as in Example~\ref{ex:th-p-comp-aulc}, $\collSem{\progp{R}^\ast}$ satisfies $1$-ALC w.r.t.\
$\tuple{\dpath{\Sign'},\dpath{\Sign'}}$.

\looseness=-1
Let now $\eta_2=\Int$. Since $\Int$ is more precise than $\Sign'$ (Example~\ref{ex:eta1eta2}),
by Corollary~\ref{cor:0-p-comp-aulc}, with
$\eta_1=\rho=\Sign'$ and $\eta_2=\Int$,
the bca $\Sign'\collSem{\progp{R}^\ast}\Int$ satisfies
$0$-partial completeness w.r.t.\
$\collSem{\progp{R}^\ast}$.
Indeed, $\Int(S)$ contains a negative value if and only if $S$ does, hence
$\Sign'(\collSem{\progp{R}^\ast}\Int(S)) = \Sign'(\collSem{\progp{R}^\ast}S)$.
Unlike $\Par$ in Example~\ref{ex:th-p-comp-aulc}, $\Int$ does not lose the sign information that $\collSem{\progp{R}^\ast}$ depends on.
$\qedex$
\end{exa}

\subsection{Partial Abstract Non-Interference}
\label{sec:pani}
Partial Abstract Non-Interference~\cite{CampionMU25} (PANI for short) is a relaxation of (abstract) non-interference~\cite{goguenmes82,GM:ANI2018} that combines both property-based and quantitative approximations on the output of a function computation. This notion can be seen as a generalization of approximate non-interference~\cite{DHW02}, originally introduced in a probabilistic process algebra,
 requiring %
 the {\em observable} behaviors of two agents to be under a similarity threshold $\varepsilon$. Specifically, PANI observes indistinguishable (distance-wise) {\em properties} of input data, while allowing a {\em bounded distance} between the observed output {\em properties}. This enables a more nuanced treatment of security policies, especially when small, bounded differences in outputs are tolerable. Under the same premises as Definition~\ref{def:alc}, $\varepsilon$-PANI w.r.t. $\tuple{\tdelta{C}^\eta,\tdelta{D}^\rho}$ has been formalized as~\cite{CampionMU25}:
\begin{equation} \label{eq:pani}
    \forall x,y \in \tC \sut \tdelta{C}^\eta(x,y) = 0 \ \Ra \ \tdelta{D}^\rho(f(x),f(y)) \leq \varepsilon
\end{equation}
Given inputs with zero property distance under $\tudelta{\eta}{C}$, $\varepsilon$-PANI w.r.t. $\tuple{\tdelta{C}^\eta,\tdelta{D}^\rho}$ allows the function to produce different outputs, potentially with different properties under $\tudelta{\rho}{D}$, as long as the variation remains bounded, i.e., not exceeding the threshold $\varepsilon$.

It turns out that ALC is a stronger property than PANI:
\begin{thm} \label{thm:alc-pani}
    Let $\tuple{C,\tpreord{C}}$ and $\tuple{D,\tpreord{D}}$ be posets, $\tuple{C,\tdelta{C}}$ and $\tuple{D,\tdelta{D}}$ pre-metric spaces, and $\eta\in\uco(C)$, $\rho\in\uco(D)$.
    If a function $f\colon \tC \ra D$ satisfies ALC w.r.t. $\tuple{\tudelta{\eta}{C}, \tudelta{\rho}{D}}$, then $f$ satisfies $\varepsilon$-PANI w.r.t. $\tuple{\tdelta{C}^\eta,\tdelta{D}^\rho}$, for any $\varepsilon \in \real_{\geq 0}$.
\end{thm}
\begin{proof}
    We prove the following implication:
    \begin{align*}
        & [\forall x,y\in C\sut \tdelta{D}^\rho( f(x), f(y)) \leq k\tdelta{C}^\eta(x,y)] \ \Rightarrow \\
        &
        [\forall x,y\in C \sut \tdelta{C}^{\eta}(x,y) = 0\ \Ra\ \tdelta{D}^\rho( f(x), f(y)) \leq 0]
    \end{align*}
    Let us assume $k$-ALC w.r.t.\ $\tuple{\tdelta{C}^\eta,\tdelta{D}^\rho}$, and let $x,y \in C$ be such that $\tdelta{C}^{\eta}(x,y) = 0$. Then, by ALC, we have $\tdelta{D}^\rho(f(x), f(y)) \leq k\tdelta{C}^\eta(x,y) = k \cdot 0 = 0$.
    Hence, $f$ satisfies $0$-PANI w.r.t. $\tuple{\tdelta{C}^\eta,\tdelta{D}^\rho}$, and thus $\varepsilon$-PANI w.r.t. $\tuple{\tdelta{C}^\eta,\tdelta{D}^\rho}$ holds for any $\varepsilon \in \real_{\geq 0}$.
\end{proof}

\begin{exa}\label{ex:alc-pani}
\looseness=-1
Consider again the semantics $\collSem{\progp{R}^\ast}$ which, on non-empty sets of integers, satisfies $1$-ALC w.r.t.\ $\tuple{\dincl{\Int},\dincl{\Int}}$ (Example~\ref{ex:R-ext}). By Theorem~\ref{thm:alc-pani}, $\collSem{\progp{R}^\ast}$ also satisfies $0$-PANI w.r.t. $\tuple{\dincl{\Int},\dincl{\Int}}$.
$\qedex$
\end{exa}

When $\tdelta{D}$ satisfies (\textit{iff-identity}), ALC implies Abstract Non-Interference (ANI)~\cite{GM:ANI2018}, which replaces distances with equalities: $\forall x,y \in \tC \sut \eta(x)=\eta(y) \ \Ra \ \rho f(x)=\rho f(y)$.
\begin{cor} \label{cor:alc-pani}
    Let $\tuple{C,\tpreord{C}}$ and $\tuple{D,\tpreord{D}}$ be posets, $\tuple{C,\tdelta{C}}$ a pre-metric space, $\tuple{D,\tdelta{D}}$ a quasisemi-metric space, and $\eta\in\uco(C)$, $\rho\in\uco(D)$.
    If a function $f\colon \tC \ra D$ satisfies ALC w.r.t. $\tuple{\tudelta{\eta}{C}, \tudelta{\rho}{D}}$, then $f$ satisfies ANI w.r.t.\ $\tuple{\eta,\rho}$.
\end{cor}
\begin{proof}
    By Theorem~\ref{thm:alc-pani}, $f$ satisfies $0$-PANI, namely: $\forall x,y\in C \colon \tdelta{C}^{\eta}(x,y) = 0\ \Ra\ \tdelta{D}^\rho( f(x), f(y)) \leq 0$.
    Let $x,y \in C$ be such that $\eta(x) = \eta(y)$. By (\textit{if-identity}) of $\tdelta{C}$, $\tdelta{C}^{\eta}(x,y) = \tdelta{C}(\eta(x),\eta(x)) = 0$, hence $\tdelta{D}^\rho(f(x), f(y)) = \tdelta{D}(\rho f(x), \rho f(y)) \leq 0$. Since $\tdelta{D}$ is a quasisemi-metric, it satisfies (\textit{iff-identity}), hence $\rho f(x) = \rho f(y)$, namely $f$ satisfies ANI w.r.t.\ $\tuple{\eta,\rho}$.
\end{proof}

\begin{exa} \label{ex:alc-ani}
Continuing Example~\ref{ex:alc-pani}, since $\dincl{}$ is a quasi-metric (Example~\ref{ex:inclusion-dist}), hence also a quasisemi-metric, Corollary~\ref{cor:alc-pani} gives that $\collSem{\progp{R}^\ast}$ satisfies ANI w.r.t.\ $\tuple{\Int,\Int}$. $\qedex$
\end{exa}

Theorem~\ref{thm:alc-pani} and Corollary~\ref{cor:alc-pani} use the same $\eta \in \uco(C)$ as the input abstraction of both ALC and PANI (or ANI).
In a security setting, however, the input abstraction of PANI is dictated by the policy, as it specifies which inputs must be (almost) indistinguishable to an observer of the output (e.g., which property of the input may be released~\cite{GM:ANI2018}), whereas ALC may be known for a different input abstraction, as discussed in Section~\ref{sec:pcompl}.
Following the generalization of Section~\ref{sec:pcompl}, we now use two different input abstractions: $\eta_1 \in \uco(C)$ for ALC, and $\eta_2 \in \uco(C)$ for PANI.
Intuitively, two inputs $x$ and $y$ that are indistinguishable through $\eta_2$ have the same abstraction $\eta_2(x) = \eta_2(y)$, and replacing each of them with it introduces an input error that ALC propagates to the output, so that their outputs are at most twice the propagated error apart.
This intuition requires the inputs that PANI deems indistinguishable to be exactly those with the same $\eta_2$-abstraction.
For this reason, Theorem~\ref{thm:p-ani-aulc} also uses two different input distances, $\tdelta{1C}$ for ALC and $\tdelta{2C}$ for PANI, which play different roles.
The distance $\tdelta{1C}$ is used quantitatively: it measures how far each input $x$ is from $\eta_2(x)$, as seen through $\eta_1$, and thus determines, together with $k$, the threshold $2k\varepsilon$.
It can be any pre-metric under which ALC is known, and it only needs to be defined on $\eta_1(C)$ (Section~\ref{sec:combining}).
The distance $\tdelta{2C}$ is instead used only qualitatively: by~\eqref{eq:pani}, only the pairs of inputs at distance $0$ matter, so $\tdelta{2C}$ only specifies which inputs must be (almost) indistinguishable.
Requiring it to be a quasisemi-metric ensures that $\tdelta{2C}^{\eta_2}(x,y) = 0$ if and only if $\eta_2(x) = \eta_2(y)$, as in ANI, whatever the choice of $\tdelta{2C}$.
The distance under which ALC is known cannot be used for PANI in general, as it may be defined on $\eta_1(C)$ only, or it may identify inputs with different $\eta_2$-abstractions, in which case the conclusion of the theorem may fail.
This issue does not arise with a single input abstraction, since ALC maps inputs at distance $0$ to outputs at distance $0$, whatever the pre-metric (Theorem~\ref{thm:alc-pani}).

\begin{thm} \label{thm:p-ani-aulc}
Let $\tuple{C,\tpreord{C}}$ and $\tuple{D,\tpreord{D}}$ be posets, $\tuple{C,\tdelta{1C}}$ a pre-metric space, $\tuple{C,\tdelta{2C}}$ a quasisemi-metric space, $\tuple{D,\tdelta{D}}$ a pseudo-metric space, and $\eta_1,\eta_2\in\uco(C)$, $\rho\in\uco(D)$. Let $k\in \real_{\geq 0}$.
If a function $f\colon C\ra D$ satisfies $k$-ALC w.r.t.\ $\tuple{\tdelta{1C}^{\eta_1},\tdelta{D}^{\rho}}$, then $f$ satisfies $2k\varepsilon$-PANI w.r.t. $\tuple{\tdelta{2C}^{\eta_2},\tdelta{D}^{\rho}}$, for $\varepsilon\definit\sup\sset{\tdelta{1C}^{\eta_1}(x,\eta_2(x))}{x\in C}$.
\end{thm}
\begin{proof}
We prove the following implication:
\begin{align*}
& [\forall x,y\in C\sut \tdelta{D}^\rho( f(x), f(y))\leq k\tdelta{1C}^{\eta_1}(x,y)] \ \Rightarrow \\
&
[\forall x,y\in C\sut \tdelta{2C}^{\eta_2}(x,y) = 0\ \Ra\ \tdelta{D}^\rho( f(x), f(y))\leq 2k\varepsilon]
\end{align*}
Let us assume $k$-ALC w.r.t.\ $\tuple{\tdelta{1C}^{\eta_1},\tdelta{D}^{\rho}}$, and let $x,y \in C$ be such that $\tdelta{2C}^{\eta_2}(x,y) = 0$. Instantiating ALC with the pairs $(x,\eta_2(x))$ and $(y,\eta_2(y))$, we have
\[
\tdelta{D}^\rho(f(x),f\eta_2(x)) \leq k\tdelta{1C}^{\eta_1}(x,\eta_2(x)) \leq k\varepsilon
\qquad\text{and}\qquad
\tdelta{D}^\rho(f(y),f\eta_2(y)) \leq k\tdelta{1C}^{\eta_1}(y,\eta_2(y)) \leq k\varepsilon
\]
by definition of $\varepsilon$. By (\textit{iff-identity}) of the quasisemi-metric $\tdelta{2C}$, we have $\eta_2(x) = \eta_2(y)$, so $f\eta_2(x) = f\eta_2(y)$ and, by (\textit{if-identity}) of $\tdelta{D}$, $\tdelta{D}^\rho(f\eta_2(x),f\eta_2(y)) = 0$. By (\textit{triangle-inequality}) of $\tdelta{D}$ (and hence of $\tdelta{D}^\rho$):
\[
\tdelta{D}^\rho(f(x),f(y)) \leq \tdelta{D}^\rho(f(x),f\eta_2(x)) + \tdelta{D}^\rho(f\eta_2(x),f\eta_2(y)) + \tdelta{D}^\rho(f\eta_2(y),f(y))
\]
and, by (\textit{symmetry}) of $\tdelta{D}$, $\tdelta{D}^\rho(f\eta_2(y),f(y)) = \tdelta{D}^\rho(f(y),f\eta_2(y)) \leq k\varepsilon$. Hence $\tdelta{D}^\rho(f(x),f(y)) \leq k\varepsilon + 0 + k\varepsilon = 2k\varepsilon$, namely $f$ satisfies $2k\varepsilon$-PANI w.r.t. $\tuple{\tdelta{2C}^{\eta_2},\tdelta{D}^{\rho}}$.
\end{proof}

\begin{exa} \label{ex:thm-p-ani-aulc}
Consider again the collecting semantics $\collSem{\progp{R}^\ast}$ on non-empty sets of integers, the sign closure $\Sign$ and the parity closure $\Par$ of Example~\ref{ex:th-p-comp-aulc}, and the undirected variant $\dupath{}$ of $\dpath{}$ on the domain of signs $\tuple{\Sign(\wp(\integer)),\subseteq}$, where paths may traverse edges of the Hasse diagram in both directions, i.e., the path-length distance of~\cite{CampionSAS23} with unit weights (cf.\ Example~\ref{ex:path-len}). Unlike $\dpath{}$, $\dupath{}$ is symmetric (hence, it is a metric):
for instance, $\dupath{}(\integer_{>0},\integer) = \dupath{}(\integer,\integer_{>0}) = 2$, whereas $\dpath{}(\integer_{>0},\integer) = 2 \neq \infty = \dpath{}(\integer,\integer_{>0})$.
By~\eqref{eq:R-star-sign}, the output distance is at most $\dupath{}(\integer_{\geq 0},\integer) = 1$, and it is $0$ when the input distance is $0$. Hence, $\collSem{\progp{R}^\ast}$ satisfies $1$-ALC w.r.t.\ $\tuple{\dupath{\Sign},\dupath{\Sign}}$.

Let now $\eta_2 = \Par$, and take as $\tdelta{2C}$ the inclusion distance $\dincl{}$, which is a quasi-metric (Example~\ref{ex:inclusion-dist}), hence also a quasisemi-metric.
As in Example~\ref{ex:th-p-comp-aulc}, $\Sign(S) \subseteq \Sign(\Par(S))$, hence $\dupath{\Sign}(S,\Par(S)) = \dpath{\Sign}(S,\Par(S)) \leq 2$, and $\varepsilon = 2$.
By Theorem~\ref{thm:p-ani-aulc}, with $\eta_1=\rho=\Sign$, $\eta_2=\Par$, and $k=1$, $\collSem{\progp{R}^\ast}$ satisfies $4$-PANI w.r.t. $\tuple{\dincl{\Par},\dupath{\Sign}}$.

Observe that the bound is not tight: as shown above, the output distance is at most $1$, which is attained, e.g., for $S_1 = \{-2\}$ and $S_2 = \{2\}$, which have the same parity but yield outputs with different signs. Hence, $\collSem{\progp{R}^\ast}$ satisfies $1$-PANI but not $0$-PANI w.r.t. $\tuple{\dincl{\Par},\dupath{\Sign}}$.

$\qedex$
\end{exa}

When $\eta_2$ is more precise than $\eta_1$ (Lemma~\ref{lem:eta1eta2}), the bound of Theorem~\ref{thm:p-ani-aulc} is $0$, and we obtain the same guarantee as Theorem~\ref{thm:alc-pani}, but w.r.t.\ $\eta_2$:
\begin{cor} \label{cor:0-p-ani-aulc}
Let $\tuple{C,\tpreord{C}}$ and $\tuple{D,\tpreord{D}}$ be posets, $\tuple{C,\tdelta{1C}}$ a pre-metric space, $\tuple{C,\tdelta{2C}}$ a quasisemi-metric space, $\tuple{D,\tdelta{D}}$ a pseudo-metric space, and $\eta_1,\eta_2\in\uco(C)$, $\rho\in\uco(D)$, such that $\eta_2 \tpreord{C} \eta_1$.
If a function $f\colon C\ra D$ satisfies ALC w.r.t.\ $\tuple{\tdelta{1C}^{\eta_1},\tdelta{D}^{\rho}}$, then $f$ satisfies $\varepsilon$-PANI w.r.t. $\tuple{\tdelta{2C}^{\eta_2},\tdelta{D}^{\rho}}$, for any $\varepsilon \in \real_{\geq 0}$.
\end{cor}
\begin{proof}
By Lemma~\ref{lem:eta1eta2}, $\eta_1\eta_2(x) = \eta_1(x)$ for every $x \in C$, so $\tdelta{1C}^{\eta_1}(x,\eta_2(x)) = \tdelta{1C}(\eta_1(x),\eta_1\eta_2(x)) = \tdelta{1C}(\eta_1(x),\eta_1(x)) = 0$ by (\textit{if-identity}). Hence $\sup\sset{\tdelta{1C}^{\eta_1}(x,\eta_2(x))}{x\in C} = 0$, and, by Theorem~\ref{thm:p-ani-aulc}, $f$ satisfies $0$-PANI w.r.t. $\tuple{\tdelta{2C}^{\eta_2},\tdelta{D}^{\rho}}$, and thus $\varepsilon$-PANI w.r.t. $\tuple{\tdelta{2C}^{\eta_2},\tdelta{D}^{\rho}}$ holds for any $\varepsilon \in \real_{\geq 0}$.
\end{proof}

\begin{exa} \label{ex:cor-0-p-ani-aulc}
Consider again the collecting semantics $\collSem{\progp{R}^\ast}$ on non-empty sets of integers, the closures $\Sign'$ and $\Int$ of Example~\ref{ex:eta1eta2}, and the size distance $\dcount{}$, which is a pseudo-metric (Example~\ref{ex:counting}).
Since all non-empty elements of $\Sign'(\wp(\integer))$ but $\{0\}$ are infinite, $\dcount{\Sign'}(S_1,S_2)$ is $0$ if $\Sign'(S_1) = \Sign'(S_2)$, and $\infty$ otherwise.
By~\eqref{eq:R-star-sign} (cf. Example~\ref{ex:cor-0-p-comp-aulc}), $\Sign'(\collSem{\progp{R}^\ast}S_1) \neq \Sign'(\collSem{\progp{R}^\ast}S_2)$ only if exactly one of $S_1, S_2$ contains a negative value, in which case also $\Sign'(S_1) \neq \Sign'(S_2)$.
Hence, it is easy to check that $\collSem{\progp{R}^\ast}$ satisfies $1$-ALC w.r.t.\ $\tuple{\dcount{\Sign'},\dcount{\Sign'}}$.

Let now $\eta_2=\Int$, which is more precise than $\Sign'$ (Example~\ref{ex:eta1eta2}), and take as $\tdelta{2C}$ the inclusion distance $\dincl{}$, which is a quasi-metric (Example~\ref{ex:inclusion-dist}), hence also a quasisemi-metric.
By Corollary~\ref{cor:0-p-ani-aulc}, with $\eta_1=\rho=\Sign'$ and $\eta_2=\Int$, the collecting semantics $\collSem{\progp{R}^\ast}$ also satisfies $0$-PANI w.r.t. $\tuple{\dincl{\Int},\dcount{\Sign'}}$: inputs with the same interval yield outputs with the same sign.
$\qedex$
\end{exa}

\subsection{Abstract Robustness} \label{sec:arob}

The idea of abstract robustness is to formally characterize the amount of error that an analyst may tolerate in a machine learning classification result~\cite{GMP24}. This error, generated by an input perturbation, either a property-based or a quantitative approximation~\cite{GMP24}, is modeled as a property-based approximation of the classification outcome, expressed through an uco $\rho$.
Formally, given a set of inputs of interest and a perturbation that maps each input to a region of inputs around it, abstract robustness requires every input in the perturbation of an input of interest $x$ to have the same abstract classification as $x$~\cite[Definition~5.2]{GMP24}.
Assuming $f$ to be a classifier, we consider the global version of abstract robustness, where every input $x \in C$ is of interest, with a quantitative input perturbation of radius $\varepsilon \in \real_{\geq 0}$, which maps $x$ to its quantitative approximations up to $\varepsilon$ (Definition~\ref{def:quantapp}), namely to the inputs $y \in C$ such that $\tdelta{C}(x,y) \leq \varepsilon$~\cite{GMP24}:
\begin{equation} \label{eq:ar}
    \forall x,y \in C \sut \tdelta{C}(x,y) \leq \varepsilon\ \Ra\ \rho f(x)=\rho f(y)
\end{equation}
In the following, we refer to~\eqref{eq:ar} as \emph{abstract robustness} w.r.t.\ $\tuple{\tdelta{C},\rho}$ \emph{for the radius} $\varepsilon$. Note that, unlike the thresholds of partial completeness and PANI, the radius $\varepsilon$ bounds the perturbation of the input, while the abstract classification of the output is not allowed to change.
If we look at this definition in terms of ALC, we observe that this notion requires that the error introduced by the perturbation of the input must be neutralized, at least for the $\rho$ observation of the output classification. In fact, if we consider a quasisemi-metric output distance, the equality $\rho f(x) = \rho f(y)$ corresponds to the null output distance guaranteed by $0$-ALC, which is thus a sufficient condition for abstract robustness, whatever the radius:
\begin{thm} \label{thm:alc-ar}
    Let $\tuple{C,\tpreord{C}}$ and $\tuple{D,\tpreord{D}}$ be posets, $\tuple{C,\tdelta{C}}$ a pre-metric space, $\tuple{D,\tdelta{D}}$ a quasisemi-metric space, $\rho\in\uco(D)$, and fix $\varepsilon \in \real_{\geq 0}$.
    If a function $f\colon C\ra D$ satisfies $0$-ALC w.r.t.\ $\tuple{\tudelta{\id}{C},\tudelta{\rho}{D}}$, then $f$ satisfies abstract robustness w.r.t.\ $\tuple{\tdelta{C},\rho}$ for the radius $\varepsilon$.
\end{thm}
\begin{proof}
    We prove the following implication:
    \begin{align*}
        & [\forall x,y\in C\sut \tdelta{D}^\rho( f(x), f(y))\leq 0 \cdot \tdelta{C}^{\id}(x,y)] \ \Rightarrow \\
        & [\forall x,y\in C\sut \tdelta{C}(x,y)\leq\varepsilon\ \Ra\ \rho f(x)=\rho f(y)]
    \end{align*}
    Let us assume $0$-ALC w.r.t.\ $\tuple{\tudelta{\id}{C},\tudelta{\rho}{D}}$, and let $x,y \in C$ be such that $\tdelta{C}(x,y) \leq \varepsilon$. Since $\tdelta{C}^{\id} = \tdelta{C}$, by Proposition~\ref{prop:alc} with $k = 0$ we have $\tdelta{D}^\rho(f(x),f(y)) \leq 0 \cdot \varepsilon = 0$, namely $\tdelta{D}(\rho f(x),\rho f(y)) = 0$ by definition of $\tdelta{D}^\rho$. Since $\tdelta{D}$ is a quasisemi-metric, it satisfies (\textit{iff-identity}), hence $\rho f(x) = \rho f(y)$.
\end{proof}
Setting the abstract Lipschitz constant to zero is necessary, in general, to stabilize the classifier's output under the same abstraction $\rho$ when $\varepsilon > 0$: with $k > 0$, inputs at distance $\varepsilon$ may be mapped to outputs at distance $k\varepsilon > 0$, hence with different abstract classifications (see Example~\ref{ex:alc-ar} below). For $\varepsilon = 0$, instead,~\eqref{eq:ar} only constrains the inputs at distance $0$, and it is implied by $k$-ALC for any $k$, as in Theorem~\ref{thm:alc-pani}. The distance $\tdelta{D}$ must also be a quasisemi-metric to ensure that output abstractions at distance zero are equal.

The converse of Theorem~\ref{thm:alc-ar} does not hold for a fixed radius $\varepsilon$: abstract robustness only constrains the inputs within distance $\varepsilon$ from each other, whereas $0$-ALC constrains all the inputs at finite distance, since $0 \cdot r = 0$ for every $r \in \real_{\geq 0}$ (cf.\ Section~\ref{sec:alc}).

\begin{exa} \label{ex:alc-ar}
Consider again the collecting semantics $\collSem{\progp{R}^\ast}$ on non-empty sets of integers, and the closures $\Int$ and $\Sign'$ of Example~\ref{ex:eta1eta2}.
As input distance we take $\tdelta{C} = \dincl{\Int}$, which is a quasipseudo-metric on $\wp(\integer)$, hence a pre-metric (Example~\ref{ex:inclusion-dist-int-axioms}): the perturbation of radius $\varepsilon$ of an input $S$ consists of the sets whose interval is obtained by adding at most $\varepsilon$ integers to $\Int(S)$.
As output abstraction we take $\rho = \Sign'$, and as output distance the path-length distance $\dpath{}$ on $\tuple{\Sign'(\wp(\integer)),\subseteq}$, which is a quasi-metric (Example~\ref{ex:path-len}), hence also a quasisemi-metric.

For any radius $\varepsilon < 1$, $\collSem{\progp{R}^\ast}$ satisfies abstract robustness w.r.t.\ $\tuple{\dincl{\Int},\Sign'}$. Indeed, since $\dincl{}$ takes values in $\nat^\infty$, $\dincl{\Int}(S_1,S_2) \leq \varepsilon$ implies $\dincl{}(\Int(S_1),\Int(S_2)) = 0$, hence $\Int(S_1) = \Int(S_2)$ by (\textit{iff-identity}) of $\dincl{}$. Thus $S_1$ contains a negative value if and only if $S_2$ does, and $\Sign'(\collSem{\progp{R}^\ast}S_1) = \Sign'(\collSem{\progp{R}^\ast}S_2)$ by~\eqref{eq:R-star-sign} (cf.\ Example~\ref{ex:cor-0-p-comp-aulc}). For instance, the perturbation of radius $\varepsilon$ of $\{0,2\}$ is $\{\{0,2\},\{0,1,2\}\}$. More generally, for $\varepsilon < 1$ the perturbation of $S$ consists of the inputs with the same interval as $S$, namely it is the perturbation induced by $\Int$~\cite[Section~7.2.1]{GMP24}, and abstract robustness amounts to ANI w.r.t.\ $\tuple{\Int,\Sign'}$ (cf.\ Example~\ref{ex:cor-0-p-ani-aulc}).

However, $\collSem{\progp{R}^\ast}$ does not satisfy $0$-ALC w.r.t.\ $\tuple{\dincl{\Int},\dpath{\Sign'}}$: since $\dincl{\Int}(\{0\},\{-1,0\}) = \dincl{}([0,0],[-1,0]) = 1$, by~\eqref{eq:R-star-sign} we have
\[
    \dpath{\Sign'}(\collSem{\progp{R}^\ast}\{0\},\collSem{\progp{R}^\ast}\{-1,0\}) = \dpath{}(\integer_{\geq 0},\integer) = 1 > 0 = 0 \cdot \dincl{\Int}(\{0\},\{-1,0\}).
\]
The same pair shows that $\collSem{\progp{R}^\ast}$ does not satisfy abstract robustness w.r.t.\ $\tuple{\dincl{\Int},\Sign'}$ for any radius $\varepsilon \geq 1$, although it satisfies $1$-ALC w.r.t.\ $\tuple{\dincl{\Int},\dpath{\Sign'}}$. Indeed, if $\dincl{\Int}(S_1,S_2) < \infty$, then $\Int(S_1) \subseteq \Int(S_2)$, so that the output signs differ only if $S_2$ contains a negative value and $S_1$ does not; in this case, $\Int(S_1) \neq \Int(S_2)$, hence the output distance $\dpath{}(\integer_{\geq 0},\integer) = 1$ is bounded by the input distance $\dincl{\Int}(S_1,S_2) \geq 1$.
$\qedex$
\end{exa}

Theorem~\ref{thm:alc-ar} and Example~\ref{ex:alc-ar} show that, in general, $0$-ALC is strictly stronger than abstract robustness for a fixed radius. The two notions coincide when abstract robustness is required for every radius $\varepsilon \in \real_{\geq 0}$, namely w.r.t.\ the perturbation that maps each input $x$ to all the inputs at finite distance from $x$:

\begin{thm} \label{thm:alc-ar-iff}
    Consider the hypotheses of Theorem~\ref{thm:alc-ar}, except for the fixed radius $\varepsilon$.
    A function $f\colon C\ra D$ satisfies $0$-ALC w.r.t.\ $\tuple{\tudelta{\id}{C},\tudelta{\rho}{D}}$ if and only if $f$ satisfies abstract robustness w.r.t.\ $\tuple{\tdelta{C},\rho}$ for the radius $\varepsilon$, for every $\varepsilon \in \real_{\geq 0}$.
\end{thm}
\begin{proof}
    ($\Ra$) By Theorem~\ref{thm:alc-ar}, applied to each $\varepsilon \in \real_{\geq 0}$.

    ($\La$) Let us assume abstract robustness w.r.t.\ $\tuple{\tdelta{C},\rho}$ for every radius $\varepsilon \in \real_{\geq 0}$, and let $x,y \in C$. If $\tdelta{C}(x,y) = \infty$, then $\tdelta{D}^\rho(f(x),f(y)) \leq \infty = 0 \cdot \tdelta{C}^{\id}(x,y)$, by the conventions of Section~\ref{sec:pre-metric}. Otherwise, $\varepsilon \definit \tdelta{C}(x,y) \in \real_{\geq 0}$ satisfies $\tdelta{C}(x,y) \leq \varepsilon$, hence $\rho f(x) = \rho f(y)$ by~\eqref{eq:ar}, and $\tdelta{D}^\rho(f(x),f(y)) = \tdelta{D}(\rho f(x),\rho f(x)) = 0 = 0 \cdot \tdelta{C}^{\id}(x,y)$ by (\textit{if-identity}) of $\tdelta{D}$. Hence, $f$ satisfies $0$-ALC w.r.t.\ $\tuple{\tudelta{\id}{C},\tudelta{\rho}{D}}$.
\end{proof}

What is interesting in these results is that abstract robustness, and therefore standard robustness, follows from the specific instance of ALC where the potential error in input does not affect in any way the classification of the output (at least in the observed output property), which corresponds to abstract robustness for perturbations of any radius. In this sense, ALC becomes a generalization of robustness by admitting an effect in the output, but this effect must be linearly bounded to the input error. From this perspective, ALC serves as a model for further weakening the robustness of ML, enabling us to better adapt it to specific contexts or application domains, for instance, by checking whether variations of input properties  affect the classification.

\section{Proving ALC for Programs}\label{sec:proof}

ALC is an extensional property: it constrains the input/output behaviour of a program
semantics and, as such, it cannot be checked directly on the program text. In this section
we introduce a syntax-directed deductive system for proving that the semantics
$\collSem{\prog} \colon C \to C$ of a program $\prog \in \imp$ satisfies $k$-ALC w.r.t.\ a
pair of abstractions $\tuple{\eta,\rho}$ and a pre-metric $\delta$ on $C$. The system is
in the tradition of the proof systems for completeness~\cite{GiacobazziLR15} and partial
completeness~\cite{CampionPOPL2026} of abstract interpretations, and of the proof systems
and type systems for the Lipschitz continuity of programs developed for program
robustness~\cite{gulwani10,Gulwani11,gulwani12} and for sensitivity analysis in
differential privacy~\cite{sensitivity}. Judgments take the form
    $\judge{k}{\tudelta{\eta}{}}{\prog}{\tudelta{\rho}{}}$
and are derived compositionally,
following the structure of $\prog$: a proof of ALC for $\prog$ is thus reduced to proofs
of ALC for its basic commands and to a small number of side conditions relating the
abstractions $\eta$ and $\rho$ through $\delta$. These side conditions are what
distinguishes ALC from Lipschitz continuity: unlike Lipschitz continuous functions,
the functions that are ALC w.r.t.\ a given pair $\tuple{\tudelta{\eta}{},\tudelta{\rho}{}}$
are not closed under composition in general, and the side conditions make explicit
sufficient assumptions under which ALC is preserved, allowing one to track them along
a derivation. Non-deterministic choice and iteration also require conditions on the
output distance (Definitions~\ref{def:+bound} and~\ref{def:wcont}), which, however,
are needed for Lipschitz continuity as well (see the end of this section). In
the spirit of the consequence rule of Hoare logic, the proposed proof system also allows one to trade precision
for provability, by weakening the abstract Lipschitz constant as well as the input and
output abstractions. The system will be proved to be sound: whenever
$\judge{k}{\tudelta{\eta}{}}{\prog}{\tudelta{\rho}{}}$ is derivable, $\collSem{\prog}$
satisfies $k$-ALC w.r.t.\ $\tuple{\tudelta{\eta}{},\tudelta{\rho}{}}$
(Theorem~\ref{thm:soundness}). It is not complete, and the derived constant may be larger
than the least abstract Lipschitz constant of $\collSem{\prog}$; the sources of this
imprecision are discussed in Section~\ref{sec:concl}. When $\eta = \rho = \id$, the
system specializes to (an extension of) the proof system of Chaudhuri et
al.~\cite{gulwani12} for the Lipschitz continuity of programs.

From now on, we fix a complete lattice $\tuple{C, \tpreord{}, \vee, \land ,\top ,\bot}$,
a pre-metric $\delta$ on $C$, and a program semantics
$\collSem{\cdot} \colon \imp \ra C \ra C$ as in Section~\ref{sec:prog}. Differently from
Section~\ref{sec:alc}, input and output abstractions are ucos on the same domain $C$:
this is forced by the language, since Kleene iteration only
makes sense for functions from $C$ to itself, whereas sequential composition alone would
also allow different, suitably matching, input and output domains. Likewise, for
readability we use a single pre-metric $\delta$ on $C$, although all the results of this section
extend verbatim to distinct pre-metrics for inputs and outputs.

Given $k \in \prealz^\infty$ and $\eta,\rho \in \uco(C)$, we write
\[
    \Cont \;\definit\; \{ f \colon C \ra C \mid \forall x,y \in C \sut
    \tudelta{\rho}{}(f(x),f(y)) \leq k \, \tudelta{\eta}{}(x,y) \}
\]
for the set of all functions satisfying $k$-ALC w.r.t.\
$\tuple{\tudelta{\eta}{},\tudelta{\rho}{}}$. Note that here we allow $k = \infty$, thus
slightly extending Definition~\ref{def:alc}: by the conventions of
Section~\ref{sec:background}, $\infty \cdot \tudelta{\eta}{}(x,y) = \infty$ for all
$x,y \in C$, hence $\Contpar{\infty}{\tudelta{\eta}{}}{\tudelta{\rho}{}}$ contains every
function. The constant $\infty$ thus plays the role of the uninformative answer, much
like $\top$ for a static analysis: it allows the deductive system to always produce a
sound, albeit possibly useless, judgment. We define:
\[
    \prog \text{ is } k\text{-ALC w.r.t.\ } \tuple{\tudelta{\eta}{},\tudelta{\rho}{}} \; \lradef \; \collSem{\prog} \in \Cont.
\]

The following lemma collects the properties of these sets on which the deductive
system relies.

\begin{lem} \label{lem:cont-set}
Let $\eta,\eta',\rho,\rho' \in \uco(C)$ and $k, k', t \in \prealz^\infty$. The following hold:
\begin{enumerate}[label={(\roman*)},itemsep=3pt,leftmargin=10mm,topsep=2pt]
    \item if $k \leq k'$ then $\Cont \subseteq \Contpar{k'}{\tudelta{\eta}{}}{\tudelta{\rho}{}}$;
    \item $\Contpar{k}{\tudelta{\id}{}}{\tudelta{\id}{}}$ is the set of all functions
    satisfying $k$-LC w.r.t.\ $\tuple{\delta,\delta}$;
    \item $\rho \in \Contpar{k}{\tudelta{\eta}{}}{\tudelta{\rho}{}}$ if and only if
    $\tudelta{\rho}{}(x,y) \leq k \, \tudelta{\eta}{}(x,y)$ for all $x,y \in C$;
    \item $\rho \in \Contpar{k}{\tudelta{\id}{}}{\tudelta{\id}{}}$ if and only if
    $\rho \in \Contpar{k}{\tudelta{\id}{}}{\tudelta{\rho}{}}$;
    \item if $f \in \Cont$, $g \in \Contpar{k'}{\tudelta{\eta'}{}}{\tudelta{\rho'}{}}$
    and $\eta' \in \Contpar{t}{\tudelta{\rho}{}}{\tudelta{\eta'}{}}$, then
    $g \comp f \in \Contpar{k \, t \, k'}{\tudelta{\eta}{}}{\tudelta{\rho'}{}}$.
\end{enumerate}
\end{lem}
\begin{proof}
(i) Multiplication is monotone on $\prealz^\infty$, hence
$k \, \tudelta{\eta}{}(x,y) \leq k' \, \tudelta{\eta}{}(x,y)$ for all $x,y \in C$.

(ii) Immediate, since $\tudelta{\id}{} = \delta$.

(iii) By idempotence of $\rho$, $\tudelta{\rho}{}(\rho(x),\rho(y)) =
\delta(\rho(\rho(x)),\rho(\rho(y))) = \tudelta{\rho}{}(x,y)$ for all $x,y \in C$, so
that $\rho \in \Contpar{k}{\tudelta{\eta}{}}{\tudelta{\rho}{}}$ amounts to
$\tudelta{\rho}{}(x,y) \leq k \, \tudelta{\eta}{}(x,y)$ for all $x,y \in C$.

(iv) By (ii), $\rho \in \Contpar{k}{\tudelta{\id}{}}{\tudelta{\id}{}}$ amounts to
$\delta(\rho(x),\rho(y)) \leq k \, \delta(x,y)$ for all $x,y \in C$; by (iii) with
$\eta = \id$, so does $\rho \in \Contpar{k}{\tudelta{\id}{}}{\tudelta{\rho}{}}$, since
$\tudelta{\rho}{}(x,y) = \delta(\rho(x),\rho(y))$ and $\tudelta{\id}{}(x,y) = \delta(x,y)$.

(v) For all $x,y \in C$ we have
$\tudelta{\rho'}{}(g(f(x)),g(f(y))) \leq k' \, \tudelta{\eta'}{}(f(x),f(y))
\leq k' \, t \, \tudelta{\rho}{}(f(x),f(y)) \leq k' \, t \, k \, \tudelta{\eta}{}(x,y)$,
where the second inequality follows from (iii) applied to $\eta'$.
\end{proof}

Item (iii) provides the reading of all the side conditions of the deductive system: a
membership $\rho \in \Contpar{k}{\tudelta{\eta}{}}{\tudelta{\rho}{}}$ is a pointwise
comparison $\tudelta{\rho}{} \leq k \, \tudelta{\eta}{}$ between two combined distances.
In particular, taking $\eta = \rho$ and $k = 1$, every closure $\rho$ belongs to
$\Contpar{1}{\tudelta{\rho}{}}{\tudelta{\rho}{}}$: indeed, a closure is an isometry
w.r.t.\ the distance it induces, $\tudelta{\rho}{}(\rho(x),\rho(y)) = \tudelta{\rho}{}(x,y)$,
which is what makes the case $\eta = \rho$ especially well behaved. Item (iv) shows that,
for a closure, being Lipschitz continuous w.r.t.\ $\delta$ is the same as being ALC with
the identity as input abstraction and the closure itself as output abstraction: again,
this is due to idempotence. Finally, item (v) is the key difference with Lipschitz
continuity: the composition of two ALC functions is ALC provided that the output
abstraction of the first and the input abstraction of the second are compatible up to
a constant $t$; when $\eta' = \rho$, item (iii) gives $t = 1$ and the constants simply
multiply, as for Lipschitz continuous functions.

\begin{figure}[t]
    \centering
    \makebox[\textwidth][c]{%
    \fbox{
        \begin{tabular}{ c c }
            \multicolumn{2}{c}{
                \infer[(\textbf{base})]{ \judge{k}{\tudelta{\eta}{}}{\code}{\tudelta{\rho}{}} }{ \collSem{\code} \in \Cont }
            }
            \\
            \\
            \multicolumn{2}{c}{
                \infer[(\textbf{weaken})]{ \judge{s t k}{\tudelta{\eta}{}}{\prog}{\tudelta{\rho}{}} }{ \judge{k}{\tudelta{\eta'}{}}{\prog}{\tudelta{\rho'}{}} & \eta' \in \Contpar{t}{\tudelta{\eta}{}}{\tudelta{\eta'}{}} & \rho \in \Contpar{s}{\tudelta{\rho'}{}}{\tudelta{\rho}{}} }
            }
            \\
            \\
            \multicolumn{2}{c}{
                \infer[(\textbf{seq})]{ \judge{k_1 t k_2}{\tudelta{\eta}{}}{\prog_1 ; \prog_2}{\tudelta{\rho}{}} }{ \judge{k_1}{\tudelta{\eta}{}}{\prog_1}{\tudelta{\rho}{}} & \judge{k_2}{\tudelta{\eta}{}}{\prog_2}{\tudelta{\rho}{}} & \eta \in \Contpar{t}{\tudelta{\rho}{}}{\tudelta{\eta}{}} }
            }
            \\
            \\
            \multicolumn{2}{c}{
                \infer[(\textbf{join})]{ \judge{k_1\bound k_2}{\tudelta{\eta}{}}{\prog_1 \pjoin \prog_2}{\tudelta{\rho}{}} }{ \judge{k_1}{\tudelta{\eta}{}}{\prog_1}{\tudelta{\rho}{}} & \judge{k_2}{\tudelta{\eta}{}}{\prog_2}{\tudelta{\rho}{}} & \+Bound(\tudelta{\rho}{},\bound) }
            }
            \\
            \\
            \multicolumn{2}{c}{
                \infer[(\textbf{bound-star})]{ \judge{K_m}{\tudelta{\eta}{}}{\prog^\ast}{\tudelta{\rho}{}} }{
                \judge{k}{\tudelta{\eta}{}}{\prog}{\tudelta{\rho}{}}
                & \deduce{ \rho \in \Contpar{v}{\tudelta{\eta}{}}{\tudelta{\rho}{}} }{ \eta \in \Contpar{t}{\tudelta{\rho}{}}{\tudelta{\eta}{}} }
                & \deduce{ \*Bound(\prog^\ast, m) }{ \+Bound(\tudelta{\rho}{},\bound) }}
            }
            \\
            \\
            \multicolumn{2}{c}{
                \infer[(\textbf{star})]{ \judge{K_\infty}{\tudelta{\eta}{}}{\prog^\ast}{\tudelta{\rho}{}} }{
                \judge{k}{\tudelta{\eta}{}}{\prog}{\tudelta{\rho}{}}
                & \deduce{ \rho \in \Contpar{v}{\tudelta{\eta}{}}{\tudelta{\rho}{}} }{ \eta \in \Contpar{t}{\tudelta{\rho}{}}{\tudelta{\eta}{}} }
                & \deduce{ \omega\text{-}\mathrm{Cont}(\tudelta{\rho}{}) }{ \+Bound(\tudelta{\rho}{},\bound) }}
            }
        \end{tabular}
    }
}
    \caption{A deductive system for proving ALC for $\imp$.}
    \label{fig:axRulesP}
\end{figure}

The deductive rules are given in Figure~\ref{fig:axRulesP}. Judgments take the form
\[
    \judge{k}{\tudelta{\eta}{}}{\prog}{\tudelta{\rho}{}}
\]
and are meant to be read as \quotes{$\prog$ is $k$-ALC w.r.t.\
$\tuple{\tudelta{\eta}{},\tudelta{\rho}{}}$}, in analogy with Hoare triples: the input
distance $\tudelta{\eta}{}$ plays the role of the precondition, the output distance
$\tudelta{\rho}{}$ that of the postcondition, and $k$ bounds the amplification of the
former into the latter. Theorem~\ref{thm:soundness} below shows that this reading is
sound, i.e., that $\collSem{\prog} \in \Cont$ for every derivable judgment
$\judge{k}{\tudelta{\eta}{}}{\prog}{\tudelta{\rho}{}}$. Let us examine the rules.

The $(\textbf{base})$ rule derives $\judge{k}{\tudelta{\eta}{}}{\code}{\tudelta{\rho}{}}$
for a basic command $\code \in \Stm$ (i.e., $\skipc$, an assignment or a Boolean guard)
from a proof that $\collSem{\code} \in \Cont$. The deductive system is thus parametric
on the ALC of basic commands, which depends on the chosen semantics, distance and
abstractions, and has to be established separately, as in the proof system of Chaudhuri
et al.~\cite{gulwani12} for Lipschitz continuity. For $\skipc$ this is immediate: since
$\collSem{\skipc}$ is the identity, $\collSem{\skipc} \in \Cont$ if and only if
$\tudelta{\rho}{}(x,y) \leq k \, \tudelta{\eta}{}(x,y)$ for all $x,y \in C$; in
particular, $\judge{1}{\tudelta{\rho}{}}{\skipc}{\tudelta{\rho}{}}$ is always derivable.

The $(\textbf{weaken})$ rule is the analogue of the consequence rule of Hoare logic:
it derives a weaker judgment from a stronger one by changing the abstractions, at the
price of a larger constant. The input abstraction $\eta'$ can be replaced by any $\eta$
such that $\eta' \in \Contpar{t}{\tudelta{\eta}{}}{\tudelta{\eta'}{}}$, i.e., by
Lemma~\ref{lem:cont-set}(iii), such that $\tudelta{\eta'}{} \leq t \, \tudelta{\eta}{}$:
the new input distance dominates the old one up to the factor $t$. Symmetrically, the
output abstraction $\rho'$ can be replaced by any $\rho$ such that
$\rho \in \Contpar{s}{\tudelta{\rho'}{}}{\tudelta{\rho}{}}$, i.e., such that
$\tudelta{\rho}{} \leq s \, \tudelta{\rho'}{}$: the new output distance is dominated by
the old one up to the factor $s$. Both replacements are paid by multiplying the
constant: indeed, for all $x,y \in C$,
$\tudelta{\rho}{}(\collSem{\prog}x,\collSem{\prog}y) \leq s \, \tudelta{\rho'}{}(\collSem{\prog}x,\collSem{\prog}y)
\leq s k \, \tudelta{\eta'}{}(x,y) \leq s k t \, \tudelta{\eta}{}(x,y)$.
The two replacements can be applied independently, by taking $\eta = \eta'$ and $t = 1$,
or $\rho = \rho'$ and $s = 1$, since $\eta' \in \Contpar{1}{\tudelta{\eta'}{}}{\tudelta{\eta'}{}}$
and $\rho' \in \Contpar{1}{\tudelta{\rho'}{}}{\tudelta{\rho'}{}}$ always hold
(Lemma~\ref{lem:cont-set}(iii)). Note that, by taking both $\eta = \eta'$ and
$\rho = \rho'$, the rule also allows one to increase the constant alone, since
$\eta' \in \Contpar{t}{\tudelta{\eta'}{}}{\tudelta{\eta'}{}}$ for every $t \geq 1$
(Lemma~\ref{lem:cont-set}(i) and~(iii)): this form of weakening is semantically valid
but never needed, since the constant in the conclusion
of every rule is a monotone function of the constants in its premises, so that the
least derivable constant for a program is obtained without ever weakening a constant
along the derivation. The weakening of the abstractions, on the other hand, is
essential: as in Hoare logic, where the consequence rule is what allows one to
establish a loop invariant, the rules for Kleene iteration require the input and
output abstractions of the body to be compatible.

Sequential composition is handled by the $(\textbf{seq})$ rule, the syntactic
counterpart of Lemma~\ref{lem:cont-set}(v). While the composition of a $k_1$-LC and a
$k_2$-LC function is $k_1 k_2$-LC, the composition of two ALC programs $\prog_1$ and
$\prog_2$ is not ALC in general: the output abstraction $\rho$ through which $\prog_1$
is observed need not be related to the input abstraction $\eta$ under which $\prog_2$
is analysed. The rule thus requires the side condition
$\eta \in \Contpar{t}{\tudelta{\rho}{}}{\tudelta{\eta}{}}$, i.e.,
$\tudelta{\eta}{} \leq t \, \tudelta{\rho}{}$ by Lemma~\ref{lem:cont-set}(iii): the
distance between the input properties observed through $\eta$ is bounded by $t$ times
the distance between the output properties observed through $\rho$, so that the bound
on the outputs of $\prog_1$ transfers to the inputs of $\prog_2$ at the cost of the
factor $t$, yielding the constant $k_1 t k_2$. The smaller $t$, the more $\rho$ has to
separate elements compared to $\eta$. The side condition can always be discharged with
$t = \infty$, which however yields the uninformative constant $\infty$: the rule is
informative only when $\eta$ and $\rho$ are comparable through $\delta$. When
$\eta = \rho$, Lemma~\ref{lem:cont-set}(iii) gives $t = 1$, and ALC turns out to be
closed under composition exactly as Lipschitz continuity is.

The $(\textbf{join})$ rule deals with non-deterministic choice. As for composition,
the join of two ALC functions is not ALC in general, this time because of the output
distance: even if $\tudelta{\rho}{}(\collSem{\prog_1}c_1,\collSem{\prog_1}c_2)$ and
$\tudelta{\rho}{}(\collSem{\prog_2}c_1,\collSem{\prog_2}c_2)$ are both bounded, nothing
guarantees a bound on
$\tudelta{\rho}{}(\collSem{\prog_1}c_1 \vee \collSem{\prog_2}c_1, \collSem{\prog_1}c_2 \vee \collSem{\prog_2}c_2)$:
the abstraction of a join may contain elements that belong to neither of the
abstractions of its operands, as is the case for the convex hull computed by $\Int$.
Following~\cite{campion2022popl,CampionPOPL2026}, we make this requirement explicit
by parameterizing the rule on a binary operation $\bound$ that combines the two
constants and bounds the effect of joins on the output distance.

\begin{defi}[$\oplus$-Bound] \label{def:+bound}
Let $\rho \in \uco(C)$. A binary operation
$\bound \colon \prealz^\infty \times \prealz^\infty \ra \prealz^\infty$ is an
\emph{$\oplus$-bound for $\tudelta{\rho}{}$}, written $\+Bound(\tudelta{\rho}{},\bound)$,
when it is associative, commutative and monotone, and it satisfies the following two conditions:
\begin{enumerate}[label={(\roman*)},itemsep=3pt,leftmargin=10mm,topsep=2pt]
    \item $\forall k_1,k_2,d \in \prealz^\infty \sut (k_1 \, d) \bound (k_2 \, d) \leq (k_1 \bound k_2) \, d$
    \item $\forall x_1,x_2,y_1,y_2 \in C \sut
    \tudelta{\rho}{}(x_1 \vee y_1, x_2 \vee y_2) \leq
    \tudelta{\rho}{}(x_1,x_2) \bound \tudelta{\rho}{}(y_1,y_2)$ $\qeddef$
\end{enumerate}
\end{defi}

The last condition is a form of subadditivity of $\tudelta{\rho}{}$ w.r.t.\ the join
of $C$: it only depends on the distance and on the output abstraction, and it is what
makes the $(\textbf{join})$ rule sound. Indeed, if
$\judge{k_1}{\tudelta{\eta}{}}{\prog_1}{\tudelta{\rho}{}}$,
$\judge{k_2}{\tudelta{\eta}{}}{\prog_2}{\tudelta{\rho}{}}$ and
$\+Bound(\tudelta{\rho}{},\bound)$ hold, then for all $c_1,c_2 \in C$:
\begin{align*}
\tudelta{\rho}{}(\collSem{\prog_1 \pjoin \prog_2}c_1, \collSem{\prog_1 \pjoin \prog_2}c_2)
&\leq \tudelta{\rho}{}(\collSem{\prog_1}c_1,\collSem{\prog_1}c_2) \bound
\tudelta{\rho}{}(\collSem{\prog_2}c_1,\collSem{\prog_2}c_2) \\
&\leq (k_1 \tudelta{\eta}{}(c_1,c_2)) \bound (k_2 \tudelta{\eta}{}(c_1,c_2))
\leq (k_1 \bound k_2) \, \tudelta{\eta}{}(c_1,c_2)
\end{align*}
by monotonicity of $\bound$ and by its compatibility with the product. The next proposition
provides a large class of $\oplus$-bounds for the path-length distance.

\begin{prop} \label{prop:incl-bound}
    Let $\rho \in \uco(\wp(L))$ be additive, and let $\dpath{}$ be the path-length distance on the poset
    $\tuple{\rho(\wp(L)),\subseteq}$ (Example~\ref{ex:path-len}). Then
    $\+Bound(\dpath{\rho},+)$ holds, where $+$ is the usual addition. In particular, $\+Bound(\dincl{},+)$ holds on $\wp(L)$.
\end{prop}
\begin{proof}
    Additivity implies $\rho(S) = \bigcup_{s \in S} \rho(\{s\})$ for all $S \in \wp(L)$.
    Let $s \sqsubseteq t$ if and only if $s \in \rho(\{t\})$: this is a preorder on $L$,
    being reflexive by extensivity and transitive by monotonicity and idempotence of
    $\rho$, and $\rho(S) = \{ s \in L \mid \exists t \in S \sut s \sqsubseteq t \}$ is the
    downward closure of $S$. Hence the elements of $\rho(\wp(L))$ are the downward closed
    sets of $\tuple{L,\sqsubseteq}$, each of which is a union of equivalence classes of
    $\sqsubseteq$; for $S \in \wp(L)$, let $\mathsf{cl}(S)$ be the set of classes
    contained in $\rho(S)$, so that $\mathsf{cl}(X \cup Y) = \mathsf{cl}(X) \cup \mathsf{cl}(Y)$
    by additivity. If $A \subsetneq A'$ are downward closed and $A' \smallsetminus A$
    contains finitely many classes, then $A \cup K$ is downward closed for every class
    $K$ that is minimal among them, and no downward closed set lies strictly between
    $A$ and $A \cup K$; conversely, every covering step adds exactly one class.
    Therefore $\dpath{\rho}(S_1,S_2) = \Count(\mathsf{cl}(S_2) \smallsetminus \mathsf{cl}(S_1))$
    if $\mathsf{cl}(S_1) \subseteq \mathsf{cl}(S_2)$, and $\dpath{\rho}(S_1,S_2) = \infty$
    otherwise. Addition on $\prealz^\infty$ is associative, commutative, monotone and
    satisfies $(k_1 d) + (k_2 d) = (k_1 + k_2) d$. Let $X_1,X_2,Y_1,Y_2 \in \wp(L)$. If
    $\mathsf{cl}(X_1) \not\subseteq \mathsf{cl}(X_2)$ or
    $\mathsf{cl}(Y_1) \not\subseteq \mathsf{cl}(Y_2)$, then the right-hand side of the
    subadditivity condition is $\infty$. Otherwise
    $\mathsf{cl}(X_1 \cup Y_1) \subseteq \mathsf{cl}(X_2 \cup Y_2)$ and
    $\mathsf{cl}(X_2 \cup Y_2) \smallsetminus \mathsf{cl}(X_1 \cup Y_1) \subseteq
    (\mathsf{cl}(X_2) \smallsetminus \mathsf{cl}(X_1)) \cup
    (\mathsf{cl}(Y_2) \smallsetminus \mathsf{cl}(Y_1))$, whence
    $\dpath{\rho}(X_1 \cup Y_1, X_2 \cup Y_2) \leq \dpath{\rho}(X_1,X_2) + \dpath{\rho}(Y_1,Y_2)$.
    $\+Bound(\dincl{},+)$ trivially holds on $\wp(L)$ since $\dincl{}$ is $\dpath{\id}$
\end{proof}

Proposition~\ref{prop:incl-bound} applies, e.g., to the domain of signs $\Sign$ of
Example~\ref{ex:sign}, to the parity abstraction~\cite{CC79}, and to the histogram abstraction of
Example~\ref{ex:byte-histogram}, all of which are induced by a partition of the
underlying set and are thus additive.

The $(\textbf{join})$ rule thus takes $\+Bound(\tudelta{\rho}{},\bound)$ as an
explicit premise, since the output abstraction may change along a derivation through
$(\textbf{weaken})$: given derivations
$\judge{k_1}{\tudelta{\eta}{}}{\prog_1}{\tudelta{\rho}{}}$ and
$\judge{k_2}{\tudelta{\eta}{}}{\prog_2}{\tudelta{\rho}{}}$, it concludes that
$\prog_1 \pjoin \prog_2$ is $(k_1 \bound k_2)$-ALC.

Kleene iteration is handled by the rules $(\textbf{bound-star})$ and $(\textbf{star})$.
Both build on the observation that, by definition, $\collSem{\prog^\ast}c$ is the join
of the iterates $\collSem{\prog}^n c$, i.e., the join of the semantics of the programs
$\skipc$, $\prog$, $\prog ; \prog$, $\dots$ with input $c$. Given a derivation
$\judge{k}{\tudelta{\eta}{}}{\prog}{\tudelta{\rho}{}}$, the side condition
$\eta \in \Contpar{t}{\tudelta{\rho}{}}{\tudelta{\eta}{}}$ of $(\textbf{seq})$ allows
us to derive the constant $k^i t^{i-1}$ for the $i$-th iterate $\prog ; \dots ; \prog$
($i \geq 1$), while the side condition $\rho \in \Contpar{v}{\tudelta{\eta}{}}{\tudelta{\rho}{}}$
provides a constant $v$ for the zero-th iterate $\skipc$, whose semantics is the
identity (cf.\ the discussion of the $(\textbf{base})$ rule). Both side conditions hold
with $t = v = 1$ when $\eta = \rho$ (Lemma~\ref{lem:cont-set}(iii)). Finally, an
$\oplus$-bound $\bound$ for $\tudelta{\rho}{}$ combines the constants of the iterates
as in the $(\textbf{join})$ rule. This leads to the following definition.

\begin{defi}[Iteration Constants] \label{def:Kn}
Let $k,t,v \in \prealz^\infty$ and let $\bound$ be an $\oplus$-bound. We define
\[
    K_0 \definit v, \qquad
    K_n \definit v \bound \biguplus_{i=1}^{n} k^i t^{i-1} \quad (n \geq 1), \qquad
    K_\infty \definit \sup_{n \in \nat} K_n,
\]
where the supremum is taken in the complete lattice $\tuple{\prealz^\infty,\leq}$.
$\qeddef$
\end{defi}

Note that $K_\infty$ always exists, although it may be $\infty$; when $0$ is neutral
for $\bound$, as for $+$ and the $\max$ function, the sequence $(K_n)_{n \in \nat}$ is increasing
by monotonicity of $\bound$, and $K_\infty$ is its limit. For finite $k$ and $t$, the
two most natural $\oplus$-bounds give the following closed forms: for $\bound = +$,
$K_\infty = v + \nicefrac{k}{(1 - kt)}$ if $kt < 1$, and $K_\infty = \infty$ if
$kt \geq 1$ and $k > 0$; for $\bound = \max$, $K_\infty = \max(v,k)$ if $kt \leq 1$,
and $K_\infty = \infty$ if $kt > 1$. In both cases, a non-expansive body ($kt \leq 1$)
is essentially what makes iteration informative (as also discussed in Section~\ref{sec:concl}).

The $(\textbf{bound-star})$ rule applies when the iteration is known to stabilize
within a bounded number of steps, as expressed by the following predicate.

\begin{defi}[$\ast$-Bound] \label{def:*bound}
Let $\prog \in \imp$ and $m \in \nat$. The predicate $\*Bound(\prog^\ast, m)$ holds
when $\collSem{\prog^{\ast}}c = \bigvee \{\collSem{\prog}^n c \mid n \leq m\}$ for every
$c \in C$. $\qeddef$
\end{defi}

In other words, on every input $c$ the result of $\collSem{\prog^\ast}c$ is obtained by
executing $\prog$ at most $m$ times. Under $\*Bound(\prog^\ast, m)$, $\prog^\ast$ has
the same semantics as the join of $m+1$ programs, and the constant $K_m$ is derived
exactly as in the $(\textbf{join})$ rule. Bounds on the number of iterations are
routinely established, e.g., by termination analyses or by manual annotations.

The $(\textbf{star})$ rule, instead, makes no assumption on the number of iterations,
and correspondingly derives the constant $K_\infty$. Here $\collSem{\prog^\ast}c$ is the
least upper bound of the increasing chain of finite joins
$\bigvee \{\collSem{\prog}^n c \mid n \leq m\}$, for $m \in \nat$, each of which
satisfies the bound $K_m \leq K_\infty$. In order to transfer these bounds to the
limit, the output distance has to be continuous w.r.t.\ least upper bounds of
chains, as defined by the following.

\begin{defi}[$\omega$-Continuity] \label{def:wcont}
A pre-metric $\delta$ on the complete lattice $\tuple{C,\tpreord{}}$ is
\emph{$\omega$-continuous}, written $\omega\text{-}\mathrm{Cont}(\delta)$, when for
all increasing chains $(a_n)_{n \in \nat}$ and $(b_n)_{n \in \nat}$ in $C$ the following inequality holds:
\[
    \delta\big(\textstyle\bigvee_{n \in \nat} a_n, \bigvee_{n \in \nat} b_n\big)
    \;\leq\; \sup_{n \in \nat} \delta(a_n,b_n). \fqeddef
\]
\end{defi}

\begin{prop} \label{prop:wcont}
The inclusion distance $\dincl{}$ on $\wp(L)$ is $\omega$-continuous, and so is
$\dpath{\rho}$ for every additive $\rho \in \uco(\wp(L))$.
\end{prop}
\begin{proof}
Let $(X_n)_{n \in \nat}$ and $(Y_n)_{n \in \nat}$ be increasing chains in $\wp(L)$
and let $s = \sup_n \dincl{}(X_n,Y_n)$. If $s = \infty$ there is nothing to
prove. Otherwise $X_n \subseteq Y_n$ for all $n$, hence
$\bigcup_n X_n \subseteq \bigcup_n Y_n$, and every
$y \in \bigcup_n Y_n \smallsetminus \bigcup_n X_n$ belongs to $Y_n \smallsetminus X_n$
for all sufficiently large $n$, since the chains are increasing. If there were
$s+1$ such elements $y$, they would all belong to $Y_n \smallsetminus X_n$ for some $n$,
contradicting $\Count(Y_n \smallsetminus X_n) \leq s$. Hence
$\dincl{}(\bigcup_n X_n, \bigcup_n Y_n) \leq s$. The proof for $\dpath{\rho}$ is similar, counting equivalence classes instead of elements, as in the proof of
Proposition~\ref{prop:incl-bound}: additivity gives $\rho(\bigcup_n X_n) = \bigcup_n \rho(X_n)$,
hence the classes contained in $\rho(\bigcup_n X_n)$ are those contained in some
$\rho(X_n)$.
\end{proof}

The following theorem shows that our deductive system is sound, namely, if
$\judge{k}{\tudelta{\eta}{}}{\prog}{\tudelta{\rho}{}}$ can be derived by applying the
rules of Figure~\ref{fig:axRulesP}, then $\collSem{\prog}$ satisfies $k$-ALC w.r.t.\
$\tuple{\tudelta{\eta}{},\tudelta{\rho}{}}$.

\begin{thm}[Soundness] \label{thm:soundness}
    $\judge{k}{\tudelta{\eta}{}}{\prog}{\tudelta{\rho}{}} \;\Rightarrow\; \collSem{\prog} \in \Cont$.
\end{thm}
\begin{proof}
    By induction on the derivation of $\judge{k}{\tudelta{\eta}{}}{\prog}{\tudelta{\rho}{}}$.
    We proceed by cases on the last rule applied, assuming, by inductive hypothesis,
    that the judgments in its premises are semantically valid, i.e., that
    $\judge{k'}{\tudelta{\eta'}{}}{\prog'}{\tudelta{\rho'}{}}$ in a premise implies
    $\collSem{\prog'} \in \Contpar{k'}{\tudelta{\eta'}{}}{\tudelta{\rho'}{}}$. In each case
    we show that $\tudelta{\rho}{}(\collSem{\prog}c_1,\collSem{\prog}c_2) \leq k \, \tudelta{\eta}{}(c_1,c_2)$
    for all $c_1,c_2 \in C$, where $k$, $\eta$ and $\rho$ are those of the conclusion.

    (\textbf{base}): immediate by the definition of $\judge{k}{\tudelta{\eta}{}}{\code}{\tudelta{\rho}{}}$ and the assumption $\collSem{\code} \in \Cont$.

    (\textbf{weaken}): Let us show the proof for weakening the input abstraction $\eta'$. Assume $\judge{k}{\tudelta{\eta'}{}}{\prog}{\tudelta{\rho}{}}$ and $\eta' \in \Contpar{t}{\tudelta{\eta}{}}{\tudelta{\eta'}{}}$.
    The second assumption can be written as $\forall c_1,c_2 \in C$: $\tudelta{\eta'}{}(\eta'(c_1), \eta'(c_2)) \leq t\tudelta{\eta}{}(c_1,c_2)$ which, by the idempotence property of $\eta'$, corresponds to $\tudelta{\eta'}{}(c_1, c_2) \leq t\tudelta{\eta}{}(c_1,c_2)$.
    We get the following derivations $\forall c_1,c_2 \in C$:
    \begin{align*}
        \tudelta{\rho}{}(\collSem{\prog}c_1, \collSem{\prog}c_2) &\;\leq\; \text{[by $\judge{k}{\tudelta{\eta'}{}}{\prog}{\tudelta{\rho}{}}$]} \\
        k\tudelta{\eta'}{}(c_1, c_2) &\;\leq\;  \text{[by $\eta' \in \Contpar{t}{\tudelta{\eta}{}}{\tudelta{\eta'}{}}$]} \\
        tk\tudelta{\eta}{}(c_1,c_2) &\;\Lra\;  \text{[by judgment definition]} \\
        \judge{tk}{\tudelta{\eta}{}}{\prog}{\tudelta{\rho}{}}
    \end{align*}
    The proof for weakening the output abstraction $\rho$ is similar and therefore omitted.

    (\textbf{seq}): Assume we have a derivation $\judge{k_1}{\tudelta{\eta}{}}{\prog_1}{\tudelta{\rho}{}}$ for program $\prog_1$, a derivation $\judge{k_2}{\tudelta{\eta}{}}{\prog_2}{\tudelta{\rho}{}}$ for program $\prog_2$, and $\eta \in \Contpar{t}{\tudelta{\rho}{}}{\tudelta{\eta}{}}$.
    By $\eta$ idempotent, the last assumption can be written as: $\forall c_1,c_2 \in C \sut \tudelta{\eta}{}(c_1, c_2) \leq t\tudelta{\rho}{}(c_1,c_2)$. Then we get the following derivations $\forall c_1,c_2 \in C$:
    \begin{align*}
        \tudelta{\rho}{}(\collSem{\prog_1;\prog_2}c_1, \collSem{\prog_1;\prog_2}c_2) &\;=\;  \text{[by definition of $\collSem{\prog_1;\prog_2}$]} \\
        \tudelta{\rho}{}(\collSem{\prog_2}\collSem{\prog_1}c_1, \collSem{\prog_2}\collSem{\prog_1}c_2) &\;\leq\;  \text{[by $\judge{k_2}{\tudelta{\eta}{}}{\prog_2}{\tudelta{\rho}{}}$]} \\
        k_2\tudelta{\eta}{}(\collSem{\prog_1}c_1, \collSem{\prog_1}c_2) &\;\leq\;  \text{[by $\eta \in \Contpar{t}{\tudelta{\rho}{}}{\tudelta{\eta}{}}$]} \\
        tk_2\tudelta{\rho}{}(\collSem{\prog_1}c_1, \collSem{\prog_1}c_2) &\;\leq\;  \text{[by $\judge{k_1}{\tudelta{\eta}{}}{\prog_1}{\tudelta{\rho}{}}$]} \\
        k_1tk_2\tudelta{\eta}{}(c_1, c_2) &\;\Lra\; \text{[by judgment definition]} \\
        \judge{k_1 t  k_2}{\tudelta{\eta}{}}{\prog_1 ; \prog_2}{\tudelta{\rho}{}}
    \end{align*}

    (\textbf{join}): Assume we have a derivation $\judge{k_1}{\tudelta{\eta}{}}{\prog_1}{\tudelta{\rho}{}}$ for program $\prog_1$, a derivation $\judge{k_2}{\tudelta{\eta}{}}{\prog_2}{\tudelta{\rho}{}}$ for program $\prog_2$, and $\+Bound(\tudelta{\rho}{},\bound)$.
    Then we get the following derivations $\forall c_1,c_2 \in C$:
    \begin{align*}
        \tudelta{\rho}{}(\collSem{\prog_1 \pjoin \prog_2}c_1, \collSem{\prog_1 \pjoin \prog_2}c_2) &\;=\;  \text{[by definition of $\collSem{\prog_1 \pjoin \prog_2}$]} \\
        \tudelta{\rho}{}(\collSem{\prog_1}c_1 \vee \collSem{\prog_2}c_1, \collSem{\prog_1}c_2 \vee \collSem{\prog_2}c_2) &\;\leq\;  \text{[by subadditivity of $\tudelta{\rho}{}$, from $\+Bound(\tudelta{\rho}{},\bound)$]} \\
        \tudelta{\rho}{}(\collSem{\prog_1}c_1, \collSem{\prog_1}c_2) \bound \tudelta{\rho}{}(\collSem{\prog_2}c_1, \collSem{\prog_2}c_2) &\;\leq\;  \text{[by $\judge{k_1}{\tudelta{\eta}{}}{\prog_1}{\tudelta{\rho}{}}$, $\judge{k_2}{\tudelta{\eta}{}}{\prog_2}{\tudelta{\rho}{}}$ and monotonicity of $\bound$]} \\
        (k_1 \tudelta{\eta}{}(c_1,c_2)) \bound (k_2 \tudelta{\eta}{}(c_1,c_2)) &\;\leq\;  \text{[by compatibility of $\bound$ with the product]} \\
        (k_1 \bound k_2) \tudelta{\eta}{}(c_1,c_2) &\;\Lra\; \text{[by judgment definition]} \\
        \judge{k_1\bound k_2}{\tudelta{\eta}{}}{\prog_1 \pjoin \prog_2}{\tudelta{\rho}{}}
    \end{align*}

    The two rules for Kleene iteration share the premises
    $\judge{k}{\tudelta{\eta}{}}{\prog}{\tudelta{\rho}{}}$,
    $\eta \in \Contpar{t}{\tudelta{\rho}{}}{\tudelta{\eta}{}}$,
    $\rho \in \Contpar{v}{\tudelta{\eta}{}}{\tudelta{\rho}{}}$ and
    $\+Bound(\tudelta{\rho}{},\bound)$, which we assume in the following. By
    Lemma~\ref{lem:cont-set}(iii), the two side conditions on the abstractions amount
    to $\tudelta{\eta}{}(c_1,c_2) \leq t\tudelta{\rho}{}(c_1,c_2)$ and
    $\tudelta{\rho}{}(c_1,c_2) \leq v\tudelta{\eta}{}(c_1,c_2)$ for all $c_1,c_2 \in C$.
    For $i \in \nat$ and $c \in C$, let
    $A_i(c) \definit \bigvee \{ \collSem{\prog}^n c \mid n \leq i \}$ denote the join of
    the first $i+1$ iterates, so that $A_0(c) = c$,
    $A_{i+1}(c) = A_i(c) \vee \collSem{\prog}^{i+1}c$, the sequence $(A_i(c))_{i \in \nat}$
    is an increasing chain, and $\collSem{\prog^\ast}c = \bigvee_{i \in \nat} A_i(c)$ by
    definition of $\collSem{\prog^\ast}$. We first establish two auxiliary lemmas.

    \begin{lem}[Iterates] \label{lem:iterates}
        For all $i \geq 1$ and $c_1,c_2 \in C$:
        $\tudelta{\rho}{}(\collSem{\prog}^i c_1, \collSem{\prog}^i c_2) \leq k^i t^{i-1} \tudelta{\eta}{}(c_1,c_2)$.
    \end{lem}
    \begin{proof}
        By induction on $i$. The case $i = 1$ is the premise
        $\judge{k}{\tudelta{\eta}{}}{\prog}{\tudelta{\rho}{}}$. For $i+1$, we get the
        following derivations $\forall c_1,c_2 \in C$:
        \begin{align*}
            \tudelta{\rho}{}(\collSem{\prog}^{i+1} c_1, \collSem{\prog}^{i+1} c_2) &\;=\; \text{[by definition of $\collSem{\prog}^{i+1}$]} \\
            \tudelta{\rho}{}(\collSem{\prog}\collSem{\prog}^{i} c_1, \collSem{\prog}\collSem{\prog}^{i} c_2) &\;\leq\; \text{[by $\judge{k}{\tudelta{\eta}{}}{\prog}{\tudelta{\rho}{}}$]} \\
            k\tudelta{\eta}{}(\collSem{\prog}^{i} c_1, \collSem{\prog}^{i} c_2) &\;\leq\; \text{[by $\eta \in \Contpar{t}{\tudelta{\rho}{}}{\tudelta{\eta}{}}$]} \\
            kt\tudelta{\rho}{}(\collSem{\prog}^{i} c_1, \collSem{\prog}^{i} c_2) &\;\leq\; \text{[by inductive hypothesis]} \\
            kt \, k^i t^{i-1} \tudelta{\eta}{}(c_1, c_2) &\;=\; k^{i+1} t^{i} \tudelta{\eta}{}(c_1, c_2) \qedhere
        \end{align*}
    \end{proof}

    \begin{lem}[Finite Joins of Iterates] \label{lem:finite-joins}
        For all $m \in \nat$ and $c_1,c_2 \in C$:
        $\tudelta{\rho}{}(A_m(c_1), A_m(c_2)) \leq K_m \tudelta{\eta}{}(c_1,c_2)$.
    \end{lem}
    \begin{proof}
        By induction on $m$. For $m = 0$, we have
        $\tudelta{\rho}{}(A_0(c_1),A_0(c_2)) = \tudelta{\rho}{}(c_1,c_2) \leq v \tudelta{\eta}{}(c_1,c_2) = K_0 \tudelta{\eta}{}(c_1,c_2)$
        by $\rho \in \Contpar{v}{\tudelta{\eta}{}}{\tudelta{\rho}{}}$. For $m+1$, we get
        the following derivations $\forall c_1,c_2 \in C$:
        \begin{align*}
            \tudelta{\rho}{}(A_{m+1}(c_1), A_{m+1}(c_2)) &\;=\; \text{[by definition of $A_{m+1}$]} \\
            \tudelta{\rho}{}(A_m(c_1) \vee \collSem{\prog}^{m+1}c_1,\, A_m(c_2) \vee \collSem{\prog}^{m+1}c_2) &\;\leq\; \text{[by subadditivity of $\tudelta{\rho}{}$, from $\+Bound(\tudelta{\rho}{},\bound)$]} \\
            \tudelta{\rho}{}(A_m(c_1), A_m(c_2)) \bound \tudelta{\rho}{}(\collSem{\prog}^{m+1}c_1, \collSem{\prog}^{m+1}c_2) &\;\leq\; \text{[by ind.\ hyp., Lemma~\ref{lem:iterates}, $\bound$ monotone]} \\
            (K_m \tudelta{\eta}{}(c_1,c_2)) \bound (k^{m+1} t^{m} \tudelta{\eta}{}(c_1,c_2)) &\;\leq\; \text{[by compatibility of $\bound$ with the product]} \\
            (K_m \bound k^{m+1} t^{m}) \tudelta{\eta}{}(c_1,c_2) &\;=\; \text{[by definition of $K_{m+1}$ and associativity of $\bound$]} \\
            K_{m+1} \tudelta{\eta}{}(c_1,c_2) &\qedhere
        \end{align*}
    \end{proof}

    (\textbf{bound-star}): Assume moreover $\*Bound(\prog^\ast, m)$. Then, for all $c_1,c_2 \in C$:
    \begin{align*}
        \tudelta{\rho}{}(\collSem{\prog^\ast} c_1, \collSem{\prog^\ast} c_2) &\;=\; \text{[by $\*Bound(\prog^\ast, m)$ and definition of $A_m$]} \\
        \tudelta{\rho}{}(A_m(c_1), A_m(c_2)) &\;\leq\; \text{[by Lemma~\ref{lem:finite-joins}]} \\
        K_m \tudelta{\eta}{}(c_1, c_2) &\;\Lra\; \text{[by judgment definition]} \\
        \judge{K_m}{\tudelta{\eta}{}}{\prog^\ast}{\tudelta{\rho}{}}
    \end{align*}

    (\textbf{star}): Assume moreover $\omega\text{-}\mathrm{Cont}(\tudelta{\rho}{})$. Then, for all $c_1,c_2 \in C$:
    \begin{align*}
        \tudelta{\rho}{}(\collSem{\prog^\ast} c_1, \collSem{\prog^\ast} c_2) &\;=\; \text{[by definition of $\collSem{\prog^\ast}$ and of $A_i$]} \\
        \tudelta{\rho}{}\big(\textstyle\bigvee_{i \in \nat} A_i(c_1), \bigvee_{i \in \nat} A_i(c_2)\big) &\;\leq\; \text{[by $\omega\text{-}\mathrm{Cont}(\tudelta{\rho}{})$, as $(A_i(c_1))_i$ and $(A_i(c_2))_i$ are increasing chains]} \\
        \textstyle\sup_{i \in \nat} \tudelta{\rho}{}(A_i(c_1), A_i(c_2)) &\;\leq\; \text{[by Lemma~\ref{lem:finite-joins}]} \\
        \textstyle\sup_{i \in \nat} K_i \tudelta{\eta}{}(c_1, c_2) &\;\leq\; \text{[by $K_i \leq K_\infty$ and monotonicity of the product]} \\
        K_\infty \tudelta{\eta}{}(c_1, c_2) &\;\Lra\; \text{[by judgment definition]} \\
        \judge{K_\infty}{\tudelta{\eta}{}}{\prog^\ast}{\tudelta{\rho}{}} &\qedhere
    \end{align*}
\end{proof}

\begin{figure}[t]
    \centering
    \resizebox{\textwidth}{!}{$
        \mathcal{D}_1 \;=\;
        \infer[(\textbf{seq})]{ \judge{1}{\dpath{\Sign}}{x < 0? \,;\, x := 0}{\dpath{\Sign}} }{
            \infer[(\textbf{base})]{ \judge{1}{\dpath{\Sign}}{x < 0?}{\dpath{\Sign}} }{ \collSem{x<0?} \in \Contpar{1}{\dpath{\Sign}}{\dpath{\Sign}} }
            &
            \infer[(\textbf{base})]{ \judge{1}{\dpath{\Sign}}{x := 0}{\dpath{\Sign}} }{ \collSem{x:=0} \in \Contpar{1}{\dpath{\Sign}}{\dpath{\Sign}} }
            &
            \Sign \in \Contpar{1}{\dpath{\Sign}}{\dpath{\Sign}}
        }
    $}
    \\[1.5em]
    \resizebox{\textwidth}{!}{$
        \mathcal{D}_2 \;=\;
        \infer[(\textbf{seq})]{ \judge{1}{\dpath{\Sign}}{x \geq 0? \,;\, \skipc}{\dpath{\Sign}} }{
            \infer[(\textbf{base})]{ \judge{1}{\dpath{\Sign}}{x \geq 0?}{\dpath{\Sign}} }{ \collSem{x \geq 0?} \in \Contpar{1}{\dpath{\Sign}}{\dpath{\Sign}} }
            &
            \infer[(\textbf{base})]{ \judge{1}{\dpath{\Sign}}{\skipc}{\dpath{\Sign}} }{ \collSem{\skipc} \in \Contpar{1}{\dpath{\Sign}}{\dpath{\Sign}} }
            &
            \Sign \in \Contpar{1}{\dpath{\Sign}}{\dpath{\Sign}}
        }
    $}
    \\[1.5em]
    $
        \infer[(\textbf{join})]{ \judge{1 + 1}{\dpath{\Sign}}{(x < 0? \,;\, x := 0) \pjoin (x \geq 0? \,;\, \skipc)}{\dpath{\Sign}} }{
            \deduce{ \judge{1}{\dpath{\Sign}}{x < 0? \,;\, x := 0}{\dpath{\Sign}} }{ \mathcal{D}_1 }
            &
            \deduce{ \judge{1}{\dpath{\Sign}}{x \geq 0? \,;\, \skipc}{\dpath{\Sign}} }{ \mathcal{D}_2 }
            &
            \+Bound(\dpath{\Sign}, +)
        }
    $
    \caption{Derivation of $\judge{2}{\dpath{\Sign}}{\progp{ReLU}}{\dpath{\Sign}}$ in Example~\ref{ex:relu2}.}
    \label{fig:relu-derivation}
\end{figure}

\begin{exa} \label{ex:relu2}
    Consider the program
    \[
        \progp{ReLU} \;=\; (x < 0? \,;\, x := 0) \pjoin (x \geq 0? \, ;\, \skipc)
    \]
    implementing the $\progp{ReLU}$ function used in artificial neural
    networks~\cite{NairH10}, which maps negative inputs to $0$ and leaves the other
    inputs unchanged. We take the domain of signs $\Sign \in \uco(\wp(\integer))$ of
    Example~\ref{ex:sign} as both input and output abstraction, $\eta = \rho = \Sign$,
    together with the path-length distance $\dpath{}$ on $\Sign(\wp(\integer))$ of
    Example~\ref{ex:path-len}. Writing $\mathsf{blocks}(S)$ for the set of blocks of
    the sign partition intersecting $S$ (cf.~Example~\ref{ex:sign}), we have that $\dpath{\Sign}(S_1,S_2)$ is the
    number of blocks in $\mathsf{blocks}(S_2) \smallsetminus \mathsf{blocks}(S_1)$ when
    $\mathsf{blocks}(S_1) \subseteq \mathsf{blocks}(S_2)$, and $\infty$ otherwise. We
    derive a judgment for the collecting semantics
    $\collSem{\progp{ReLU}} \colon \wp(\integer) \ra \wp(\integer)$ w.r.t.\
    $\tuple{\dpath{\Sign},\dpath{\Sign}}$. The full derivation is shown in Figure~\ref{fig:relu-derivation}.

    \emph{Basic commands.} The guard $x < 0?$ removes from its input all the values
    that are not negative, hence
    $\mathsf{blocks}(\collSem{x<0?}S) = \mathsf{blocks}(S) \cap \{\integer_{<0}\}$.
    If $\mathsf{blocks}(S_1) \subseteq \mathsf{blocks}(S_2)$, the only block that
    $\collSem{x<0?}S_2$ may have in addition to $\collSem{x<0?}S_1$ is $\integer_{<0}$,
    and in that case $\integer_{<0}$ is also a block that $S_2$ has in addition to
    $S_1$. Therefore $\collSem{x<0?} \in \Contpar{1}{\dpath{\Sign}}{\dpath{\Sign}}$
    and, by $(\textbf{base})$, $\judge{1}{\dpath{\Sign}}{x<0?}{\dpath{\Sign}}$. The
    same argument, with the blocks $\{0\}$ and $\integer_{>0}$, gives
    $\judge{1}{\dpath{\Sign}}{x \geq 0?}{\dpath{\Sign}}$.
    The assignment $x := 0$ maps every non-empty
    set to $\{0\}$ and $\varnothing$ to $\varnothing$: the only pairs of inputs at
    finite distance with different outputs are $\varnothing$ and a non-empty $S$, for
    which $\dpath{\Sign}(\varnothing,\{0\}) = 1 \leq \dpath{\Sign}(\varnothing,S)$.
    Hence $\judge{1}{\dpath{\Sign}}{x := 0}{\dpath{\Sign}}$ by $(\textbf{base})$, and
    $\judge{1}{\dpath{\Sign}}{\skipc}{\dpath{\Sign}}$ as observed after the
    $(\textbf{base})$ rule.

    \emph{Composition.} Since $\eta = \rho$, Lemma~\ref{lem:cont-set}(iii) gives
    $\Sign \in \Contpar{1}{\dpath{\Sign}}{\dpath{\Sign}}$, so that $(\textbf{seq})$
    derives $\judge{1}{\dpath{\Sign}}{x<0? \,;\, x:=0}{\dpath{\Sign}}$ and
    $\judge{1}{\dpath{\Sign}}{x\geq 0? \,;\, \skipc}{\dpath{\Sign}}$, with constant
    $1 \cdot 1 \cdot 1 = 1$.

    \emph{Join.} By Proposition~\ref{prop:incl-bound}, $\+Bound(\dpath{\Sign},+)$ holds, hence
    $(\textbf{join})$ derives $\judge{2}{\dpath{\Sign}}{\progp{ReLU}}{\dpath{\Sign}}$
    and, by Theorem~\ref{thm:soundness}, $\collSem{\progp{ReLU}}$ satisfies $2$-ALC
    w.r.t.\ $\tuple{\dpath{\Sign},\dpath{\Sign}}$.

    The derived constant is sound but not the least one: $\collSem{\progp{ReLU}}$
    actually satisfies $1$-ALC. Indeed, $\mathsf{blocks}(\collSem{\progp{ReLU}}S)$
    contains $\{0\}$ if and only if $S \cap \integer_{\leq 0} \neq \varnothing$, and
    contains $\integer_{>0}$ if and only if $S \cap \integer_{>0} \neq \varnothing$;
    hence, when $\mathsf{blocks}(S_1) \subseteq \mathsf{blocks}(S_2)$, each block gained
    by the output on $S_2$ w.r.t.\ the output on $S_1$ corresponds to a distinct block
    gained by $S_2$ w.r.t.\ $S_1$. The imprecision is due to the $(\textbf{join})$
    rule, which adds up the constants of the two branches: the two branches are
    sensitive to disjoint sets of input blocks ($\integer_{<0}$ for the first one,
    $\{0\}$ and $\integer_{>0}$ for the second one), so that an additional input
    block can change the output of at most one of them, a fact that the uniform bound
    $+$ cannot exploit.
    $\qedex$
\end{exa}

\begin{figure}[t]
    \centering
    $
        \infer[(\textbf{star})]{ \judge{K_\infty}{\delta_{\mathrm{H}}^{\Int}}{\progp{S}^\ast}{\delta_{\mathrm{H}}^{\Int}} }{
            \infer[(\textbf{base})]{ \judge{\nicefrac{1}{4}}{\delta_{\mathrm{H}}^{\Int}}{\progp{S}}{\delta_{\mathrm{H}}^{\Int}} }{ \collSem{\progp{S}} \in \Contpar{\nicefrac{1}{4}}{\delta_{\mathrm{H}}^{\Int}}{\delta_{\mathrm{H}}^{\Int}} }
            &
            \deduce{ \Int \in \Contpar{1}{\delta_{\mathrm{H}}^{\Int}}{\delta_{\mathrm{H}}^{\Int}} }{ \Int \in \Contpar{1}{\delta_{\mathrm{H}}^{\Int}}{\delta_{\mathrm{H}}^{\Int}} }
            &
            \deduce{ \omega\text{-}\mathrm{Cont}(\delta_{\mathrm{H}}^{\Int}) }{ \+Bound(\delta_{\mathrm{H}}^{\Int}, \max) }
        }
    $
    \\[1em]
    where $K_\infty = \max(1, \sup_{i \geq 1} (\nicefrac{1}{4})^i) = 1$.
    \caption{Derivation of $\judge{1}{\delta_{\mathrm{H}}^{\Int}}{\progp{S}^\ast}{\delta_{\mathrm{H}}^{\Int}}$ in Example~\ref{ex:scale-star}.}
    \label{fig:scale-derivation}
\end{figure}

\begin{exa}[Iterated Scaling] \label{ex:scale-star}
    Consider the program
    \[
        \progp{S} =\; x := x / 4
    \]
    which scales its input by a
    factor $\nicefrac{1}{4}$, and its Kleene closure $\progp{S}^\ast$, which iterates
    the scaling an arbitrary number of times. In order to have a domain of values
    closed under division, we take $\mathbb{V} = \raz$, so that $\progp{S}$ is
    expressible in $\imp$ as $x := \nicefrac{1}{4} * x$, and, as in
    Example~\ref{ex:R-ext}, we let $C$ be the set of non-empty subsets of $\raz$.
    As input and output abstraction we consider here the interval abstraction over rationals
     $\Int(T) = \{ q \in \raz \mid \inf T \leq q \leq \sup T \}$,
    where infima and suprema are taken in $\real \cup \{-\infty,+\infty\}$, so that
    an element of $\Int(C)$ is determined by its two bounds. As distance we take
    the Hausdorff distance between intervals, defined on $\Int(\wp(\raz))$ by
    \[
        \delta_{\mathrm{H}}(I_1,I_2) \definit \max(|\inf I_1 - \inf I_2|, |\sup I_1 - \sup I_2|)
    \]
    with the conventions $|\infty - \infty| = 0$ and $|r - \infty| = \infty$ for
    $r \in \real$, which is a metric on $\Int(\wp(\raz))$. Hence
    $\delta_{\mathrm{H}}^{\Int}(T_1,T_2)$ measures the largest displacement of the two
    bounds of the interval abstractions of $T_1$ and $T_2$. The derivation of a
    judgment for $\collSem{\progp{S}^\ast}$ is shown in Figure~\ref{fig:scale-derivation},
    and described as follows.

    \emph{Basic command.}
    Since $\collSem{\progp{S}}T = \{ \nicefrac{q}{4} \mid q \in T \}$,
    we have $\inf \collSem{\progp{S}}T = (\inf T)/4$ and
    $\sup \collSem{\progp{S}}T = (\sup T)/4$, hence
    $\delta_{\mathrm{H}}^{\Int}(\collSem{\progp{S}}T_1, \collSem{\progp{S}}T_2) =
    \nicefrac{1}{4} \, \delta_{\mathrm{H}}^{\Int}(T_1,T_2)$ for all $T_1,T_2 \in C$:
    $\progp{S}$ is a contraction, and $(\textbf{base})$ derives
    $\judge{\nicefrac{1}{4}}{\delta_{\mathrm{H}}^{\Int}}{\progp{S}}{\delta_{\mathrm{H}}^{\Int}}$.

    \emph{Side conditions.} Since $\eta = \rho = \Int$, Lemma~\ref{lem:cont-set}(iii)
    gives $t = v = 1$. The operation $\max$ is an $\oplus$-bound for
    $\delta_{\mathrm{H}}^{\Int}$: it is associative, commutative and monotone, it
    satisfies $\max(k_1 d, k_2 d) = \max(k_1,k_2) \, d$, and it is subadditive since
    $\inf (X \cup Y) = \min(\inf X, \inf Y)$ and $\sup (X \cup Y) = \max(\sup X, \sup Y)$,
    where $|\min(a,b) - \min(a',b')| \leq \max(|a - a'|, |b - b'|)$ and likewise for
    $\max$. Finally, $\delta_{\mathrm{H}}^{\Int}$ is $\omega$-continuous: for an
    increasing chain $(X_n)_{n \in \nat}$ in $C$, $\inf \bigcup_n X_n = \lim_n \inf X_n$
    and $\sup \bigcup_n X_n = \lim_n \sup X_n$, so that the bounds of
    $\Int(\bigcup_n X_n)$ are the limits of the bounds of $\Int(X_n)$, and
    $\delta_{\mathrm{H}}^{\Int}(\bigcup_n X_n, \bigcup_n Y_n) \leq
    \sup_n \delta_{\mathrm{H}}^{\Int}(X_n,Y_n)$ follows by continuity of the absolute
    value.

    \emph{Iteration.} By Definition~\ref{def:Kn} with $\bound = \max$, $k = \nicefrac{1}{4}$
    and $t = v = 1$, we get $K_\infty = \max(1, \sup_{i \geq 1} (\nicefrac{1}{4})^i) = 1$,
    hence $(\textbf{star})$ derives
    $\judge{1}{\delta_{\mathrm{H}}^{\Int}}{\progp{S}^\ast}{\delta_{\mathrm{H}}^{\Int}}$
    and, by Theorem~\ref{thm:soundness}, $\collSem{\progp{S}^\ast}$ satisfies $1$-ALC
    w.r.t.\ $\tuple{\delta_{\mathrm{H}}^{\Int},\delta_{\mathrm{H}}^{\Int}}$. Note that this
    derived abstract Lipschitz constant is the least one: for $\Int(T) = [a,b]$ we have
    $\Int(\collSem{\progp{S}^\ast} T) = [\min(a,0), \max(b,0)]$, since the iterates
    $\nicefrac{q}{4^n}$ approach $0$ from the side of $q$, and
    $\delta_{\mathrm{H}}^{\Int}(\collSem{\progp{S}^\ast}\{1\}, \collSem{\progp{S}^\ast}\{2\})
    = \delta_{\mathrm{H}}([0,1],[0,2]) = 1 = \delta_{\mathrm{H}}^{\Int}(\{1\},\{2\})$.
    Moreover it is worth noting that $+$ is an $\oplus$-bound for $\delta_{\mathrm{H}}^{\Int}$ as well, since
    $\max(a,b) \leq a + b$, but with $\bound = +$ the same derivation yields the
    geometric series $K_\infty = 1 + \sum_{i \geq 1} (\nicefrac{1}{4})^i = \nicefrac{4}{3}$,
    a sound but not optimal constant. Finally, $(\textbf{bound-star})$ is not
    applicable here, since no $m \in \nat$ satisfies $\*Bound(\progp{S}^\ast, m)$. %
    $\qedex$
\end{exa}

\begin{figure}[t]
    \centering
    \resizebox{\textwidth}{!}{$
        \mathcal{D} \;=\;
        \infer[(\textbf{seq})]{ \judge{1}{\dpath{\Sign}}{x > 0? \,;\, x := -x}{\dpath{\Sign}} }{
            \infer[(\textbf{base})]{ \judge{1}{\dpath{\Sign}}{x > 0?}{\dpath{\Sign}} }{ \collSem{x>0?} \in \Contpar{1}{\dpath{\Sign}}{\dpath{\Sign}} }
            &
            \infer[(\textbf{base})]{ \judge{1}{\dpath{\Sign}}{x := -x}{\dpath{\Sign}} }{ \collSem{x:=-x} \in \Contpar{1}{\dpath{\Sign}}{\dpath{\Sign}} }
            &
            \Sign \in \Contpar{1}{\dpath{\Sign}}{\dpath{\Sign}}
        }
    $}
    \\[1.5em]
    $
        \infer[(\textbf{bound-star})]{ \judge{K_1}{\dpath{\Sign}}{\progp{N}^\ast}{\dpath{\Sign}} }{
            \deduce{ \judge{1}{\dpath{\Sign}}{x > 0? \,;\, x := -x}{\dpath{\Sign}} }{ \mathcal{D} }
            &
            \deduce{ \Sign \in \Contpar{1}{\dpath{\Sign}}{\dpath{\Sign}} }{ \Sign \in \Contpar{1}{\dpath{\Sign}}{\dpath{\Sign}} }
            &
            \deduce{ \*Bound(\progp{N}^\ast, 1) }{ \+Bound(\dpath{\Sign}, +) }
        }
    $
    \\[1em]
    where $K_1 = 1 + 1 = 2$.
    \caption{Derivation of $\judge{2}{\dpath{\Sign}}{\progp{N}^\ast}{\dpath{\Sign}}$ in Example~\ref{ex:neg-bound-star}.}
    \label{fig:neg-derivation}
\end{figure}

\begin{exa}[Bounded Iteration] \label{ex:neg-bound-star}
    In the setting of Example~\ref{ex:relu2}, namely $\eta = \rho = \Sign$ on
    $\wp(\integer)$ with the path-length distance $\dpath{\Sign}$, consider the
    program
    \[
        \progp{N} \;=\; (x > 0? \,;\, x := -x)
    \]
    which negates the positive
    values of its input and discards the other ones, and its Kleene closure
    $\progp{N}^\ast$. The derivation of a judgment for $\collSem{\progp{N}^\ast}$ is
    shown in Figure~\ref{fig:neg-derivation}.

    \emph{Basic commands.} The guard $x > 0?$ is $1$-ALC by the same argument used
    for the guards of Example~\ref{ex:relu2}.
    The assignment $x := -x$ swaps the blocks $\integer_{<0}$ and
    $\integer_{>0}$ and leaves $\{0\}$ unchanged, hence
    $\dpath{\Sign}(\collSem{x:=-x}S_1, \collSem{x:=-x}S_2) = \dpath{\Sign}(S_1,S_2)$:
     it is $1$-ALC. By $(\textbf{seq})$ with
    $t = 1$, we get $\judge{1}{\dpath{\Sign}}{\progp{N}}{\dpath{\Sign}}$.

    \emph{Bounded iteration.} Since $\collSem{\progp{N}}S \subseteq \integer_{<0}$ for
    every $S \in \wp(\integer)$, a second iteration of $\progp{N}$ yields the empty set:
    $\collSem{\progp{N}}^2 S = \varnothing$, and therefore
    $\collSem{\progp{N}^\ast}S = S \cup \collSem{\progp{N}}S$, i.e.,
    $\*Bound(\progp{N}^\ast, 1)$ holds. With $\bound = +$, which is an $\oplus$-bound
    for $\dpath{\Sign}$ by Proposition~\ref{prop:incl-bound}, and $t = v = 1$,
    Definition~\ref{def:Kn} gives $K_1 = v + k = 2$, hence $(\textbf{bound-star})$
    derives $\judge{2}{\dpath{\Sign}}{\progp{N}^\ast}{\dpath{\Sign}}$ and, by
    Theorem~\ref{thm:soundness}, $\collSem{\progp{N}^\ast}$ satisfies $2$-ALC w.r.t.\
    $\tuple{\dpath{\Sign},\dpath{\Sign}}$. The derived Lipchitz constant $2$ is precise:
    $\dpath{\Sign}(\varnothing,\{1\}) = 1$, whereas
    $\collSem{\progp{N}^\ast}\{1\} = \{1,-1\}$ has two blocks more than
    $\collSem{\progp{N}^\ast}\varnothing = \varnothing$. Two remarks are in order.
    First, the $(\textbf{star})$ rule would be applicable as well, but uninformative:
    since $kt = 1$, it yields $K_\infty = 1 + \sum_{i \geq 1} 1 = \infty$, which shows
    the value of the information carried by $\*Bound$. Second, differently from
    Example~\ref{ex:scale-star}, $\max$ is not an $\oplus$-bound for $\dpath{\Sign}$:
    the blocks added by two joined sets add up, e.g.,
    $\dpath{\Sign}(\varnothing,\{-2\}) = \dpath{\Sign}(\varnothing,\{0\}) = 1$ whereas
    $\dpath{\Sign}(\varnothing \cup \varnothing, \{-2\} \cup \{0\}) = 2$. The suitable
    $\oplus$-bound thus depends on the distance at hand: $+$ for distances that count
    elements or blocks, such as $\dincl{}$ and $\dpath{}$, and $\max$ for distances
    that measure displacements, such as $\delta_{\mathrm{H}}$.
    $\qedex$
\end{exa}

As a direct consequence of Theorem~\ref{thm:soundness}, when both abstractions are
the identity, $\eta = \rho = \id$, the deductive system of Figure~\ref{fig:axRulesP}
becomes a proof system for the Lipschitz continuity of programs w.r.t.\
$\tuple{\delta,\delta}$ (Definition~\ref{def:lc}), since
$\Contpar{k}{\tudelta{\id}{}}{\tudelta{\id}{}}$ is the set of all $k$-LC functions
(Lemma~\ref{lem:cont-set}(ii)). In this case the side conditions on the abstractions
become trivial: $\id \in \Contpar{1}{\tudelta{\id}{}}{\tudelta{\id}{}}$ holds by
Lemma~\ref{lem:cont-set}(iii), hence $t = v = 1$ in $(\textbf{seq})$,
$(\textbf{bound-star})$ and $(\textbf{star})$, and $(\textbf{weaken})$ is never
needed as long as derivations only involve the identity abstraction. The side
conditions on the distance, namely $\+Bound(\delta,\bound)$ and
$\omega\text{-}\mathrm{Cont}(\delta)$, as well as the bound $\*Bound(\prog^\ast,m)$ on
the number of iterations, are instead still required: they express what is needed to
compose Lipschitz constants across non-deterministic choice and iteration,
independently of any abstraction.

\begin{cor}
        $\judge{k}{\tudelta{\id}{}}{\prog}{\tudelta{\id}{}} \;\Ra\; \collSem{\prog} \mbox{ is $k$-LC w.r.t. } \tuple{\delta,\delta}$
\end{cor}
\begin{proof}
    By Theorem~\ref{thm:soundness} and Lemma~\ref{lem:cont-set}(ii).
\end{proof}

\section{Conclusion} \label{sec:concl}

Abstract Lipschitz continuity (ALC) requires the properties of the outputs of a function, e.g., a program semantics, to change at most linearly with the properties of its inputs: the distance between the abstractions of two outputs is at most $k$ times the distance between the abstractions of the corresponding inputs. Like Lipschitz continuity, it is a stability guarantee, but on the properties of data: input variations that do not affect the property of interest, such as reordering the bytes of a file in Example~\ref{ex:byte-histogram}, do not count as perturbations, whereas the relevant ones are measured by a distance tailored to the abstract domain, which need not be a metric. Keeping the property-based and the quantitative approximations apart has two advantages. First, a single proof of ALC, obtained compositionally through the deductive system of Section~\ref{sec:proof}, yields guarantees for extensional properties from different areas, namely partial completeness, partial abstract non-interference and abstract robustness (Section~\ref{sec:alc4prog}), as well as Lipschitz continuity when no abstraction is involved. Second, it makes explicit the assumptions on abstractions and distances these guarantees rely on, such as (\textit{iff-identity}), which turns bounds on distances into equalities of properties.

Among the most important consequences, we showed that every monotone function satisfying ALC w.r.t.\ an abstraction $\rho$, in particular every program semantics, admits a complete best correct approximation on $\rho$ for suitable distances, namely $\tpreord{}$-compatible ones satisfying (\textit{iff-identity}) on the abstract domain (Corollary~\ref{cor:alc-compl}): ALC programs admit an analysis on $\rho$ that introduces no imprecision. Moreover, when the output distance satisfies (\textit{iff-identity}), ALC functions satisfy abstract non-interference w.r.t.\ the same abstractions (Corollary~\ref{cor:alc-pani}): inputs with the same property are mapped to outputs with the same property, so that nothing beyond the input property flows to the observed output. Both facts stem from ALC mapping inputs at distance $0$ to outputs at distance $0$, whatever the constant $k$; with different input abstractions, ALC still guarantees partial completeness and partial abstract non-interference, with an imprecision proportional to $k$. Abstract robustness of classifiers also follows from ALC, for any perturbation radius, provided that the constant is null, $0$-ALC being in general strictly stronger. We conclude by discussing some directions for future work.

As stated in Section~\ref{sec:prog}, basic commands could be instantiated
differently, for instance with probabilistic or non-deterministic assignments. In
this work we considered deterministic assignments only, and $\pjoin$ is interpreted
as the semantic join. Extending the language with probabilistic choice, and the deductive system with a
corresponding rule, is a natural next step: we expect the $(\textbf{join})$ rule to
carry over, with an $\oplus$-bound tailored to standard distances between
distributions such as the total variation or the Kantorovich metric, whereas the
identification of meaningful abstractions of distributions, i.e., ucos on spaces of
distributions playing the role of $\rho$, is an open question.

The proposed ALC notion is a \emph{global} property, in the sense that it is
universally quantified over all inputs. As a future work, we plan to formalize its
\emph{local} version, requiring ALC on a strict subset of inputs only, and to study
its relation with other local properties in the context of abstract
interpretation~\cite{BruniGGR23,Campion2024popl,CampionPOPL2026}.
Dropping the universal quantification may
invalidate the correlations established in Sections~\ref{sec:pcompl}
and~\ref{sec:arob} for the global notions, and reasoning about local properties
is more challenging for the deductive system, whose judgments would have to carry
assumptions on the inputs reaching each program point, in the style of local
completeness logic~\cite{BruniGGR23}.
A related question is whether ALC can be established on a restricted set of
canonical inputs. In~\cite{CampionDGMU26}, (partial) completeness is reduced to the
generators of a generator-dense abstraction, namely the minimal inputs having each
abstract property: it would be interesting to understand under which conditions ALC can be established on generators only.

Another direction consists in considering weaker abstraction notions such as weak closures~\cite{MastroeniP23} or $\alpha$-relations~\cite{LemerreSAS2026}, which are able to
formalize properties that do not admit a best abstraction function, such as the
domain of convex polyhedra~\cite{Grunbaum67}.
In~\cite{MastroeniP23}, a correspondence is established, under
suitable assumptions, between completeness in abstract interpretation and abstract
non-interference in language-based
security~\cite{GiacobazziM04,GM:ANI2018,M05Aplas,GMadj10}.
Whether the correspondence extends to weak closures, and to other quantitative program properties such as (quantitative) input data usage~\cite{Urban18,MazzucatoCU24,SASMazzucatoCU24} and the notions in~\cite{Mastroeni24,imvmcai25,MPsac19,MP25}, deserves further investigation.

As another interesting line of work, we aim to investigate the correlation between ALC and \emph{differential privacy}~\cite{Dwork-ICALP-2006}, a standard approach to ensure privacy in data-driven software. To enforce differential privacy, traditional methodologies, e.g., based on type systems, first track \emph{sensitivity}~\cite{sensitivity}. Once an upper bound on sensitivity is established on a program, the appropriate amount of random noise is injected into a randomized function to guarantee compliance with differential privacy~\cite{Dwork-TC-2006}. Because sensitivity in its general form corresponds to Lipschitz Continuity, differential privacy can be naturally modeled and analyzed within ALC. Indeed, ALC can provide a semantic framework to automatically reason about probability distributions and sensitivity bounds by combining quantitative metrics with abstract interpretation. Crucially, by integrating property-based abstractions, ALC can allow us to define an \emph{abstract version of differential privacy} that accounts for a weaker notion of the attacker model with limited observation power, where adversaries can only perceive coarse-grained or property-based outputs rather than exact values. Still, incorporating abstractions into a distance metric on probability spaces is not immediate. We leave for future work the modeling of differential privacy inside the ALC framework and the formalizaztion of differential privacy accounting for input/output abstractions.

The deductive system of Section~\ref{sec:proof} isolates two properties of the
output distance that make ALC compositional across non-deterministic choice and
iteration: the existence of an $\oplus$-bound (Definition~\ref{def:+bound}) and
$\omega$-continuity (Definition~\ref{def:wcont}). We proved that both hold for the
inclusion and path-length distances combined with additive abstractions
(Propositions~\ref{prop:incl-bound} and~\ref{prop:wcont}), and
Examples~\ref{ex:scale-star} and~\ref{ex:neg-bound-star} show that the suitable
$\oplus$-bound depends on the nature of the distance ($+$ for distances that count
elements, $\max$ for distances that measure displacements).
A general characterization of the pre-metrics
and abstractions admitting $\oplus$-bounds and $\omega$-continuity, possibly in
terms of the lattice structure of the abstract domain, would clarify the scope of
the deductive system and is left for future work.

Finally, Theorem~\ref{thm:soundness} shows that the deductive system is sound, but
it is not complete: the least derivable abstract Lipschitz constant may be strictly
greater than the actual one, or no finite constant may be derivable at all. We
identified three sources of imprecision. First, the $(\textbf{base})$ rule requires
basic commands to be ALC, which for Boolean guards holds only when the abstraction
is complete for the guard, as for the domain of signs in Example~\ref{ex:relu2},
and fails for the interval abstraction. Second, the $(\textbf{join})$ rule combines
the constants of the two branches through a uniform $\oplus$-bound, which cannot
exploit the fact that the branches may be sensitive to disjoint parts of the input,
as in Example~\ref{ex:relu2}. Third, the rules for Kleene iteration are informative
only when the body is non-expansive or a bound on the number of iterations is
available, as shown by Examples~\ref{ex:scale-star} and~\ref{ex:neg-bound-star}.
Natural directions towards a (relatively) complete deductive system, together with
the automatic inference of abstract Lipschitz constants, are refined rules: a
$(\textbf{seq})$ rule allowing different intermediate abstractions, as in
Lemma~\ref{lem:cont-set}(v), instead of reaching common ones through
$(\textbf{weaken})$, and $\oplus$-bounds sensitive to the guards of the branches.

\bibliographystyle{alphaurl}
\bibliography{biblio}

\end{document}